\documentclass[fleqn,usenatbib]{mnras}
\usepackage{newtxtext,newtxmath}
\usepackage[T1]{fontenc}
\usepackage{ae,aecompl}
\usepackage{epsfig,rotating}
\usepackage{graphicx}
\usepackage{amsmath, amssymb}
\usepackage{amsfonts}
\usepackage{mathrsfs}
\usepackage{mathtools}
\usepackage{hyperref}
\usepackage[utf8]{inputenc}

\title[Does slow and steady win the race?]{Does slow and steady win the race? Investigating feedback processes in giant molecular clouds}
\author[Garratt-Smithson et al.]{
Lilian Garratt-Smithson$^{1,2}$\thanks{lilian.garratt-smithson@uwa.edu.au},
Graham A. Wynn$^{1}$,
Chris Power$^{2}$ and
C. J. Nixon$^{1}$\\
$^{1}$Theoretical Astrophysics Group, Department of Physics \& Astronomy, University of Leicester, Leicester, LE1 7RH, UK \\
$^{2}$International Centre for Radio Astronomy Research, University of Western Australia, 35 Stirling Highway, Crawley, \\ 
Western Australia 6009, Australia
} 

\date{\today.}
\pubyear{2018}
\pagerange{\pageref{firstpage}--\pageref{lastpage}}
\begin{document}
\label{firstpage}
\maketitle

\begin{abstract}
We investigate the effects of gradual heating on the evolution of turbulent molecular clouds of mass $2\times 10^6$ M$_\odot$ and virial parameters ranging between $0.7-1.2$. This gradual heating represents the energy output from processes such as winds from massive stars or feedback from High Mass X-ray binaries (HMXBs), contrasting the impulsive energy injection from supernovae (SNe). For stars with a mass high enough that their lifetime is shorter than the life of the cloud, we include a SN feedback prescription. Including both effects, we investigate the interplay between slow and fast forms of feedback and their effectiveness at triggering/suppressing star formation. We find that SN feedback can carve low density chimneys in the gas, offering a path of least resistance for the energy to escape. Once this occurs the more stable, but less energetic, gradual feedback is able to keep the chimneys open. By funneling the hot destructive gas away from the centre of the cloud, chimneys can have a positive effect on both the efficiency and duration of star formation. Moreover, the critical factor is the number of high mass stars and SNe (and any subsequent HMXBs) active within the free-fall time of each cloud. This can vary from cloud to cloud due to the stochasticity of SN delay times and in HMXB formation. However, the defining factor in our simulations is the efficiency of the cooling, which can alter the Jeans mass required for sink particle formation, along with the number of massive stars in the cloud.
\end{abstract}

\begin{keywords}
 hydrodynamics --- stars: formation --- ISM: clouds --- ISM: kinematics and dynamics
\end{keywords}

\section{Introduction}\label{sec:intro}
In this paper we focus on the interplay between different feedback processes in GMCs (Giant Molecular Clouds). Stellar feedback is integral to galaxy evolution; it affects both the formation of individual stars, as well as the global gas reservoir in a galaxy. This large dynamic range makes disentangling the complex mix of positive and negative stellar feedback effects a difficult task. Cosmological simulations often invoke subgrid prescriptions to encompass the global effects of an entire stellar population, focusing on the larger scale impacts of stellar feedback. In this paper we are interested in studying stellar feedback on resolved scales, instead looking at the effects of individual massive stars on the star formation rate and the state of the gas in isolated GMCs. In this way we can study the complex feedback effects that would be missed on larger scales. However, through these smaller scale interactions between different stellar feedback processes we will show there can be larger global impacts on the galaxy. 
\newline \indent GMCs are cold, molecular gas reservoirs and are the main site for star formation in galaxies such as the Milky Way \citep[e.g.][for a review on the subject see \citealt{Heyer2015}]{Williams1997,Rosolowsky2005,Stark2006}. GMCs are dynamic environments with gas exhibiting supersonic motions which is commonly attributed to turbulence \citep{Larson1981}. The majority of the molecular gas in the Milky Way is found in clouds with a mass greater than $10^5$ M$_\odot$, while there also exists an upper mass limit of $\sim 6\times 10^6$ M$_\odot$ \citep{Williams1997,Rosolowsky2005}. The lifetime of GMCs is a heavily debated topic; predicted lifetimes range from very short ($< 3$\,Myr) to $10^8$\,yrs (for a discussion on this see \citealt{Heyer2015}). Arguments for smaller lifetimes include the lack of old ($> 3$\,Myr) stars associated with molecular gas in the Large Magellanic Cloud \citep{Elmegreen2000} as well as in clouds in the Milky Way \citep{Heyer2015}. However, there also exists observational evidence for longer/intermediate cloud lifetimes ($\sim$ tens of Myr; e.g. \citealt{Kawamura2009,Murray2011}).

Due to the large range of scales involved, there are several approaches to numerical models in this context. For example, modelling the formation of GMCs in galaxies \citep[e.g.][]{Dobbs2011} where the mass of each resolution element is much more than a single star but small enough to resolve the GMC, to modelling the assembly of individual stars on smaller scales \citep[e.g.][]{Bate2009} where the properties of individual stars are robust but only a few hundred solar masses of gas can be simulated. \cite{Dale2015} provides a coherent review of the current methods for numerical modelling of stellar feedback processes inside molecular clouds. Galactic-scale simulations of (spiral) disc galaxies are advantageous as they allow the environmental effects of the host galaxy to be included self-consistently. However, given the lower mass resolution of these simulations, it is harder to model the effects of stellar feedback on the interstellar medium (ISM). For example, the modelling of SN on galactic scales commonly results in the over-cooling problem \citep[e.g.][]{DallaVecchia2012}. This is the artificial cooling enhancement which results from heating a large gas mass when supernova feedback energy is injected in low resolution simulations. The onset of this cooling enhancement can be physically interpreted as the point at which the energy-conserving Sedov-Taylor phase of the supernova remnant can no longer be resolved. Solutions to this problem have been proposed by various authors \citep[e.g.][]{Springel2003, Scannapieco2006, Hopkins2012, Gatto2015}, often invoking sub-grid models with additional parameters which represent the star formation efficiency of collapsing gas, along with the efficiency with which the energy liberated during a SN event couples to the ISM.

Computational work that includes the large-scale environment of molecular clouds ranges from modelling a shearing box (in order to mimic conditions in a spiral arm) \citep[e.g.][]{Kim2013}, to simulating whole galaxies \citep[e.g.][]{Springel2003, Agertz2011, Dobbs2011, Dobbs2013, Renaud2013}. These simulations are able to investigate global trends, such as the offset between molecular gas clouds and star formation, along with the effect of feedback on the overall stellar population. There have also been numerous simulations of star formation and feedback on GMC scales or lower. Often this involves applying a fractal-like initial density field onto either a gas cloud or periodic box of gas, using either a turbulent velocity spectrum \citep[e.g.][]{Dale2012} or by design \citep[e.g.][]{Walch2013}. Furthermore, scales range from resolving individual stars \citep[e.g.][]{Bate2009, Clark2011} to modelling stellar populations, often as sink particles (e.g. \citealt{Dale2012}). 

GMCs are highly inefficient at forming stars ($\sim 2$ per cent of their mass is converted into stars; \citealt{Murray2011}) and this is commonly attributed to stellar feedback effects. As well as this, stellar feedback processes are often investigated as the driving mechanism behind the observed turbulence. Recent papers have focused on the energy and momentum injection by stellar winds from massive O-stars \citep[e.g.][]{Ntormousi2011, Ngoumou2015}, the radiation from low to high mass stars \citep[e.g.][]{Matzner2002, Gritschneder2010}, SN explosions \citep[e.g.][]{Walch2014, Iffrig2015} and protostellar jets \citep[][]{Matzner2007, Federrath2014}. Other works investigate how these processes interplay \citep[e.g][]{Freyer2003, Krumholz2012, Pelupessy2012, Rogers2013, Fierlinger2016, Dale2018}.

In this paper we investigate high mass GMCs of mass $\sim 10^6$ M$_\odot$. The clouds have a large enough stellar mass to support multiple high mass stars (although this is dependent on the star formation efficiency of the cloud gas) and hence they offer an interesting laboratory for investigating the effects of stellar feedback. Moreover, the free-fall timescales of these clouds are over $10\,$Myr, which is enough time for the most massive stars to leave the main sequence (which occurs beyond $3\,$Myr). Furthermore, a typical OB-type star will have left the main sequence by $30\,$Myr, which represents the timescale of the simulations we run. 

Molecular clouds can in principle be disrupted due to other stellar feedback processes prior to the onset of SN feedback. One of the processes that can lead to molecular cloud disruption is the expansion of HII regions, ionised by stellar radiation. However, \cite{Dale2012} find that clouds of mass $\sim 10^6$ M$_\odot$ are not significantly disrupted by photo-ionising winds prior to the onset of SN feedback. \cite{Rogers2013} find that molecular clouds are able to survive the stellar winds from O-stars by forming low-density pillar-like chimneys in the cloud, through which hot gas can escape, allowing the dense star forming material to survive until SN feedback begins. Also, taking the galactic environment into account, \cite{Dobbs2013}, find typical GMC lifetimes ranging between $4-25\,$Myr for clouds with masses greater than $\sim 10^{5}$ M$_\odot$. This puts them in the regime where SN feedback can become important. 

We are particularly interested in the interplay between gradual feedback mechanisms and instantaneous energy injections. Our paper follows on from work by \cite{Rogers2013}; in this case the `gradual feedback' we refer to would be stellar winds, which represents a constant (comparatively low power) heat source for the ISM, while SN feedback is an instantaneous injection of the canonical $10^{51}$ ergs of energy. However, rather than focusing on stellar winds as a gradual heating source, we instead investigate the mechanical energy input from a population of HMXBs (High Mass X-ray Binaries). 

\subsection{High Mass X-ray Binaries (HMXBs)}\label{sec:HMXBs}
HMXBs consist of an OB-type stellar companion orbiting a neutron star or black hole. Observations of HMXBs are rarer than their low mass counterparts and they are typically associated with areas of star formation (e.g. \citealt{Fabbiano2006} and references therein; \citealt{Mineo2012}) due to their relatively short formation time of approximately $1-10$ Myr  (set by the main sequence lifetime of the massive primary star, which is usually short). The compact object (or primary star) in HMXBs is fed either by winds or Roche lobe overflow from the companion star. The winds associated with OB stars are such that, despite a very small capture fraction ($\sim 10^{-2}-10^{-1}$ \%), they still transfer mass at a high enough rate to produce high luminosity from accretion on to the compact object. Ultra-Luminous X-ray Sources (ULXs) are thought to be HMXBs in a high mass transfer rate phase. For example, the HMXB SS433 is thought to be a ULX on its side \citep{Begelman2006} and would look brighter if viewed along the outflow axis. Its high luminosity is thought to be the result of the companion star filling its Roche lobe and transferring mass to the primary on a thermal timescale \citep{King2000}.

Jets are common in HMXBs and recent work suggests there exists a universal relation between the radio luminosity and the kinetic power of jets, spanning supermassive black holes down to stellar mass black holes \citep{Fender2016}. The properties of jets in HMXBs are broadly split into two categories. Firstly, persistent jets in the low-luminosity state; where the X-ray spectrum is predominantly hard and the jet is prolonged with Lorentz factors $\sim 1.4$ \citep{Fender2004}. Secondly, powerful ballistic jets in the high-luminosity high-variability state, which describes the transition from the X-ray spectrum being dominated by hard to soft X-rays and is associated with Lorentz factors of $\sim 2$ \citep{Fender2004}. An example of an HMXB that shows transitions from a hard state to a soft state is Cygnus X-1. This is a highly luminous system consisting of a $14.8 \pm 1.0$ M$_\odot$ black hole accreting material via stellar winds from a super-giant O-type companion of mass $19.2 \pm 1.9$ M$_{\odot}$ \citep{Orosz2011}. It has been extensively investigated in multiple wavelengths; for example \citet[][radio]{Gallo2005}, \citet[][X-ray]{Sell2015} and \citet[][optical]{Russell2007}. Associated with Cygnus X-1 is an inflated radio lobe, which is surrounded by a ring-like shock approximately 5 pc in scale \citep{Gallo2005}. This lobe is thought to have been inflated by a steady jet and \cite{Gallo2005} used the lobe to conclude the kinetic power of Cygnus X-1 could be as high as the total X-ray luminosity of the system. Moreover, recently \cite{Fender2006} reported a transient, extended radio jet from Cygnus X-1. 

Recent work on the relation between the mechanical power of relativistic compact objects compared with their bolometric X-ray luminosities has concluded the former could be equal, if not greater, than the X-ray luminosity. For example, observations of SS433 indicate relativistic jets ($\sim 0.26\,c$) with a mechanical energy of $> 10^{39}$\,erg/s \citep[e.g.][]{Blundell2001,Mirabel2011,Goodall2011}. These jets are thought to have inflated the surrounding W50 nebula and also have interacted with the preceding supernova remnant \citep{Lockman2007,Goodall2011}. Moreover, \cite{Pakull2010} reported a jet-inflated bubble, with a diameter of 300\,pc, surrounding the microquasar S26 in the galaxy NGC 7793. \cite{Pakull2010} also reported S26 has a greater mechanical power output than SS433, while the jets were found to be $10^4$ times more energetic than its associated X-ray emission. Given these rates of kinetic energy injection into the ISM, depending on the lifetime of the HMXB and the mode of accretion, it is possible their jets release ten times the amount of energy associated with a single SN event across their lifetime (assuming the canonical value of $10^{51}$ erg). This makes them energetically significant, particularly in GMCs with low gravitational binding energies. 

There has also been a great deal of interest in jets associated with ULXs (Ultra Luminous X-ray Sources); for example \cite{Justham2012} investigated the potential impact of a ULX population in models/simulations of galaxy formation. They argued the stochasticity of ULX events, coupled with their significant energetic contribution to the ISM, could result in a variety of different star formation histories, particularly in dwarf galaxies. Moreover, \cite{Justham2012} discussed the intriguing possibilities that may result from the interplay between SN and XRB (X-ray Binary) feedback. One such possibility is that the XRB feedback dominates initially, resulting in a warm, heated but not unbound ISM, which stops star formation and decreases the efficiency of SN feedback at unbinding the gas in the galaxy.

Moreover, \cite{Artale2015} also focussed on the interplay between SNe and BH-HMXBs (HMXBs containing a black hole) feedback and their impact on the early evolution of a dwarf galaxy. They investigated this using SPH simulations and concluded that although BH-HMXBs acted to reduce the star formation rate earlier on in their simulations, the overall star formation efficiencies in their simulated galaxies were increased (particularly in low mass galaxies). 

In this paper we investigate the interplay of SN feedback and jet mechanical feedback from a HMXB population and the resulting effect on star formation in a region of bound/marginally unbound, cold molecular gas. In Section \ref{sec:nummodel}, we discuss our numerical model. In Section \ref{sec:results} we present our results, which we further discuss in Section \ref{sec:discussion}. Finally in Section \ref{sec:conclusions} we provide our conclusions.

\section{Numerical model}\label{sec:nummodel}
\subsection{Basics of the model}
We use GADGET-3; a hybrid N-body/SPH (smoothed particle hydrodynamics) code that is an updated version of the publicly available code GADGET-2 \citep{Springel2005}. We use the SPHS extension described in \cite{Read2012}, which is designed to model mixing of multiphase gas. We also use a Wendland-2 kernel \citep{Wendland1995,Dehnen2012} with 100 neighbours. In all our simulations we use an ideal equation of state, $P=(\gamma-1)\rho u$, where $P$ is the gas pressure, $\gamma$ is the adiabatic constant, set to $5/3$ in our simulations, $u$ is gas internal energy and $\rho$ is particle density. For temperatures above $10^4$\,K, we implement gas cooling using the look-up tables generated by the MAPPINGS III code \citep{Sutherland1993}. Below $10^4$\,K we use the metal line cooling scheme described in \cite{Mashchenko2008}. We set a maximum temperature for the gas of $\sim 10^8$\,K, in order to avoid prohibitively small timesteps. We note that typically only $\sim 10$ particles in our simulations are at this temperature limit, so our results are unaffected by this. Whilst we vary parameters between different simulations, our canonical values are a total mass of $2\times 10^6$ M$_\odot$ modeled using $5$ million SPH particles, meaning the mass of each individual particle is $\approx 0.4$ M$_\odot$. We employ both adaptive smoothing lengths and softening lengths for the SPH particles with a minimum value of $0.1$\,pc. For convenience, any unbound gas particles beyond an outer radius of $5\,$kpc are removed from the simulation. 

In our simulations the minimum resolvable mass is given by $2N_{\rm neigh}m_{\rm p}$, where $N_{\rm neigh}$ is the number of neighbours within a smoothing kernel and $m_{\rm p}$ is the mass of an SPH particle \citep{Bate1997}. Therefore we replace gas particles with sink particles if they reach a density corresponding to the Jeans density for this minimum resolvable mass, i.e.
\begin{equation}\label{Jeans_rho}
\rho_{\rm J} = \left(\frac{3}{4\pi}\right)\left(\frac{5 k_{\rm B} T}{G\mu m_{\rm H}}\right)^{3} \frac{1}{(2N_{\rm neigh}m_{\rm p})^{2}}
\end{equation}
where $k_{\rm B}$ is the Boltzmann constant, $G$ is Newton's gravitational constant, $\mu$ is the mean molecular weight and $m_{\rm H}$ is the mass of Hydrogen. We obtained this equation using the virial theorem; i.e. by setting 2E$_{kin}$ + E$_{pot}$ = 0. We also require that the gas must be converging (i.e. $\nabla \cdot {\bf v} < 0$) and that the temperature of the gas particle must be $< 500$\,K. In calculating $\rho_{\rm J}$ we use approximate values of $\mu$ (1.291 for $Z_\odot$ and 1.242 for $Z\sim 0.001$) for simplicity \citep{Sutherland1993}. However, for calculating the cooling rates, $\mu$ is calculated self-consistently using the electron fraction. Once the sink particles are formed, they accrete any gas particles that enter within a radius of $0.5$\,pc with a kinetic energy less than the gravitational potential energy with respect to the sink particle. This avoids gas particles being assigned prohibitively small timesteps within the simulation and also allows sink particles, which represent unresolved star forming regions, to grow in mass.

Once a sink particle accretes enough gas to reach a mass of 180 M$_\odot$, we consider this sink to contain 1 massive ($>$ 8 M$_\odot$) star, which in a fraction of cases is situated in a binary system with a second high mass star - see Section \ref{Bin_Pop} for details. The value of 180 M$_\odot$ lies in the range between 95 M$_\odot$ and 210 M$_\odot$, which correspond to the masses inferred from a Kroupa IMF (initial mass function) \citep{Kroupa2001} without/ with the contribution from unresolved binaries, evaluated in the mass range 0.01 M$_\odot$ -- 100 M$_\odot$ . These values are arrived upon using the mean masses, along with the number fractions, given in supplementary table 2 in \citealt{Kroupa2002}. The number fraction we use, 0.2$\%$, is conservative compared with that found in \citealt{Power2009}; which was 1.1$\%$.

Furthermore, we assume all stars with a mass $>$8 M$_\odot$ are in binaries, which is consistent with the high multiplicity fractions ($>$50 $\%$) observed for massive stars \citep[e.g.][]{Sana2011,Chini2012}. The properties of the binary system within the sink particle are determined using a Monte-Carlo approach described below. Furthermore, a physical selection process is used to determine which systems go on to become HMXBs. These processes are described in detail in the next section (Section~\ref{Bin_Pop}).

We restrict our attention to only one HMXB system ($1$-$2$ linked SN events) per sink particle, as the likelihood of a second within the time window of our $33$\,Myr simulations is small. For our purposes, this method results in a random distribution of SN/HMXB events in likely locations of dense star formation. Moreover, using this method we obtain a number of SN/HMXB events which is roughly consistent with star forming mass. 

It is important to note that throughout our simulations, the binary properties of the stellar population are considered separately to the properties of the sink particle within the simulation. In other words, each sink particle has a mass from which its gravity in the simulation is calculated, and an associated `primary star mass' of the binary which is a virtual property solely used to set the HMXB feedback properties of the sink particle.

\subsection{Binary Population Synthesis}\label{Bin_Pop}
The relevant parameters when considering our population of binaries are: the initial mass of the primary star, the mass ratio, $q$, of the primary to the secondary star and the lifetime of both stars. Since a HMXB consists of a neutron star or black hole accreting material from an O or B companion star, the primary star needs to exceed $\sim$ 8 M$_\odot$ to form a neutron star or 20 M$_\odot$ to form a black hole. Therefore, as soon as a sink particle exceeds the minimum mass of 180 M$_\odot$, a Kroupa IMF is sampled between 8 M$_\odot$ and 100 M$_\odot$ to determine the primary star mass. Next, the binary mass ratio is sampled uniformly between 0 to 1 since the distribution of $q$ values is still considered to be largely flat (e.g. \citealt{Sana2013}). If this sampling results in a secondary mass of less than 8 M$_\odot$, then the system is discounted as a HMXB progenitor since the secondary is required to be a massive OB type companion.

We then determine a lifetime of the primary star and, in the case when the secondary mass meets the criteria, a companion stellar lifetime. The primary lifetime will determine the time of the first Type II supernova and the secondary lifetime sets a limit on the lifetime of the HMXB feedback phase, along with the time of the companion supernova. In order to obtain the lifetime of the primary star and its companion we use the same method as in \cite{Power2009}. This method uses a lookup table which lists mass versus lifetime in order to interpolate the stellar lifetimes. For a metallicity of [Fe/H]$=-1.2$ ($Z \sim 0.001$) we use Table~46 from \cite{Schaller1992} and calculate lifetimes of massive stars by making the approximation $t_{\rm life} \sim t_{\rm H} + t_{\rm He}$ (where $t_{\rm H}$ is the lifetime of the Hydrogen burning phase of the star and $t_{\rm He}$ is the same for Helium). For solar metallicity we use results from Table 1 of \cite{Meynet2000} and Table~45 of \cite{Schaller1992}. The mean delay time between a sink particle being created and accreting enough gas to reach 180 M$_\odot$, is $\sim$ 0.8 Myr. Furthermore, sink particles begin to be created at $\sim$ $5$-$6$ Myr in all runs (see figures \ref{fig:Zsol_SFE} and \ref{fig:Zlow_SFE}). This delay time depends on the initial velocities of the gas particles (e.g. comparing Runs A and E in Fig. \ref{fig:Zsol_SFE}), along with the size of the cloud (for example the sink particles are formed earlier in Run V - a larger cloud - than Run A; see Fig. \ref{fig:Zsol_SFE}). We take into account the time since the particle was first formed - which we refer to as the age of the sink particle - when determining the delay time before the first SN. If the age of the sink particle is longer than the lifetime of the primary star, the sink particle immediately undergoes SN feedback. Therefore, given the minimum lifetime of a massive star in our simulations is $\sim$ 3 Myr (corresponding to 100 M$_{\odot}$), this results in the first SN and HMXB events occurring at $\sim$ $8$-$9$ Myr (see Fig. \ref{fig:Zsol_HMO}). 

It is worth noting a limitation to our method is that we do not include the effect of binarity on the evolution of the massive stars. Recent work has concluded interactions in close binary systems (such as Roche-lobe overflow) can significantly affect the evolution of both the primary star and its companion \citep[e.g.][]{Eldridge2008, Yoon2010, Song2016}. Focusing on stellar lifetimes, mass accretion onto the primary star in a binary system could prolong its lifetime by providing additional nuclear fuel, however the increase in stellar mass also acts to speed up the nuclear burning process. Moreover, the loss of mass from the secondary star can also prevent it undergoing core-collapse at the end of its life. The interplay between these two conflicting processes was explored in  \citealt{Zapartas2017}, where the delay times of core-collapse SNe (or SNe rate versus time) was compared between a single-star stellar population and a population containing binary stars. The paper found the largest differences between the distribution of delay times occurred beyond 20 Myr. Prior to this the two distributions were very similar - indicating there are no significant divergences in SNe rate between single star and binary star populations in the time-frame we are interested in.

A fraction of the possible HMXB progenitors will not survive the supernova of the primary, becoming unbound when mass is lost. In order to estimate this fraction, as in \cite{Power2009}, we make the assumption the binary system becomes unbound if over half its mass is lost in the supernova. A significant factor in determining the surviving remnant mass is the amount of `fallback' that occurs during the SN explosion, a quantity that is metal-dependent. For example, \cite{Zhang2008} found the supernova of massive stars at lower metallicity are more likely to result in a black hole, compared with those at solar metallicity. In order to calculate the mass of the system post-supernova, we use look-up tables with remnant masses taken from Table 4 of \cite{Maeder1992} for [Fe/H]$=-1.2$ and a combination of Tables~4-5 from \cite{Sukhbold2016}, Table~4 of \cite{Maeder1992} and Table~4 from \cite{Zhang2008} for [Fe/H]$=0$. \cite{Zhang2008} also explore the difference between explosion mechanisms; altering piston locations and explosion energies. For this paper, we average over the results for each initial mass at $Z=Z_\odot$. From \cite{Zubovas2013} we expect $>93$ per cent of all (including low mass) binary systems to be disrupted. The remaining systems then switch on their HMXB phase immediately and this lasts for the lifetime of the secondary star. At the end of the secondary lifetime the sink particle switches off HMXB feedback and undergoes SN feedback for a second time. After this the sink particle ceases to produce feedback. The lifetime of the HMXB feedback is typically of the order of $10$ Myr, while the timestep of the simulation is $\sim 10^4$\,yrs. This ensures the gradual heating of the HMXB feedback is resolved. 

We have implemented this binary population synthesis method into GADGET in such a way as to tie the random seed to the particle identity and the number of the current time step, neither of which should vary between runs with identical initial conditions prior to the onset of feedback. In this way, for runs with identical initial conditions, prior to the first HMXB event, the gas will have the same phase/morphology and the massive stars will occur at the same locations. For example, Runs A and C will be identical prior to the first HMXB event, when they will diverge. At this point the underlying stochasticity of massive star feedback will play a role. The same is true of Runs E and F, V and W and so on. This facilitates easy comparison between different feedback effects.

\subsection{The Metallicity of our Simulations}
We chose to run two sets of simulations at metallicities of $Z=Z_\odot$ (or [Fe/H] $= 0$) and $Z \sim 0.001$ (or [Fe/H] $= -1.2$) in order to compare the effects of the altered cooling regimes and HMXB populations on the gas cloud. In particular, the $Z \sim 0.001$ value was chosen since this is beyond the `critical metallicity' for Population II stars, hence the gas in the simulations has been metal enriched via the SNe of Population III stars. This allows cooling below $200$\,K via metal-line cooling, which is more efficient than at zero metallicity with cooling by $H_2$ molecules (\citealt{Larson2005}). Our resolution is such that the gas in the [Fe/H]$=-1.2$ runs can cool enough to meet the Jeans density criterion and undergo star formation, something that was not possible at zero metallicity.

\subsection{Feedback Mechanisms}
Throughout our simulations we implement supernova feedback by injecting the canonical value of $10^{51}$\,erg of thermal energy into the surrounding $\sim 100$ neighbouring gas particles. The amount of energy each gas particle receives is kernel weighted, hence those particles closest to the star particle receive the most energy. Thermal supernova feedback commonly results in the `over-cooling' problem. This is a consequence of low resolution simulations, where energy is initially injected into a large gas mass, causing the gas to cool artificially quickly and stall feedback \citep{Springel2003,Stinson2006,Creasey2011}. However, the mass resolution of our simulations is such that we can capture the Sedov-Taylor phase of the shock expansion and hence are not significantly affected by over-cooling (see Appendix~\ref{appendix:Sedov}). 

We explore HMXB feedback with two different implementations; the first is using thermal energy injection and the second is using kinetic energy injection. For the thermal case the internal energy of the particles is increased, while for the kinetic case the energy is delivered as a radial velocity kick. Similarly to the SN feedback, the HMXB schemes weight the amount of energy given to their neighbouring particles using the SPH kernel. Here we have assumed the shock-heating of the jets has isotropically raised the temperature of the surrounding gas. This assumption is dependent on factors such as the angle of jet precession and cone angle \citep{Goodall2011}, along with jet power and the density of the surrounding ISM \citep{Abolmasov2011}. These factors are beyond the scope of this paper, however they would be interesting to investigate in future work. Moreover, there are multiple observations of super-bubbles around ULXs , which have been collisionally energized by jets \citep[e.g.][]{Russell2011}, which support the inclusion of HMXB feedback as an isotropic heating of the surrounding gas. 

Both the thermal and kinetic HMXB feedback schemes inject the same amount of energy across the total HMXB lifetime. We set this value as $10^{52}$\,erg and the amount of energy a sink particle injects is proportional to its time-step. This results in a mean energy injection rate of $3\times 10^{37}\,$erg/s for a lifetime of 10\,Myr. We find lifetimes between $5-35\,$Myr in the simulations we report below. We have chosen this rate of energy injection as it is consistent with rates observed for HMXBs. For example, the wind-fed jet in Cygnus X-1 is estimated to input $\sim 10^{35-37}$\,erg/s into the ISM \citep{Gallo2005}, and SS433 inputs $\sim 10^{39}$\,erg/s into the ISM (\citealt{Brinkmann2005}). Thus our value of injected energy falls in the observed range. Similar numbers were used by \cite{Artale2015}. HMXBs exhibit time variability in their luminosity and energy injection rate, associated with a spectral transition from hard to soft X-ray (as is seen in Cygnus X-1), however this is on a much smaller timescale than the time resolution of our simulation where a single time-step roughly corresponds to $10^4$\,yrs.

We note that massive star winds inject between $\sim 10^{35}$\,erg/s -- $10^{36}$\,erg/s of energy into the ISM (based on a mass loss rate of between $10^{-6}\,$M$_\odot$/yr -- $10^{-5}\,$M$_\odot$/yr, along with wind velocities of $1,000$\,km/s; \citealt{Repolust2004} and \citealt{Leitherer1992} respectively). Thus the energy input from a single massive star wind is expected to be smaller than the energy input we assume for an HMXB, owing to the fundamental fact jets are produced further into the potential well of the stellar object than line-driven winds, along with the fact observations indicate much of the accretion energy in XRBs is channeled into the mechanical power of the jets \citep{Gallo2005, Sell2010, Soria2010}. However, there are many massive stars for every HMXB. Very crudely, if we assume all massive stars are in binaries, and we expect only $1$-$10$\% of these systems to form HMXBs, then we would expect $10$-$100$ massive stars per HMXB. Thus the energy injected by massive stars would be comparable to that injected by HMXBs. We therefore anticipate that our results will be qualitatively similar to considering feedback from massive star winds. Although including HMXBs as an anisotropic source of heating would most likely affect this comparison. Moreover, stellar winds would be present prior to the first SN event, which could affect the efficiency of the SN feedback by lowering the density of the surrounding ISM (e.g. as in \citealt{Rogers2013}). We will discuss this further in section \ref{sec:discussion}.

\subsection{Initial Conditions}\label{sec:ICs}
In this paper we model giant molecular clouds that range from being globally bound to marginally unbound. We measure the boundedness of each cloud using the virial parameter $\alpha_{\rm vir} = |E_{\rm kin} + E_{\rm therm}|/|E_{\rm pot}|$, which we vary between 0.7 and 1.2 (the cloud is considered virialised at $\alpha_{\rm vir} =$ 0.5). We take the cloud mass to be $2\times 10^6\,$M$_\odot$, with some additional simulations run with $2 \times 10^5$ M$_\odot$ for comparison. The clouds are initially seeded with a non-driven turbulent velocity spectrum based on \cite{Dubinski1995}. We use a Kolmogorov power spectrum with $P(k) \sim k^{-11/3}$ to generate a velocity field with homogeneous, incompressible (divergence-free) turbulence. This is achieved by defining the velocity field as the curl of a vector potential $\textbf{A}$. The Fourier transform of this vector potential is then taken and each k mode is assigned an amplitude drawn from a Rayleigh distribution of variance $\sim k^{- (11/3 + 2)/2}$; and an associated phase is drawn uniformly from $0$ to $2\pi$. We then take the inverse Fourier transform of the curl of $\textbf{A}$ in order to obtain the real velocity components. As we take the inverse Fourier transform, we sum over k values between $k_{\rm max}$ and $k_{\rm min}$. These limiting values are set by the minimum length scale and maximum length scale of the simulation respectively, for our simulation $k_{min} = 2\pi/R_{\rm out}$, where $R_{\rm out}$ is the outer radius of the gaseous sphere and $k_{\rm max} = N^{1/3}/2 R_{\rm out}$, corresponding to the Nyquist frequency (which is set by the number of particles, $N$, in the initial conditions). 

Table \ref{tab:Runs} summarises the different parameters used in our simulations. We vary the gas metallicity, the mass and size of the cloud and the virial parameter. We chose our initial conditions in accordance with the parameter space used by \citealt{Dale2012}, which in turn was derived from a catalogue of 158 galactic molecular clouds collated by \citealt{Heyer2009}. We perform simulations with the two (kinetic and thermal) HMXB feedback schemes. We also performed control simulations containing no HMXB feedback, however with SN feedback still included and others with no feedback at all. We present the results of these simulations in the next section. Finally we explore the effects of numerical resolution on the results in Appendix~\ref{appendix:Res}.

\begin{table*}
  \caption{The different simulations run through the course of this paper. Here $M_0$ is the initial gas mass (in M$_\odot$) of each simulation, [Fe/H] is the log$_{10}$ of the ratio between the metal content of the cloud compared with that of our sun, $R_0$ is the initial cloud radius in pc, $T_i$ is the initial temperature in K, $\alpha_{\rm vir}$ is the initial virial parameter (|$E_{\rm kin}+E_{\rm therm}|/|E_{\rm pot}|$), $t_{\rm ff}$ is the free-fall time of the cloud, $m_{\rm pcl}$ is the gas particle mass in each simulation, and HMXB is the type of HMXB feedback present (either kinetic or thermal).}
  \begin{tabular}{|l|l|l|l|l|l|l|l|l|l|l|}\label{tab:Runs}
    Run&[Fe/H]&$M_0$ ($\times\,10^{6}$ M$_\odot$)&$R_{0}$(pc)&$T_{i}$ (K)&$\alpha_{\rm vir}$&Mach No.&$t_{\rm ff}$ (Myr)&$m_{\rm pcl}$ (M$_{\odot}$)&HMXB&SNe (Y/N)\\
    \hline
    A&0&2&100&50&0.7&14.9&11.7&0.4&Therm&Y\\
    \hline
    B&0&2&100&50&0.7&14.9&11.7&0.4&Kin&Y\\
    \hline
    C&0&2&100&50&0.7&14.9&11.7&0.4&None&Y\\
    \hline
    D&0&2&100&50&0.7&14.9&11.7&0.4&None&N\\
    \hline
    E&0&2&100&50&1.2&19.5&11.7&0.4&Therm&Y\\
    \hline
    F&0&2&100&50&1.2&19.5&11.7&0.4&None&Y\\
    \hline
    G&0&2&100&50&1.2&19.5&11.7&0.4&None&N\\
    \hline
    H&-1.2&2&100&150&0.7&8.3&11.7&0.4&Therm&Y\\
    \hline
    I&-1.2&2&100&150&0.7&8.3&11.7&0.4&None&Y\\
    \hline
    J&-1.2&2&100&150&0.7&8.3&11.7&0.4&None&N\\
    \hline
    K&-1.2&2&100&150&1.2&11.0&11.7&0.4&Therm&Y\\
    \hline
    L&-1.2&2&100&150&1.2&11.0&11.7&0.4&None&Y\\
    \hline
    M&-1.2&2&100&150&1.2&11.0&11.7&0.4&None&N\\
    \hline
    R&0&0.5&65&50&0.7&9.1&12.3&0.1&Therm&Y\\
    \hline
    S&0&0.5&65&50&0.7&9.1&12.3&0.1&None&Y\\
    \hline
    T&-1.2&0.5&65&150&0.7&5.0&12.3&0.1&Therm&Y\\
    \hline
    U&-1.2&0.5&65&150&0.7&5.0&12.3&0.1&None&Y\\
    \hline
    V&0&5&150&50&0.7&19.3&13.6&1&Therm&Y\\
    \hline
    W&0&5&150&50&0.7&19.3&13.6&1&None&Y\\
    \hline
    X&-1.2&5&150&150&0.7&10.8&13.6&1&Therm&Y\\
    \hline
    Y&-1.2&5&150&150&0.7&10.8&13.6&1&None&Y\\
    \hline		
  \end{tabular}
\end{table*}

\section{Results}\label{sec:results}
Here we present the main results of the paper. We split these into two main categories; simulations run with [Fe/H] $= 0$  and those run at [Fe/H] $= -1.2$. However, firstly we discuss the results of the simulations with no feedback at both metallicities, for comparison in later sections. 

On a general note, during our results and discussion section we refer to `star-forming gas' and the `star formation efficiency'. In the context of this paper, we are actually referring to the mass contained in sink particles and the efficiency of the gas particle to sink particle conversion. Once the gas has been accreted onto a sink particle we have no further information on its fate and it is assumed that a fraction of the mass contained in sink particles will actually be involved in star formation. However, this is beyond the scope of the simulations in this paper. Furthermore, once a sink particle has reached 180\,M$_\odot$ the properties of the massive star(s) that are considered to be located there are calculated and further accretion is ignored. However, our simulations show the mean mass accreted by a sink particle with a mass of 180\,M$_\odot$ or higher is 41 M$_\odot$ across all runs which include stellar feedback, hence we consider this to have negligible impact on the massive star population in our simulations and therefore on our results. 

\subsection{Runs with no feedback: D, G, J, M}\label{sec:Nofb}
\begin{figure} 
  \psfig{file=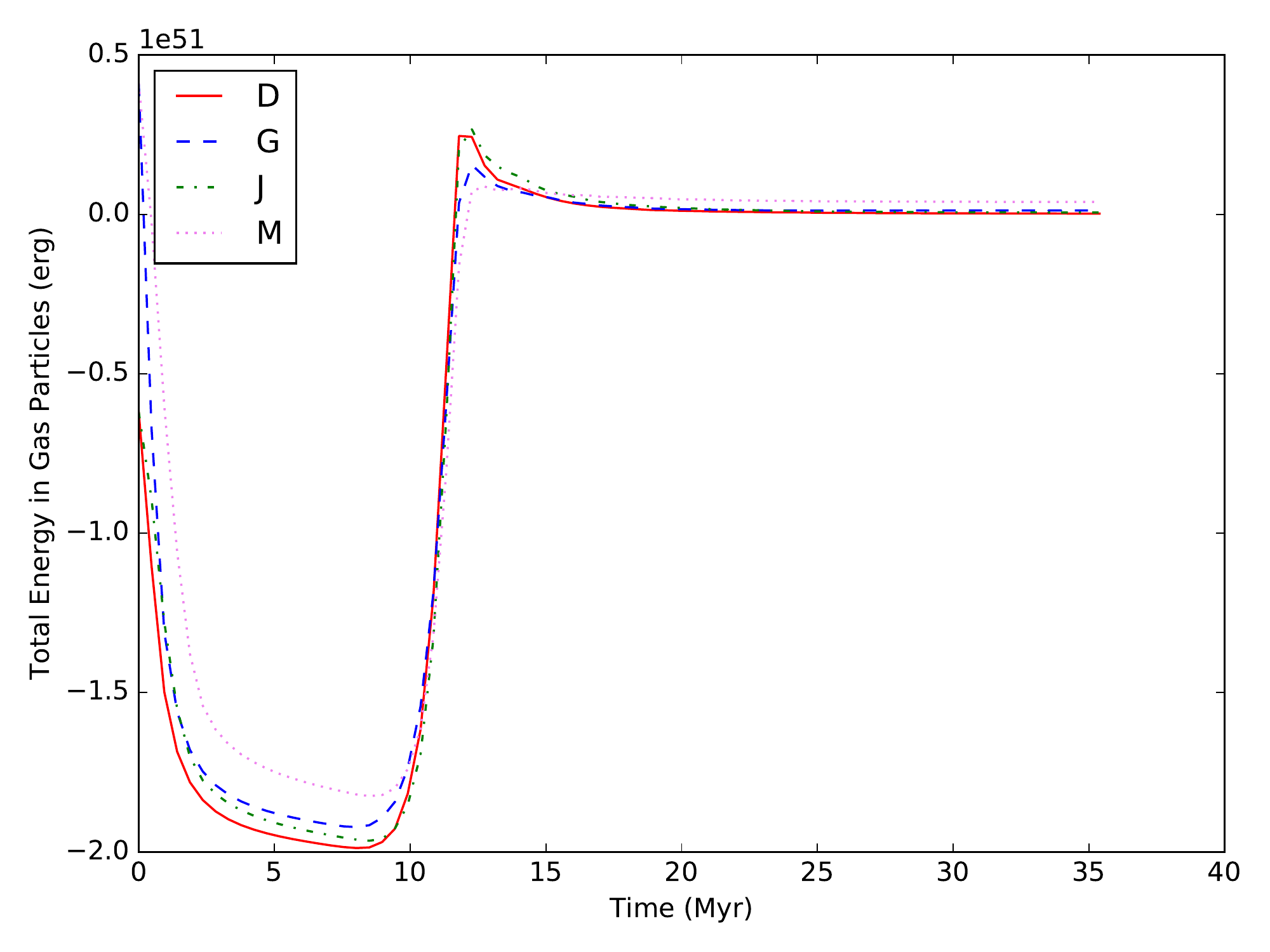,width=0.5\textwidth,angle=0,scale=1.}
  \caption{The time evolution of the total energy (calculated as the sum of the total potential, kinetic and thermal energies of the gas particles (i.e. minus sink particle mass) in the simulation) of the clouds with no feedback included; D, G, J and M.}
  \label{fig:Nofb_Etot} 
\end{figure}

\begin{figure} 
  \psfig{file=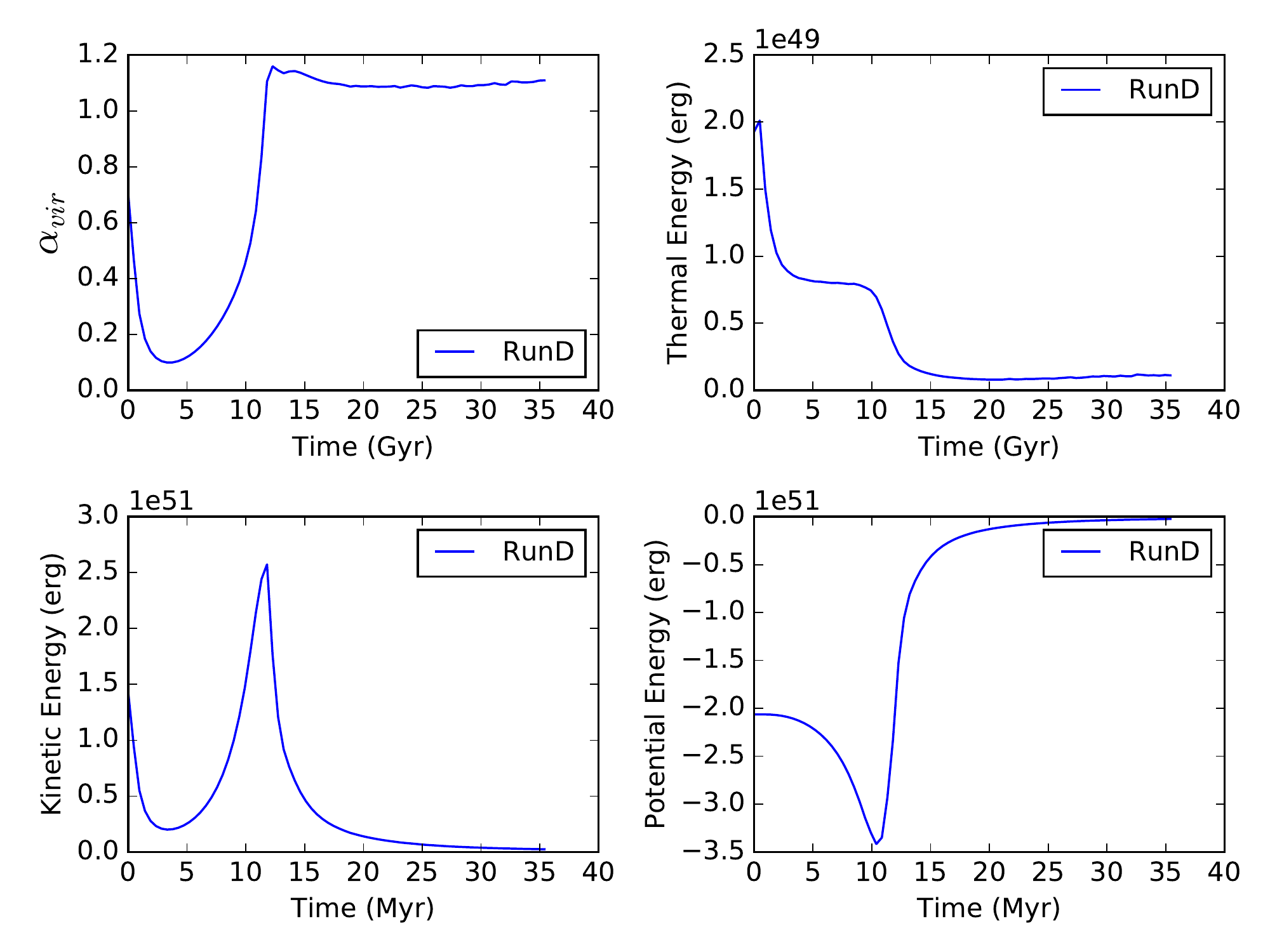,width=0.5\textwidth,angle=0,scale=1.}
  \caption{An example of the time evolution of the virial parameter (upper left plot), thermal energy (upper right), kinetic energy (lower left) and potential energy (lower right) of the gas particles in a run without feedback (Run D).}
  \label{fig:Nofb_Eplots} 
\end{figure}

\begin{figure} 
  \psfig{file=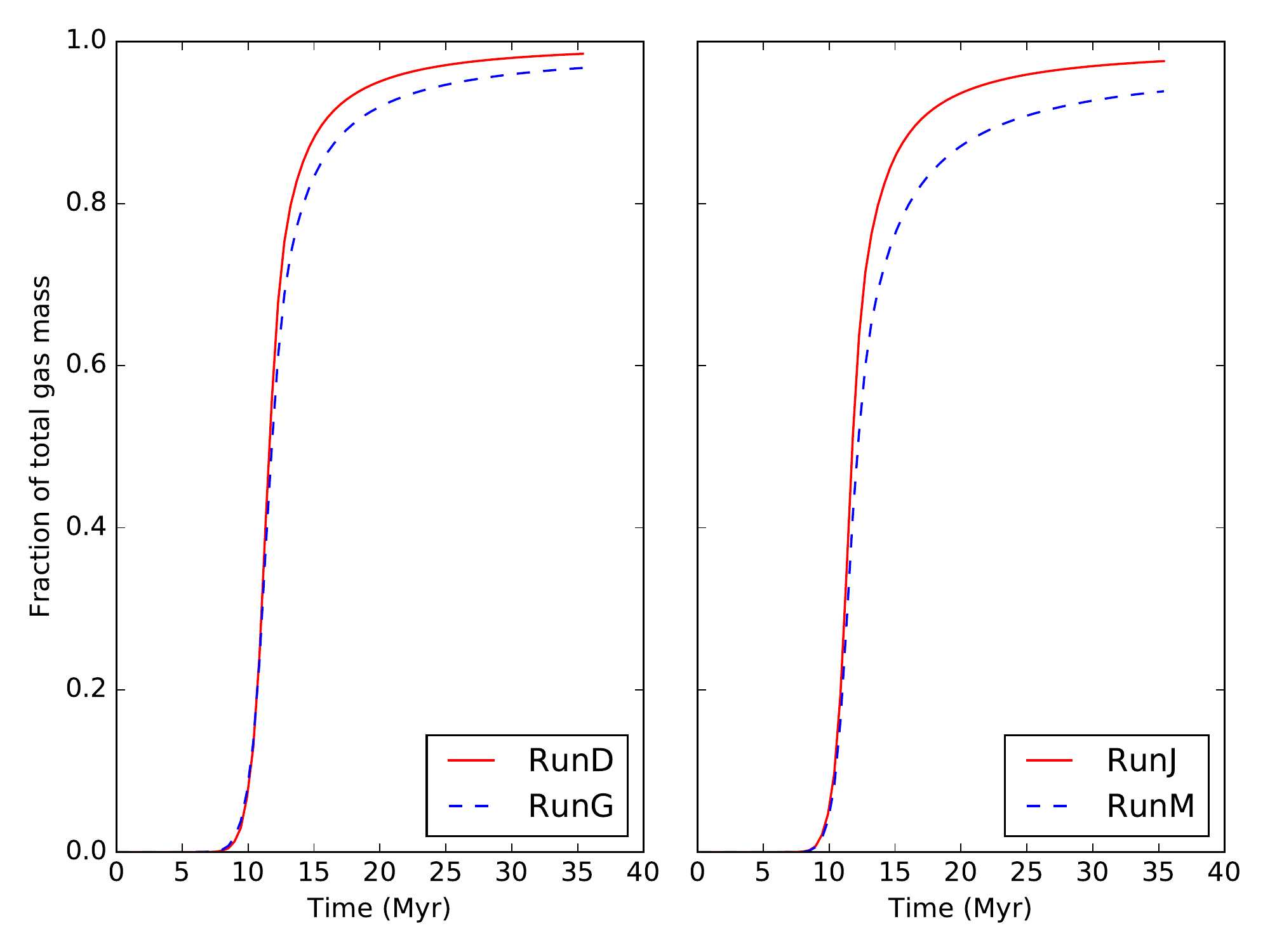,width=0.5\textwidth,angle=0}
  \caption{The time evolution of the fraction of the initial gas mass contained in sink (f$_{\rm sink}$) particles for the runs with no feedback: D, G, J and M.}
  \label{fig:Nofb_sink} 
\end{figure}

Fig. \ref{fig:Nofb_Etot} plots the total energy of the gas in clouds (or Runs) D, G, J and M. None of these runs include feedback, instead the cloud is allowed to evolve under the action of the initial turbulent velocity field and collapse under its own gravity. Furthermore, Fig. \ref{fig:Nofb_Eplots} gives an example of the time evolution of $\alpha_{\rm vir}$ along with the thermal, kinetic and potential energies of the gas for a run with no feedback (in this case Run D). From Fig. \ref{fig:Nofb_Etot} and Fig. \ref{fig:Nofb_Eplots}, we can see the initial kinetic energy given to the gas particles at the beginning of the simulation is quickly converted into thermal energy and radiated away prior to the first snapshot time. The cloud then collapses and gravitational potential energy is converted to kinetic energy until 10-13 Myr (which is roughly the free-fall time for each cloud - see Table \ref{tab:Runs}). At this point the majority ($>$ 90 $\%$) of the gas is converted into sink particles. The high sink particle mass and resulting low gas mass in the simulation beyond this point leads to a high gravitational potential (see Fig. \ref{fig:Nofb_Eplots}), which explains the rapid jump in total energy seen in Fig. \ref{fig:Nofb_Etot}. This can also be seen in Fig. \ref{fig:Nofb_sink}, which plots the time evolution of the fraction of the initial gas mass contained in sink particles. Beyond the free-fall time, the remaining gas in each cloud continues to collapse, increasing the kinetic energy of the gas as the gravitational potential energy decreases. 

From Fig. \ref{fig:Nofb_sink} it is noticeable the mass fraction contained in sink particles is less in the runs at a metallicity [Fe/H]$=-1.2$; G and M. This is expected as cooling is less efficient at lower metallicity below $10^4$ K, due to a lower amount of ions such as OII, OIII and CII \citep{Mo2010}. This relatively inefficient cooling means the temperature of the gas in the low metallicity runs is higher, resulting in a higher Jeans density and a reduction in the number of particles reaching this required density to form stars.  However, the sink mass fraction still reaches $90$\% by $35$\,Myr in the low metallicity runs, indicating the majority of the gas is able to become star-forming within the time frame of the simulation.
\newline \indent Overall, despite the initial velocity field imposed on each cloud, the majority of the gas is able to cool and collapse to form sink particles in the free-fall time. Furthermore, the free-fall time of the clouds corresponds to the lower end of the lifetimes of massive stars (which can be as short as $3$\,Myr). However, in this suite of simulations, sink particles do not start being formed until $5-6$\,Myr. As a result, we expect the number and locations of feedback events between $8-13$\,Myr to be crucial in determining the star formation efficiency of each cloud. 

\subsection{[Fe/H]$=$0 Runs}
\subsubsection{The injection of HMXB: Runs A (HMXBs and SNe), C (Just SNe)}
\begin{figure*} 

 \includegraphics[trim={0 0 14cm 0},clip, width=\textwidth]{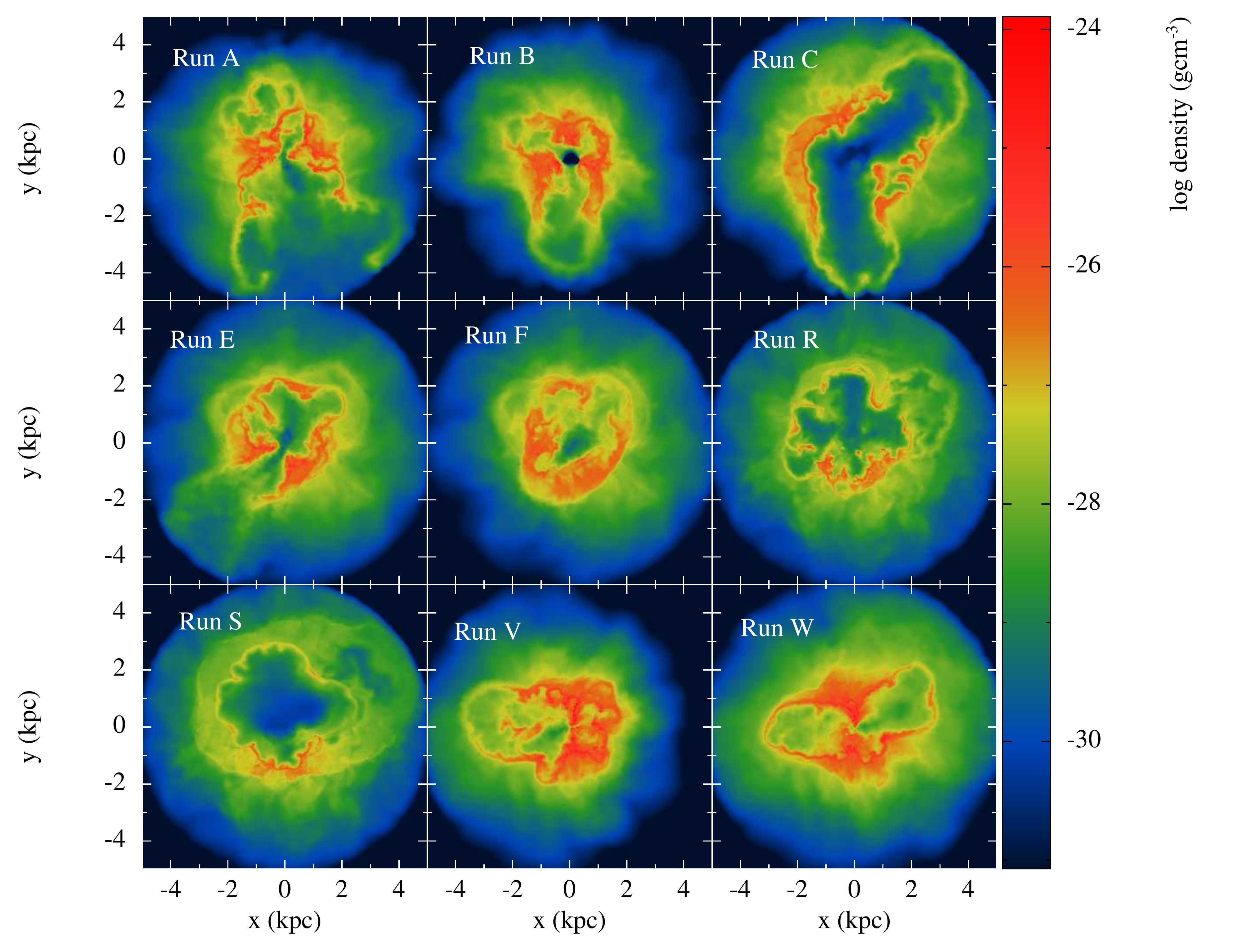}
  \caption{Density slices in the x-y plane at z=0, for the simulations run at solar metallicity, taken $t = 35$\,Myr into each simulation.}
  \label{fig:Dens_Zsol} 
\end{figure*}
\begin{figure*} 
 \includegraphics[trim={0 0 14cm 0},clip, width=\textwidth]{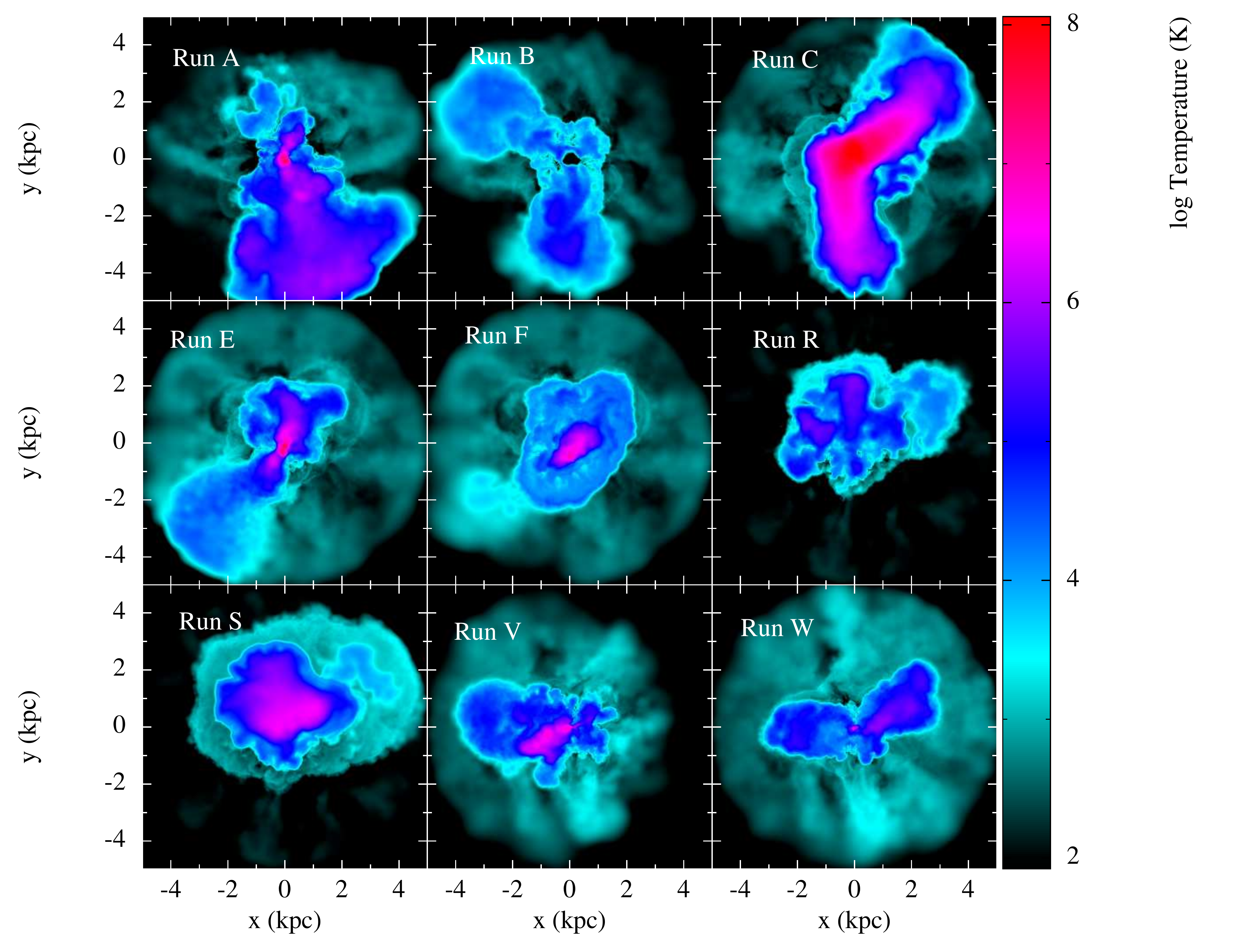}
  \caption{Temperature slices in the x-y plane at z=0, for the simulations run at solar metallicity, taken $t = 35$\,Myr into each simulation.}
  \label{fig:Temp_Zsol} 
\end{figure*}
\begin{figure} 
 \psfig{file=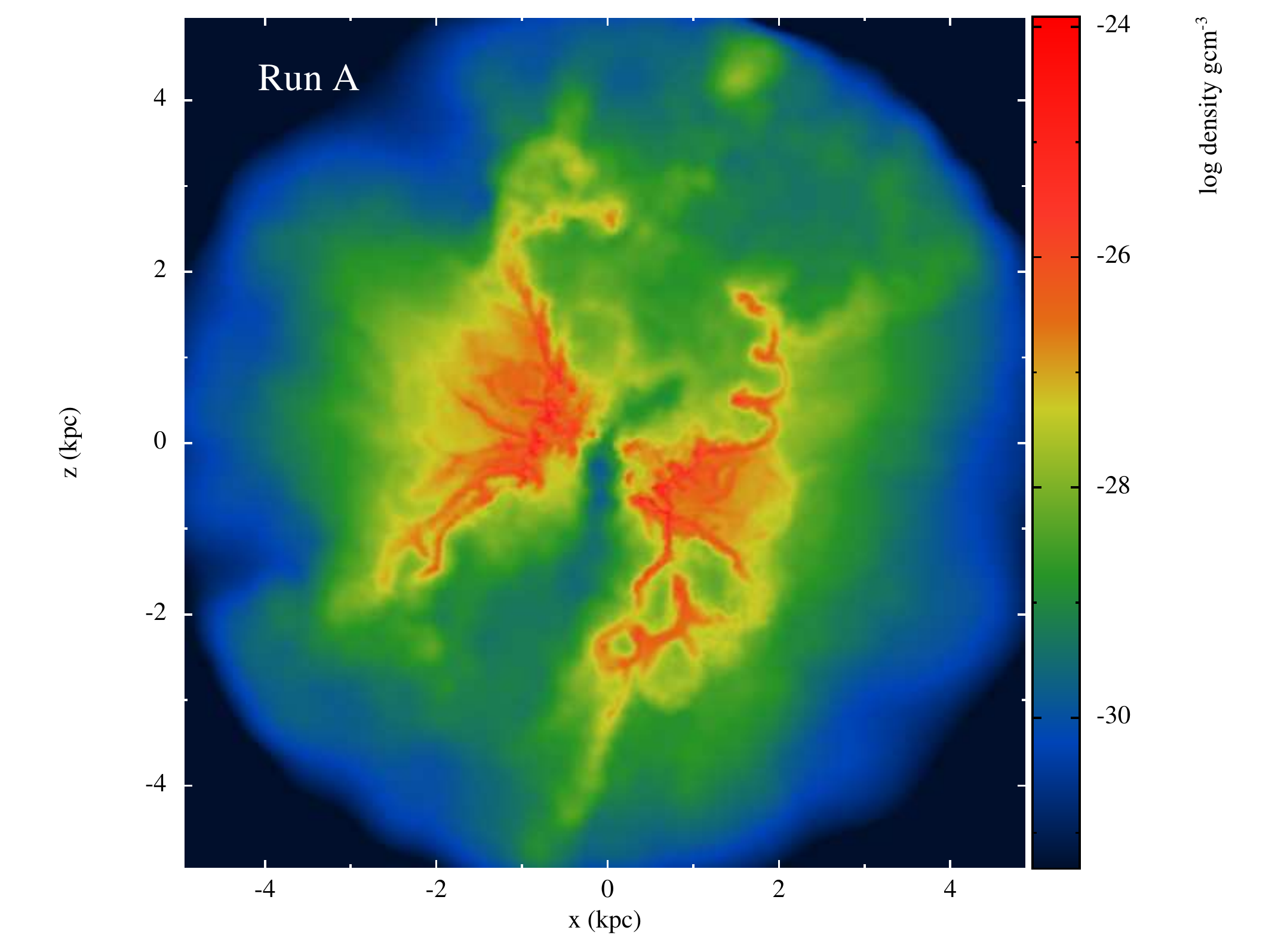,width=0.5\textwidth,angle=0}
  \caption{Density slice in the x-z plane at y=0, for Run A at $t = 35$\,Myr. }
  \label{fig:RunA_xz} 
\end{figure}
Density slices for the $Z=Z_\odot$ runs, taken at $t = 35$\,Myr into each simulation, are shown in Fig. \ref{fig:Dens_Zsol}. Focusing on Runs A (HMXB and SN feedback) and C (just SN feedback), the addition of HMXB feedback on top of SN feedback has resulted in a larger amount of high density (10$^{-24}$ gcm$^{-3}$) gas inside the inner 2 kpc. Moreover, a lobe of lower density gas can be seen to extend from the inner $\sim$ 100 pc out to 5\,kpc of cloud A. This low density lobe represents a possible `chimney' -- a region of lower density gas through which hot, feedback-heated gas can expand and escape the core of the molecular cloud. Looking instead at the temperature slice for Run A (Fig. \ref{fig:Temp_Zsol}), we see the lower right chimney does indeed contain hot gas, extending to the outer regions of the simulation ($\sim$ 5\,kpc). These chimneys are also present in the corresponding x-z density slice (see Fig. \ref{fig:RunA_xz}). For Run C in Fig. \ref{fig:Dens_Zsol}, the dominant feature is the multi-lobed super-bubble. It is also evident the central kpc of the molecular cloud has been efficiently cleared of gas (this can also be verified in the corresponding x-z and y-z plane density slices). Moreover, this isotropic heating of the central kpc can also be seen in the corresponding temperature slice for Run C (Fig. \ref{fig:Temp_Zsol}), where the inner kpc is filled with gas heated to $\sim$ 10$^8$ K. There is no obvious correlation between the super-bubble lobes visible in Run C and the chimneys in Run A at the end of each simulation. However, both the chimney and the super-bubbles are surrounded by arcs of cold gas (seen in Fig. \ref{fig:Temp_Zsol}).
\begin{figure*} 
 \psfig{file=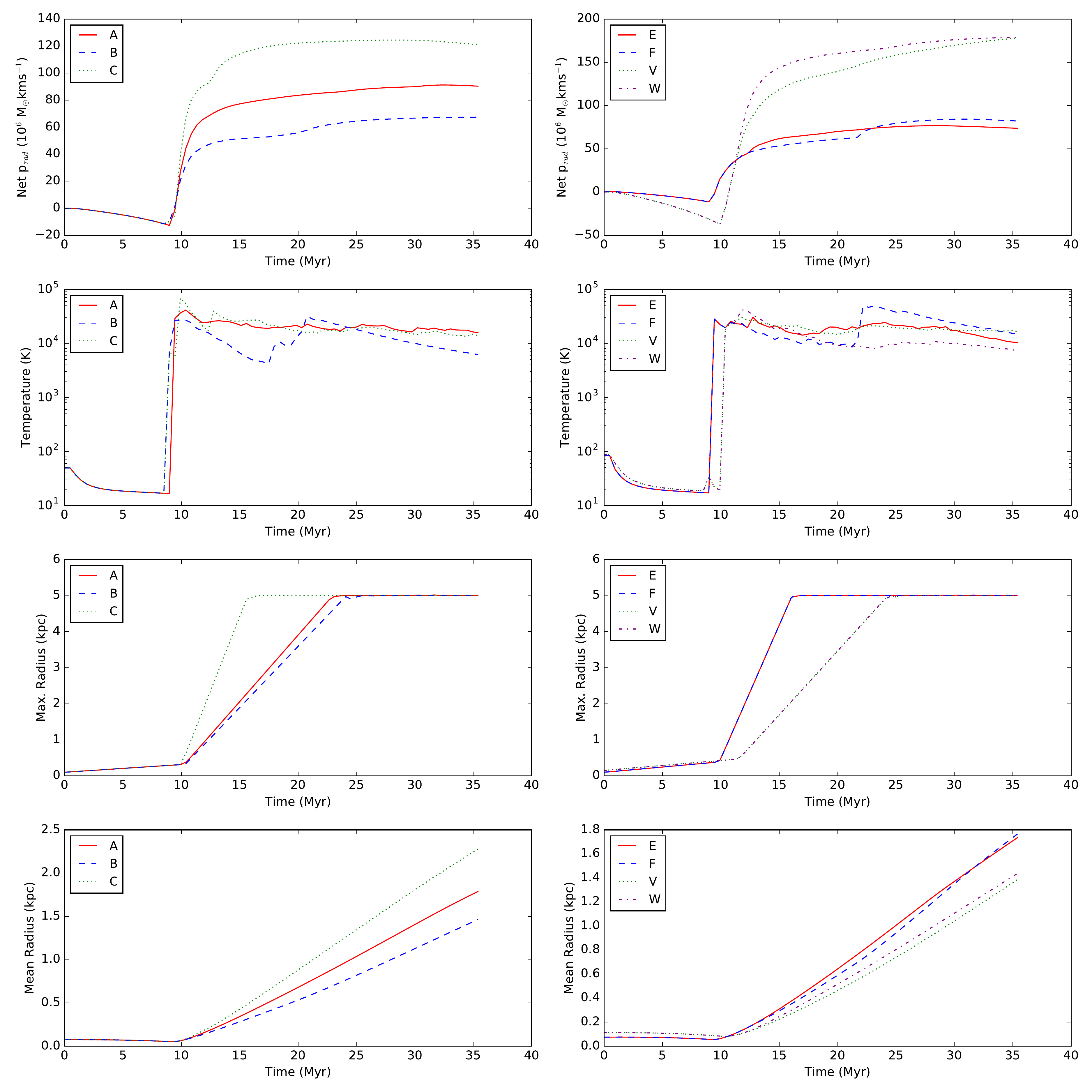,width=\textwidth,angle=0}
  \caption{Upper row - the net radial momentum of the gas across Runs A,B,C,E,F,V and W (evaluated at each snapshot time). Second row - the time evolution of the mean temperature of the gas across each simulation. Third row - the time evolution of the maximum radius of the gas. Bottom row - the time evolution of the mean radius of the gas particles.}
  \label{fig:Zsol_pnet} 
\end{figure*}
\begin{figure*} 
 \psfig{file=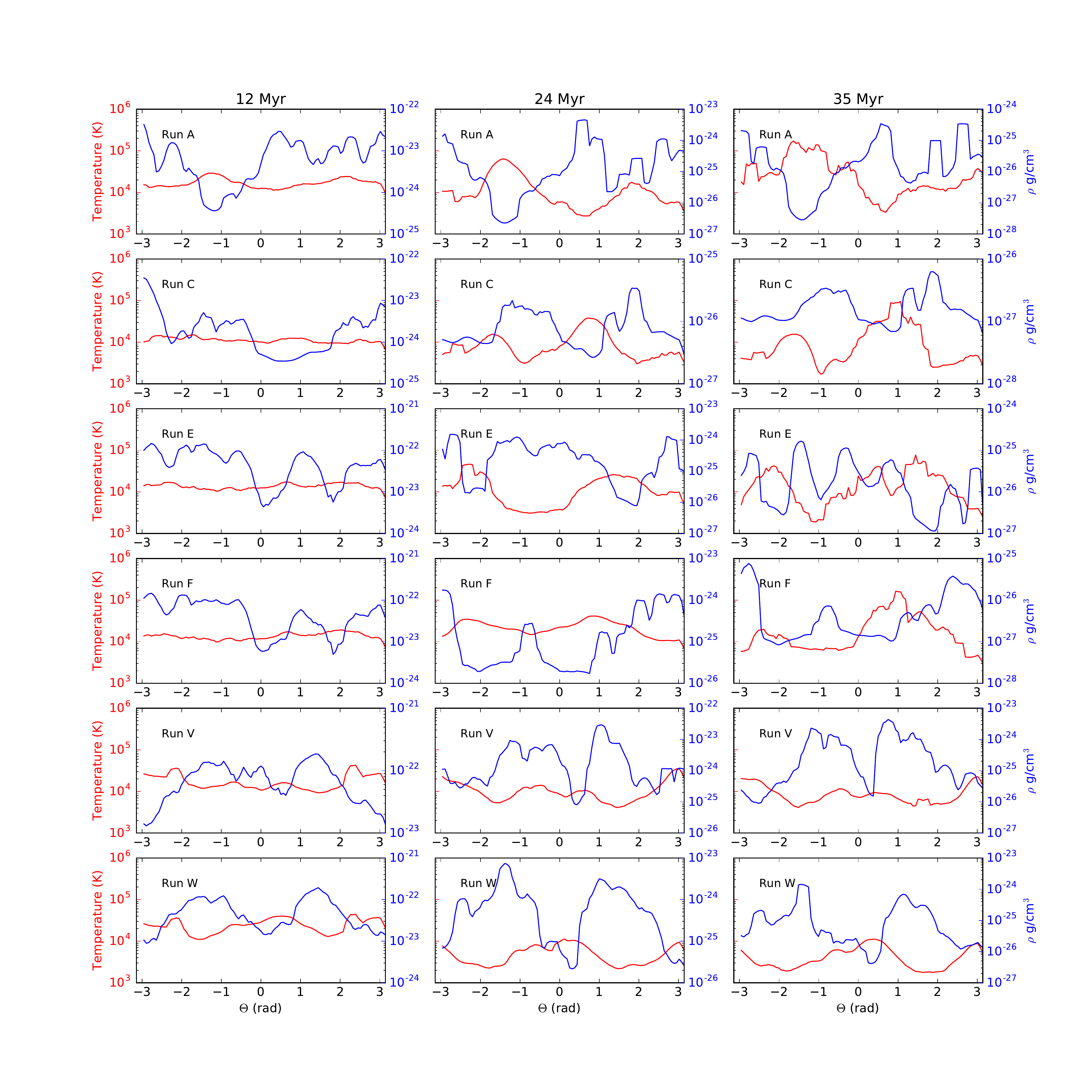,width=\textwidth,angle=0}
  \caption{Plots to show the mean temperature (red) and density (blue) in $\theta$ bins ranging from -$\pi$ to $\pi$ radians. The maximal radius of each $\theta$ bin is set to the maximum radius of the gas in the individual bin.} The left column is for snapshots taken at 12 Myr, the middle column is for 20 Myr and the right column shows snapshots at 32 Myr. The name of the corresponding run is in the upper left hand corner of each plot. 
  \label{fig:theta_bins} 
\end{figure*}

To ascertain the difference between the global properties of the gas in Runs A and C, we plot the net radial momentum of the gas, along with the mean temperature, mean radius and maximum radius in Fig. \ref{fig:Zsol_pnet}. From this plot we can see both the net (outwards) radial momentum and mean radius of the gas in Run C are consistently higher than in Run A. Moreover, the mean temperature of Runs A and C converge beyond $\sim 10\,$Myr, at a value of $\sim$ 10$^{4.4}$ K. This indicates the mean temperature of both runs is dominated by the $10^4$ K gas, where cooling via collisional excitation is highly inefficient.

To investigate the formation of the low density chimneys of Run A further, we binned the density and temperature data in spherical polar azimuthal angle $\theta$ (ranging from -$\pi$ to $\pi$), using a maximal radius set by the largest radius of each $\theta$ bin, and plotted the mean temperature and density in each bin (Fig. $\ref{fig:theta_bins}$) at 3 different snapshot times; 12 ($\sim$ a free-fall time) Myr, 24 Myr and 35 Myr ($\sim$ 2 and 3 times the free-fall time of the cloud respectively). It should be noted the temperature seen in Fig. $\ref{fig:theta_bins}$ is expected to be lower than that seen in Fig. $\ref{fig:Temp_Zsol}$ since it represents an average value, encompassing all radii within each $\theta$ bin. On Fig. \ref{fig:theta_bins}, a chimney manifests as a peak in the temperature at a specific $\theta$, corresponding with a trough in density at the same $\theta$ value. The steepness of the temperature and density gradient indicates the prominence and efficiency of the chimney. Using Fig. \ref{fig:theta_bins} we can therefore investigate when and where chimneys develop. We chose the x-y plane since this corresponds with the density and temperature slices in figures \ref{fig:Dens_Zsol} and \ref{fig:Temp_Zsol}. 

At 12 Myr, the $\theta$-density profiles of A and C have already diverged, despite the initial conditions prior to the first HMXB feedback event (inside Run A) being identical. The density is consistently lower in Run C than Run A, across all $\theta$ bins. This indicates gas has been heated isotropically (as was seen in the inner kpc of Run C in Fig. \ref{fig:Temp_Zsol}) and that the maximal radii of the gas in each $\theta$ bin is larger in Run C than Run A (as verified in Fig. \ref{fig:Zsol_pnet}). There is also a two orders of magnitude density drop at $\theta \sim -2.5$ rad and no corresponding positive density gradient until beyond 2 rad. On the other hand, Run A contains both a two orders of magnitude density drop and rise within -2 to 0 rad. This indicates a clearly defined low density chimney. Moreover, comparing the location of the chimney in Run A with the corresponding location in Run C, we can see there is a density peak in Run C spanning a single order of magnitude. Again, given that Run A and Run C are identical (with respect to both the gas density inhomogeneities and the locations of massive stars) prior to the onset of HMXB feedback, this suggests the gas in Run C was able to cool efficiently, preventing a chimney forming. On the other hand, the addition of HMXB feedback in Run A kept the gas in the chimney hot and retained the low density channel, through which hot gas can be funneled. 

The chimney in Run A (seen in Fig. \ref{fig:theta_bins}) develops in prominence from 24-35 Myr, finally spanning approximately 3 orders of magnitude in density and one order of magnitude in temperature. Looking at Fig. \ref{fig:Zsol_pnet}, we can see by 35 Myr the maximum radius is the same in Runs A and C, hence the higher density peaks in Run A indicate mass clustering in particular directions. Chimney-like features can also be seen to develop in Run C, however the density gradient across these features only spans 1 order of magnitude, while the temperature of the gas associated with them is consistently lower than in the chimney in Run A. It is also worth noting the x-z and y-z planes were also investigated, which showed the same behavior as the x-y plane; i.e. the density in each $\theta$ bin was lower in Run C than Run A, while there were hot, low density chimneys present in Run A which were not present in Run C. 

\begin{figure*} 
 \psfig{file=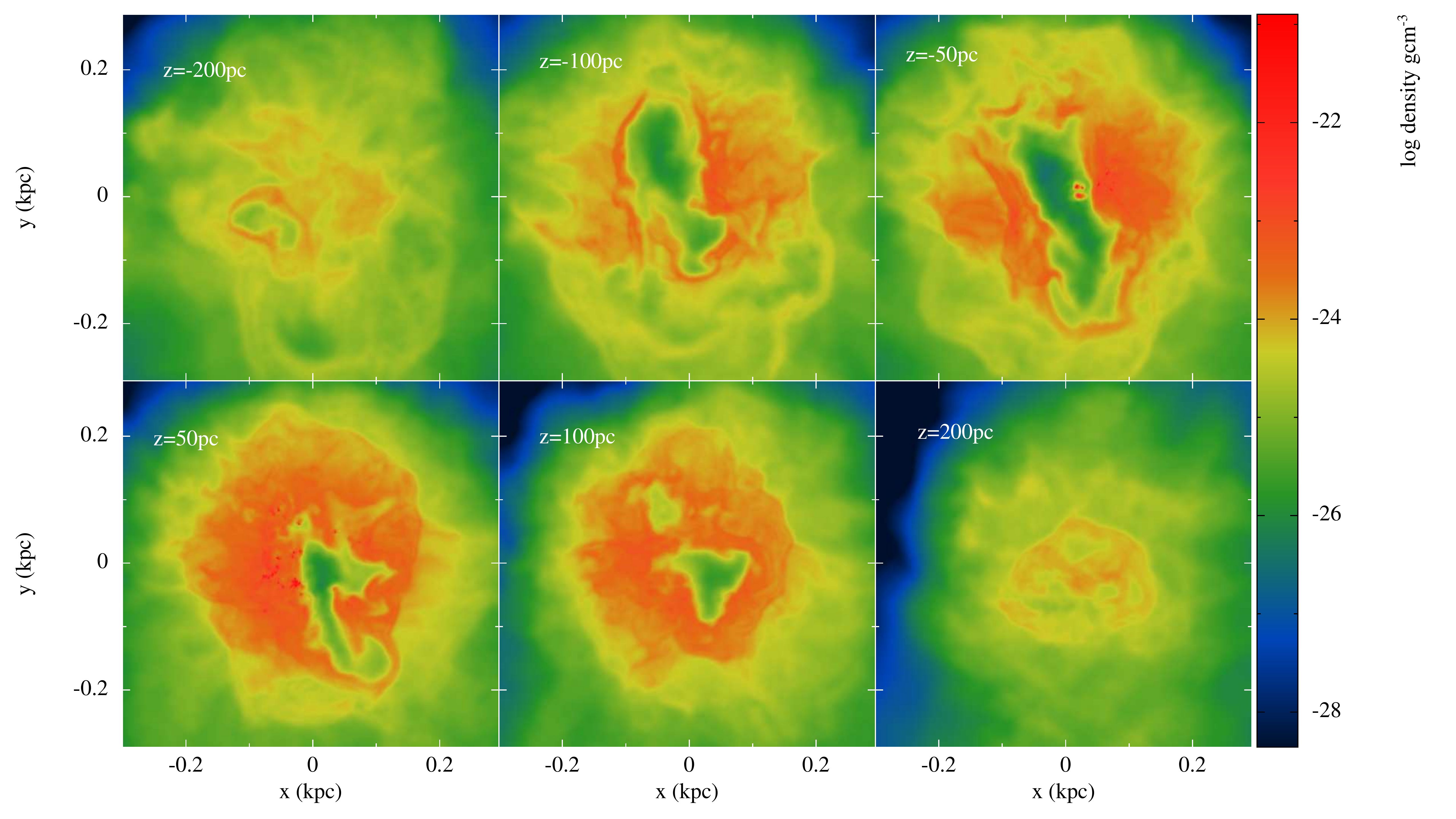,width=\textwidth,angle=0}
  \caption{Density slices in the x-y plane of Run A at z=-0.2, -0.1, -0.05, 0.05, 0.1 and 0.2 kpc at 12 Myr into the simulation ($\sim$ 1 free-fall time.)  }
  \label{fig:Dens_Zslices_A} 
\end{figure*}

\begin{figure*} 
\psfig{file=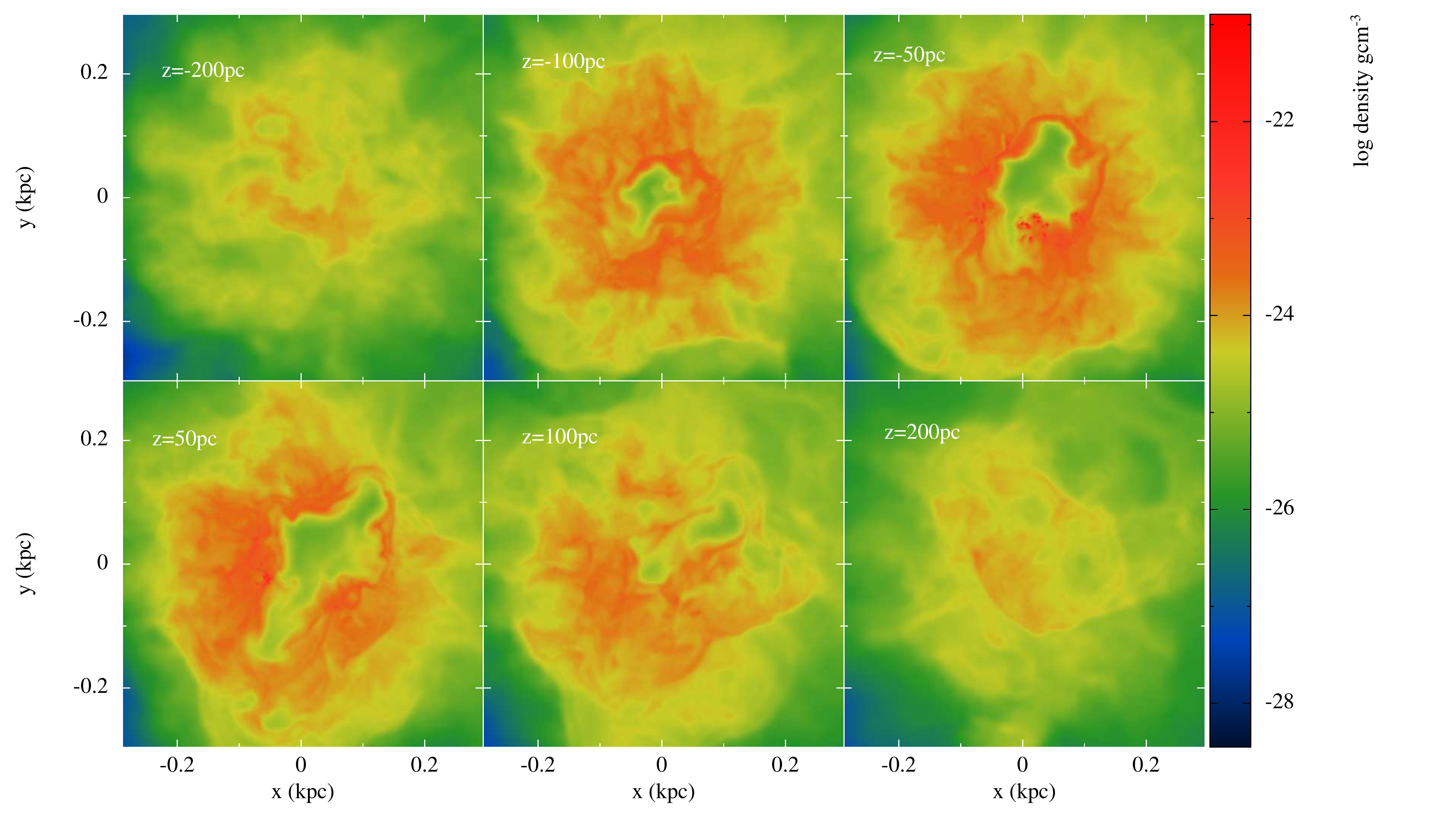,width=\textwidth,angle=0}
  \caption{Density slices in the x-y plane of Run C at z=-0.2, -0.1, -0.05, 0.05, 0.1 and 0.2 kpc at 12 Myr into the simulation ($\sim$ 1 free-fall time.)  }
  \label{fig:Dens_Zslices_C} 
\end{figure*}

\begin{figure*} 
\psfig{file=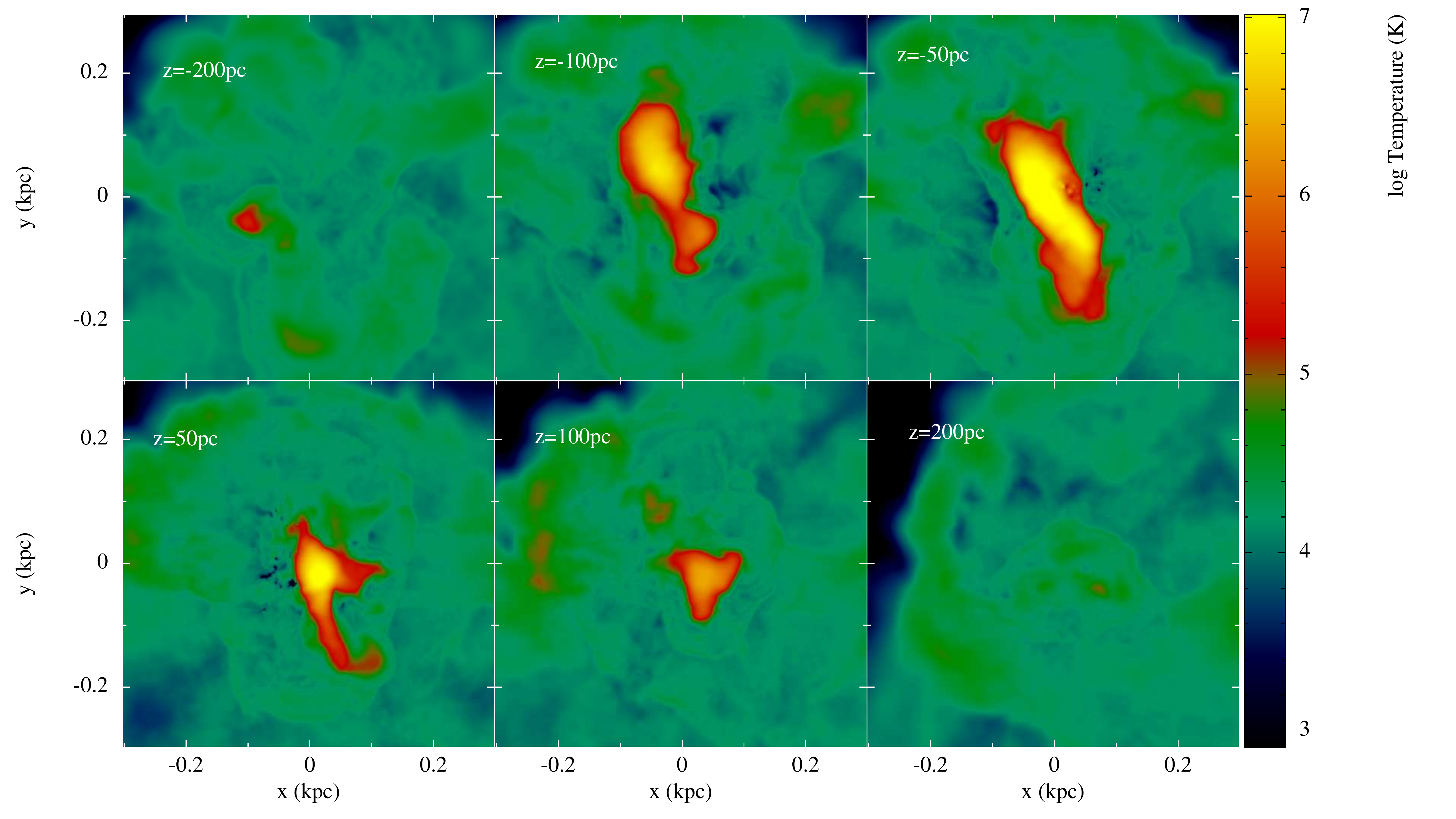,width=\textwidth,angle=0}
  \caption{Temperature slices in the x-y plane of Run A at z=-0.2, -0.1, -0.05, 0.05, 0.1 and 0.2 kpc at 12 Myr into the simulation ($\sim$ 1 free-fall time.)  }
  \label{fig:Temp_Zslices_A} 
\end{figure*}

\begin{figure*} 
	 \includegraphics[trim={0 4cm 0 0},clip, width=\textwidth]{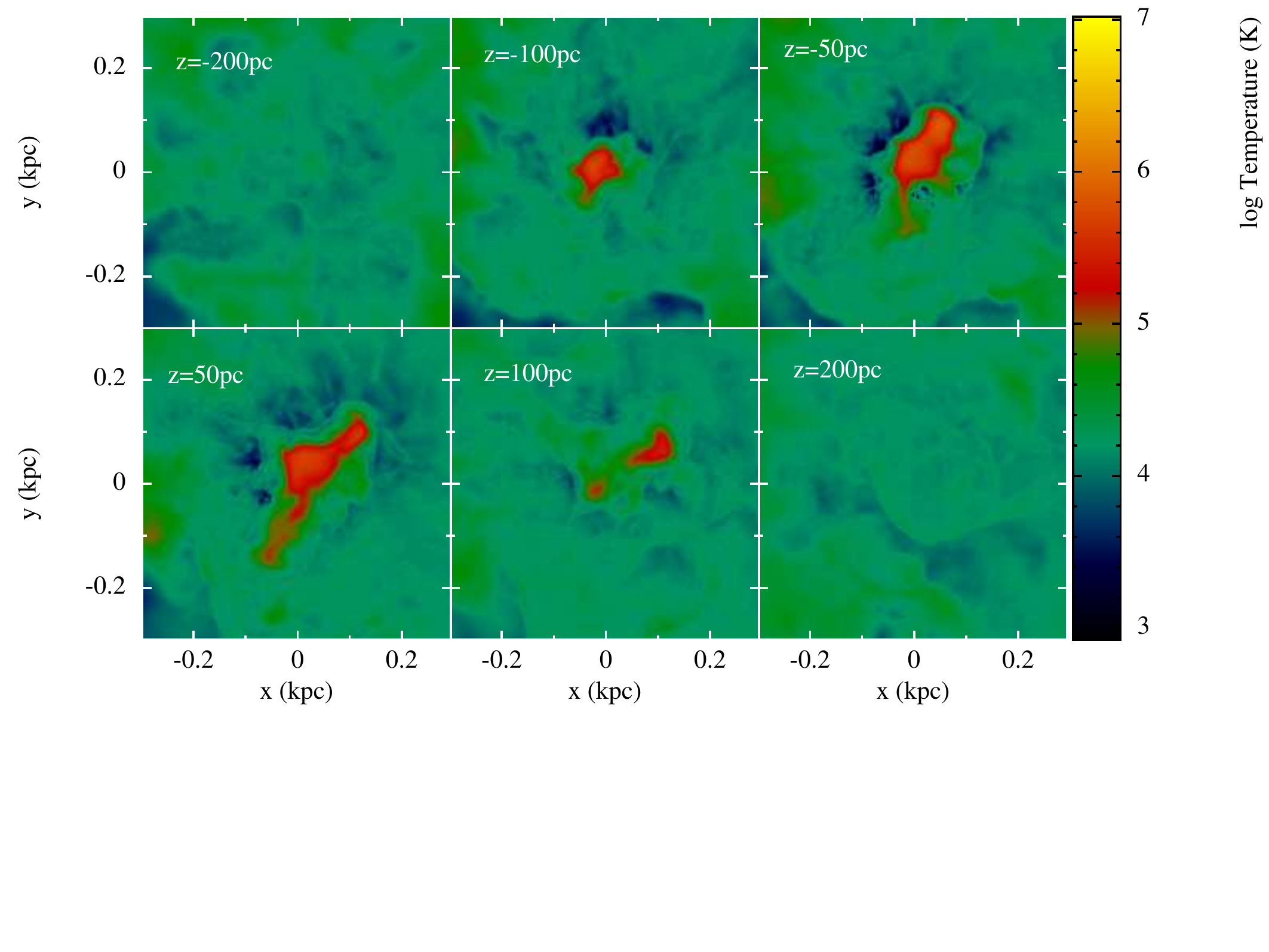}
  \caption{Temperature slices in the x-y plane of Run C at z=-0.2, -0.1, -0.05, 0.05, 0.1 and 0.2 kpc at 12 Myr into the simulation ($\sim$ 1 free-fall time.)  }
  \label{fig:Temp_Zslices_C} 
\end{figure*}

In order to visualise the developing chimneys in 3D we plotted the density and temperature slices across z slices spanning z = -0.2 kpc to 0.2 kpc for both Runs A (figures \ref{fig:Dens_Zslices_A} and \ref{fig:Temp_Zslices_A}) and C (\ref{fig:Dens_Zslices_C} and \ref{fig:Temp_Zslices_C}) at 12 Myr ($\sim$ 1 free-fall time). Comparing figures \ref{fig:Dens_Zslices_A} and \ref{fig:Dens_Zslices_C}, we can see there are chimneys present in both Runs, however those in Run A are filled with lower density gas ($<$ 10$^{-26}$ gcm$^{-3}$). They also appear to be more spatially extended, particularly in the z=-50\,pc slice. 

Looking at figures \ref{fig:Temp_Zslices_A} and \ref{fig:Temp_Zslices_C}, we can see the gas inside the chimneys in Run A is also an order of magnitude hotter than in Run C. The hot gas in Run C reaches a peak in temperature of $\sim$ 10$^6$ K, which corresponds to an enhanced region of the cooling function for solar metallicity gas; primarily due to the collisional excitation of elements such as neon and iron. However, the hot gas in the chimneys in Run A peaks around 10$^7$ K, corresponding to Bremsstrahlung dominated cooling, which is relatively inefficient, particularly due to the low densities in the chimneys. In this way, the HMXB feedback has acted to increase the temperature of the gas in the chimneys and in so doing has reduced its ability to cool efficiently. The chimneys can then continue to provide a `path of least resistance' through which hot gas can escape, allowing star formation to continue in other regions of the cloud.

\begin{figure} 
 \psfig{file=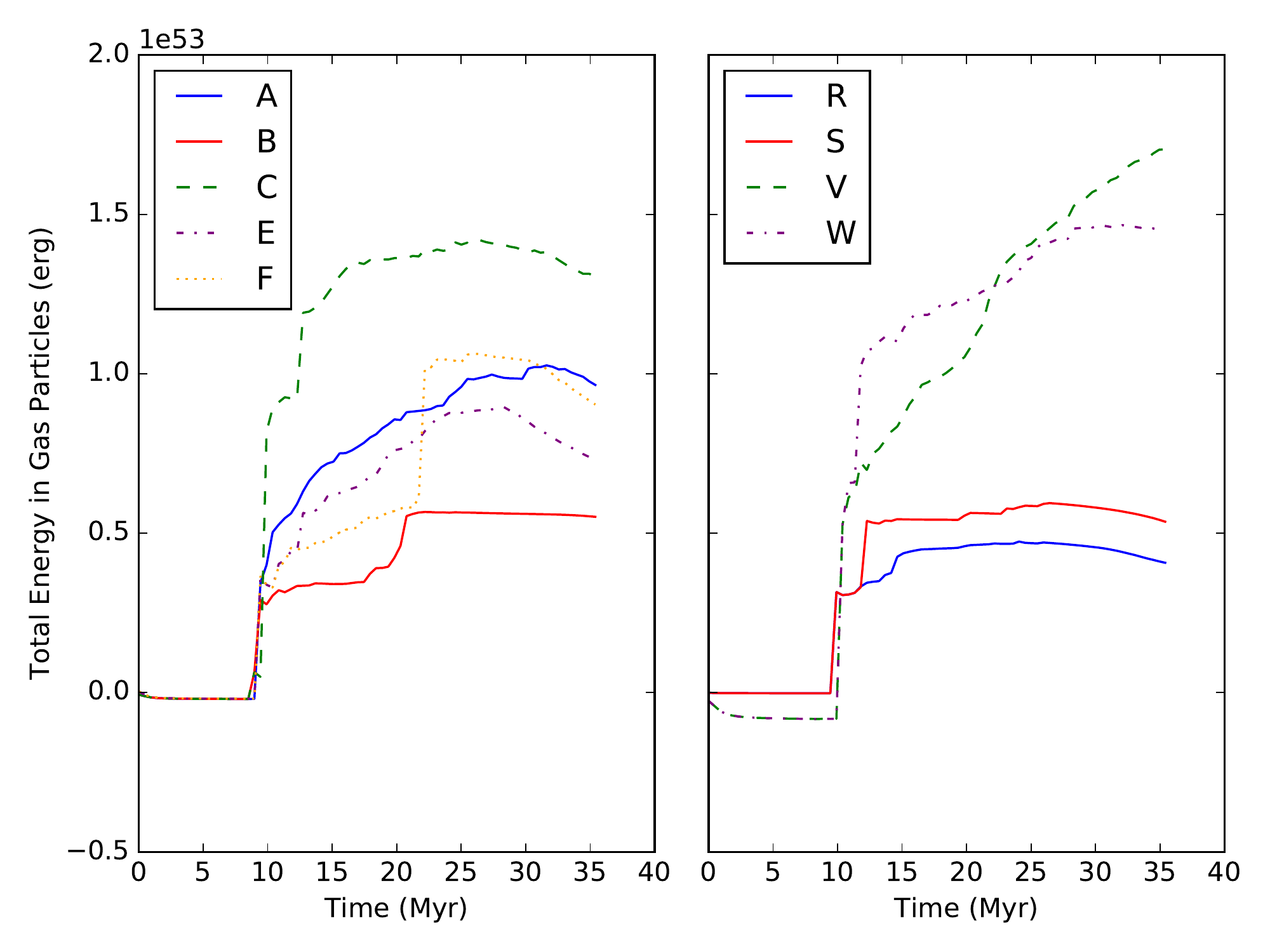,width=0.5\textwidth,angle=0}
  \caption{Plots to show the time evolution of the total energy of Runs A, B,C,E,F,R,S,V and W. The total energy was found by summing the total thermal, kinetic and potential energies across all gas particles in each simulation.}
  \label{fig:Etot_Zsol} 
\end{figure}

\begin{figure} 
\psfig{file=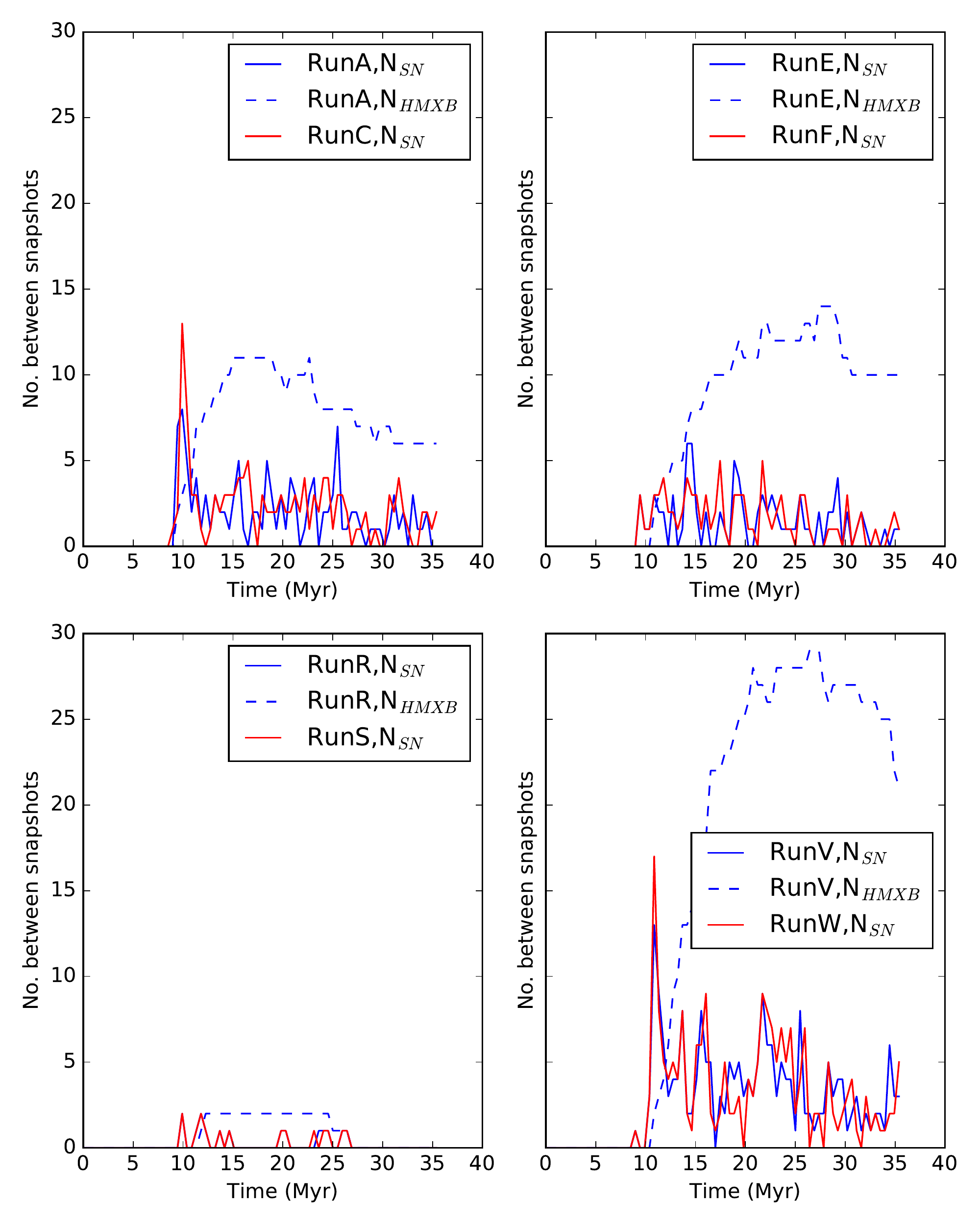,width=0.5\textwidth,angle=0}
  \caption{Plots to show the number of SNe (solid lines) and HMXBs (dashed lines) active between snapshots for Runs A,C,E,F,R,S,V and W.}
  \label{fig:Zsol_HMO} 
\end{figure}

\begin{figure}
 \psfig{file=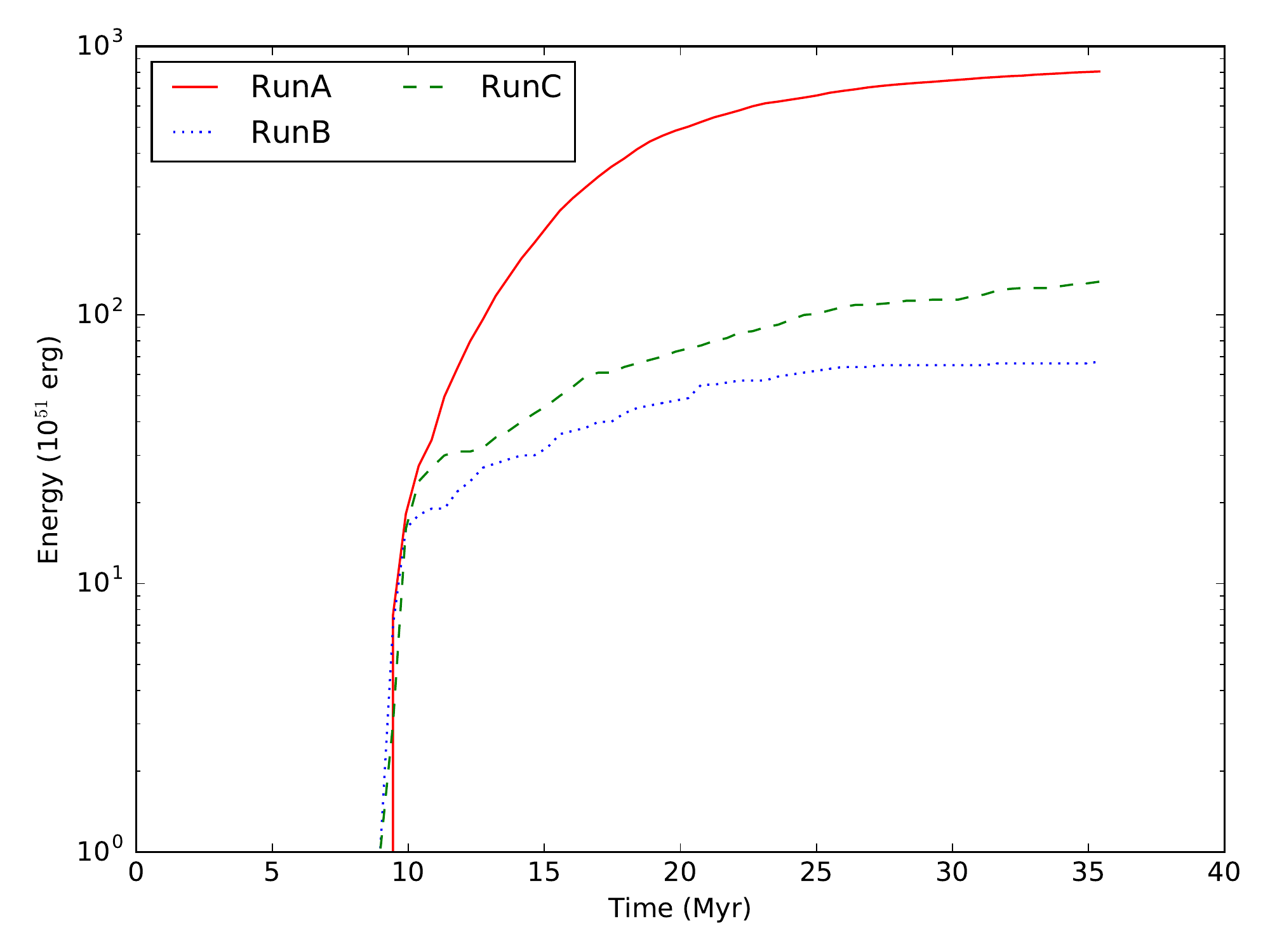,width=0.5\textwidth,angle=0}
  \caption{Figure to show the time evolution of the cumulative injected energy for Runs A,B and C.}
  \label{fig:Zsol_Einj} 
\end{figure}

In Fig. \ref{fig:Etot_Zsol} we plot the time evolution of the total energy; $E_{\rm tot}$ (where $E_{\rm tot}$ is the sum of the total thermal, kinetic and potential energy of the gas in the system)  of selected solar metallicity runs. Again, focusing on the results for Runs A and C, we see the addition of HMXB feedback on top of SN feedback (Run A) has resulted in a lower total energy of the system. Moreover, Fig. \ref{fig:Zsol_HMO} plots the number of SNe and HMXBs active between snapshot times, where we see Runs A and C have similar numbers of SNe throughout the simulation, however Run A has an additional energy source from the $\sim$ 10 extra HMXB feedback events occurring at at any one time. Furthermore, Fig. \ref{fig:Zsol_Einj} plots the cumulative injected energy from both HMXB feedback in Run A and just SN feedback in Run C and it is clear the injected energy for Run A is greater than Run C. 

\begin{figure} 
\psfig{file=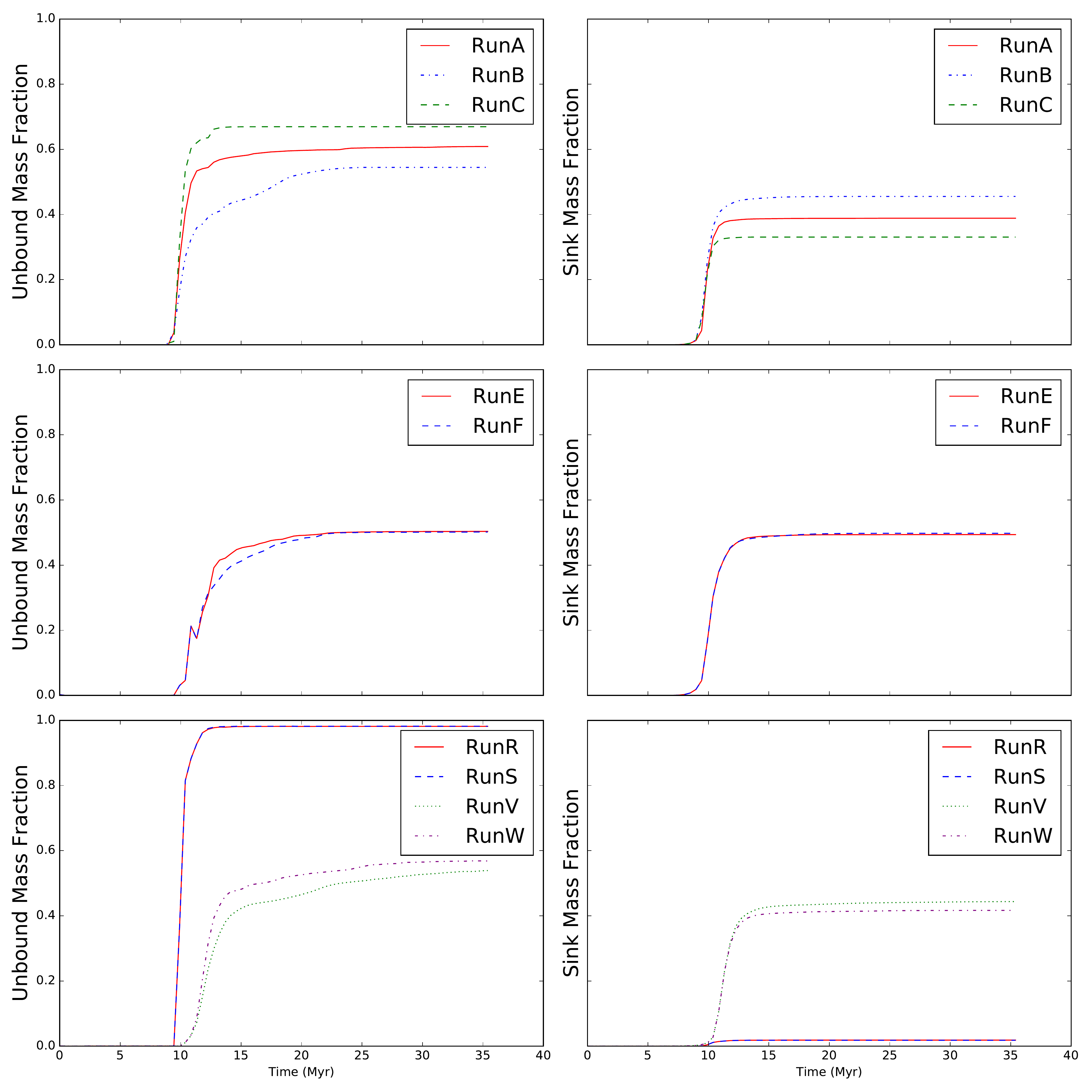,width=0.5\textwidth,angle=0}
  \caption{Plots to show the fraction of the initial mass that has (a) been unbound (first column) or (b) either been accreted by, or become sink particles (second column). The runs plotted are:  A,B,C,E,F,R,S,V and W (which are all at solar metallicity).}
  \label{fig:Zsol_unbound} 
\end{figure}

In Fig. \ref{fig:Zsol_unbound} we plot the fraction of the original gas mass in each run that has; (a) become unbound (first column) or (b) ended up in sink particles (second column). Fig. \ref{fig:Zsol_unbound} shows 5\% less gas has been unbound in Run A compared with Run C. This gas has instead become star forming, adding to the total sink particle mass in the simulation. Furthermore, as expected from Fig. \ref{fig:Etot_Zsol}, fractionally more gas has been ejected from the simulation domain in Run C than in Run A. From the sink mass fraction, we can also see the majority the star formation has occurred in the free-fall time of the cloud, making the first 11.7 Myr crucial in determining the star formation efficiency of the cloud. 

\begin{figure} 
\psfig{file=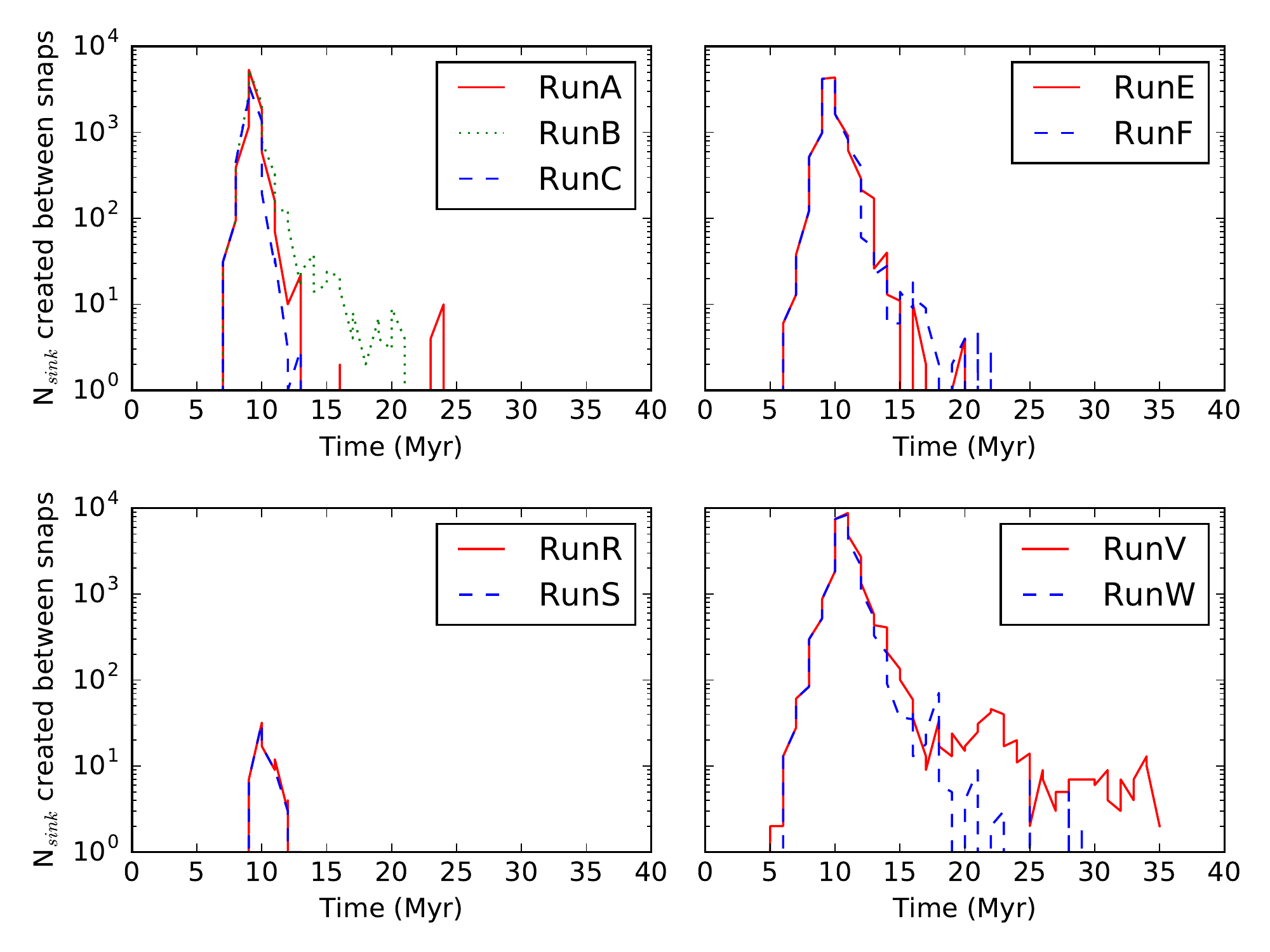,width=0.5\textwidth,angle=0}
  \caption{Plots to show the number of sink particles produced between timesteps for Runs A,B,C,E,F,R,S,V and W (solar metallicity).} 
  \label{fig:Zsol_SFE} 
\end{figure}

Fig. \ref{fig:Zsol_SFE} plots the number of sink particles created between snapshots versus time into the simulation. The chimney seen in Run A, e.g. Fig. \ref{fig:Dens_Zsol}, has acted to keep a fraction of the gas cool and dense enough to become star forming beyond 11.7 Myr. This has resulted in additional sink particles being formed at 16 Myr and 23-24 Myr ($\sim$ 2 free-fall time-scales). In contrast, the gas in Run C ceases to produce sink particles beyond 14 Myr (just over 1 free-fall time of the molecular cloud). 

Comparing Fig. \ref{fig:Zsol_SFE} to Fig. \ref{fig:Nofb_sink}, we can see the addition of feedback in Runs A and C has had little impact on when the majority of sink particles are formed inside the cloud. Most of the sink particles in each simulation are produced in a single burst of star formation between 6-13 Myr.

\begin{figure} 
\psfig{file=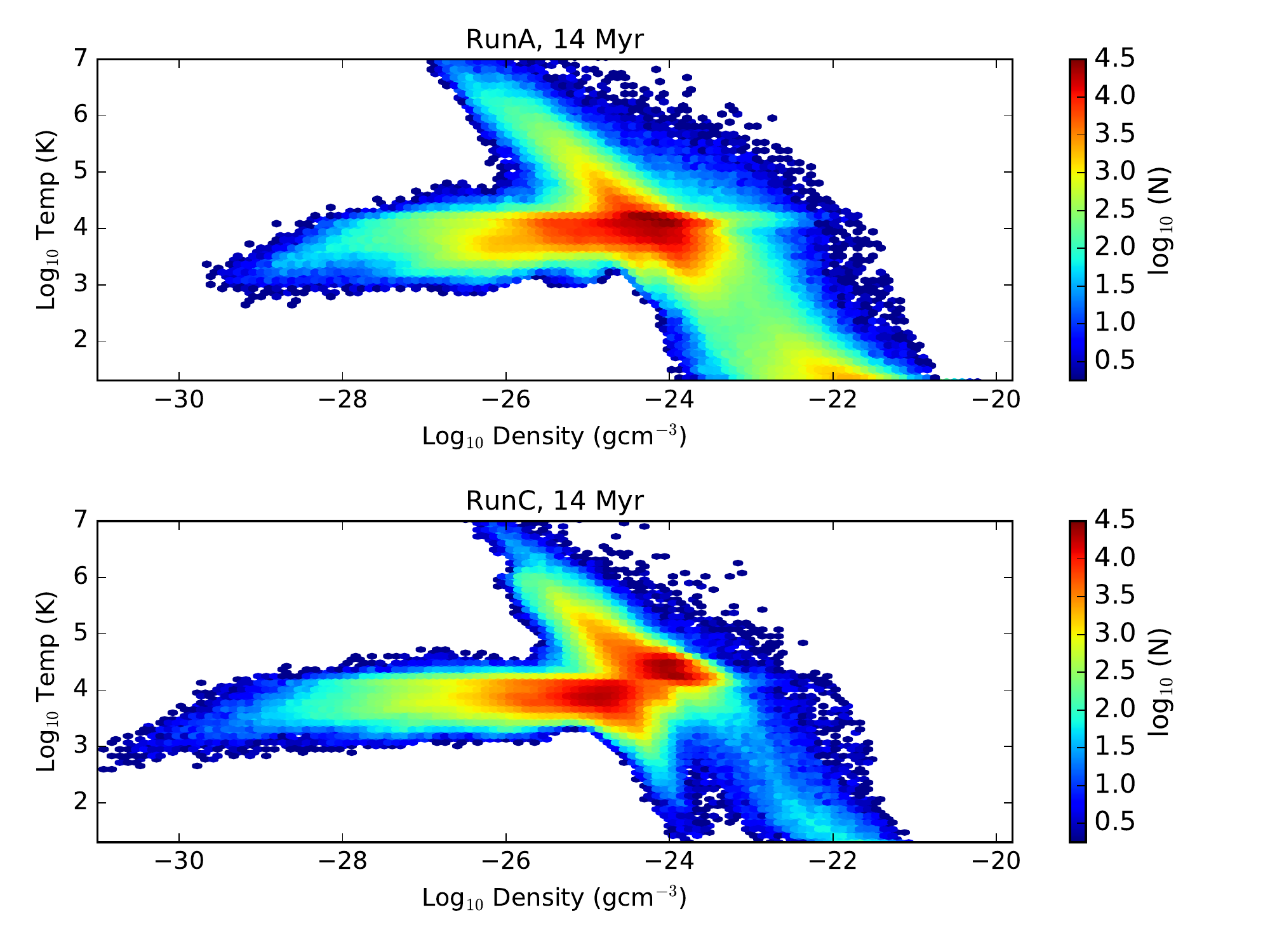,width=0.5\textwidth,angle=0}
  \caption{Plots to show the temperature versus density of the gas in Runs A (top) and C (bottom), 14 Myr into each simulation. The plot is rendered according to the number of gas particles.} 
  \label{fig:Zsol_TvsD} 
\end{figure}

In Fig. \ref{fig:Zsol_TvsD} we investigate the state of the gas in Runs A and C at t = 14 Myr, in order to ascertain why Run A continues to form sink particles beyond this point and Run C does not. We plot the temperature versus density of the gas in the simulation, rendered according to the number of particles. From Fig. \ref{fig:Zsol_TvsD}, we can see both Runs A and C have a `tail' of low temperature gas, formed primarily by fine structure cooling of the heavier elements (e.g. C II, O I), however this `tail' contains more gas particles in Run A, resulting in the extended sink particle formation seen in Fig. \ref{fig:Zsol_SFE}. 

Overall, by comparing Runs A and C we have found that adding HMXB feedback on top of SN feedback can result in an increase in star formation efficiency and period, despite more energy being injected into the ISM. This result arises primarily through the action of low density `chimneys', funneling hot gas from the inner regions, maintaining enough cool, dense gas to fuel further star formation. These chimneys are also present when just SN feedback is present (see Fig. \ref{fig:Temp_Zslices_C}), however the gradual heating from HMXBs acts to increase the temperature of the hot gas in these chimneys and reduces its ability to cool efficiently -- enhancing their effectiveness at funneling hot, destructive gas away from the inner regions of the cloud. Despite this, the majority of star formation in clouds A and C still occurs within the free-fall time of the cloud, as in Run D, making the first $\sim$ 11.7 Myr pivotal in determining the star formation efficiency and rate of these molecular clouds.

\subsubsection{Changing the virial parameter, $\alpha_{vir}$ $=$ 1.2: Runs E,F}\label{sec:EF}
Looking at the density slices for Runs E (including HMXB feedback) and Run F (just SN feedback) in Fig. \ref{fig:Dens_Zsol}, the differences are less apparent than between Runs A and C. Immediately, we can see making the molecular cloud marginally unbound has particularly affected the run containing just SN feedback (F). Cloud F has retained a larger amount of higher density ($>$ 10$^{-24}$ gcm$^{-3}$) gas inside the inner region than Run C. This is likely the work of chimneys; a few cavities of low density ($\sim$ 10$^{-26}$ to 10$^{-27}$ gcm$^{-3}$) gas can be seen in the inner 1 kpc of Run F. 

On the other hand, comparing Runs E and F, we still see a larger fraction of higher density gas towards the centre of Run E, as well as two clearly defined chimneys. Indeed, the mean density of the gas inside a radius of 500\,pc is 1.5 $\times =\,10^{-24}$ gcm$^{-3}$ for Run E, compared with 5.6 $\times\,10^{-25}$ gcm$^{-3}$ in Run F. Moreover, the total gas mass inside the inner 500 pc of Run E is 3.8 $\times\,10^4$ M$_\odot$, compared with 3.3 $\times\,10^4$ M$_\odot$ for Run F. The fact that the mean density inside this radius varies more widely than the total gas mass between the two simulations, suggests the gas is more clustered inside the inner 500 pc of Run E than Run F. The chimneys visible in the x-y plane of Run E (Fig. \ref{fig:Dens_Zsol}) are also apparent on the corresponding temperature slice (Fig. \ref{fig:Temp_Zsol}), which shows two lobes of high temperature gas expanding from the centre of the gas cloud through the chimneys. One chimney of low density gas is visible in cloud F in Fig. \ref{fig:Dens_Zsol} and the hot gas filling it can be seen in Fig. \ref{fig:Temp_Zsol}. The chimney seen in Run F roughly corresponds to the top-right chimney seen in Run E. This suggests the early SNe (i.e. those initialised prior to the first feedback event and hence shared between Run E and Run F) were primarily responsible for determining the location of this chimney in both clouds. Furthermore, the x-z and y-z planes were also considered; with both clouds showing a hot, dense chimney in the upper right corner of the y-z plane, of similar spatial extent and both containing gas heated to $\sim$ 10$^{8}$ K. Moreover, the x-z plane showed very similar density and temperature trends to the x-y plane, with Run E containing two chimneys (one in the upper right corner and one in the lower left corner) and Run F containing one in the upper right corner. This is indicative of the 3-d structure of the chimneys visible in the x-y plane.

\begin{figure} 
 	\psfig{file=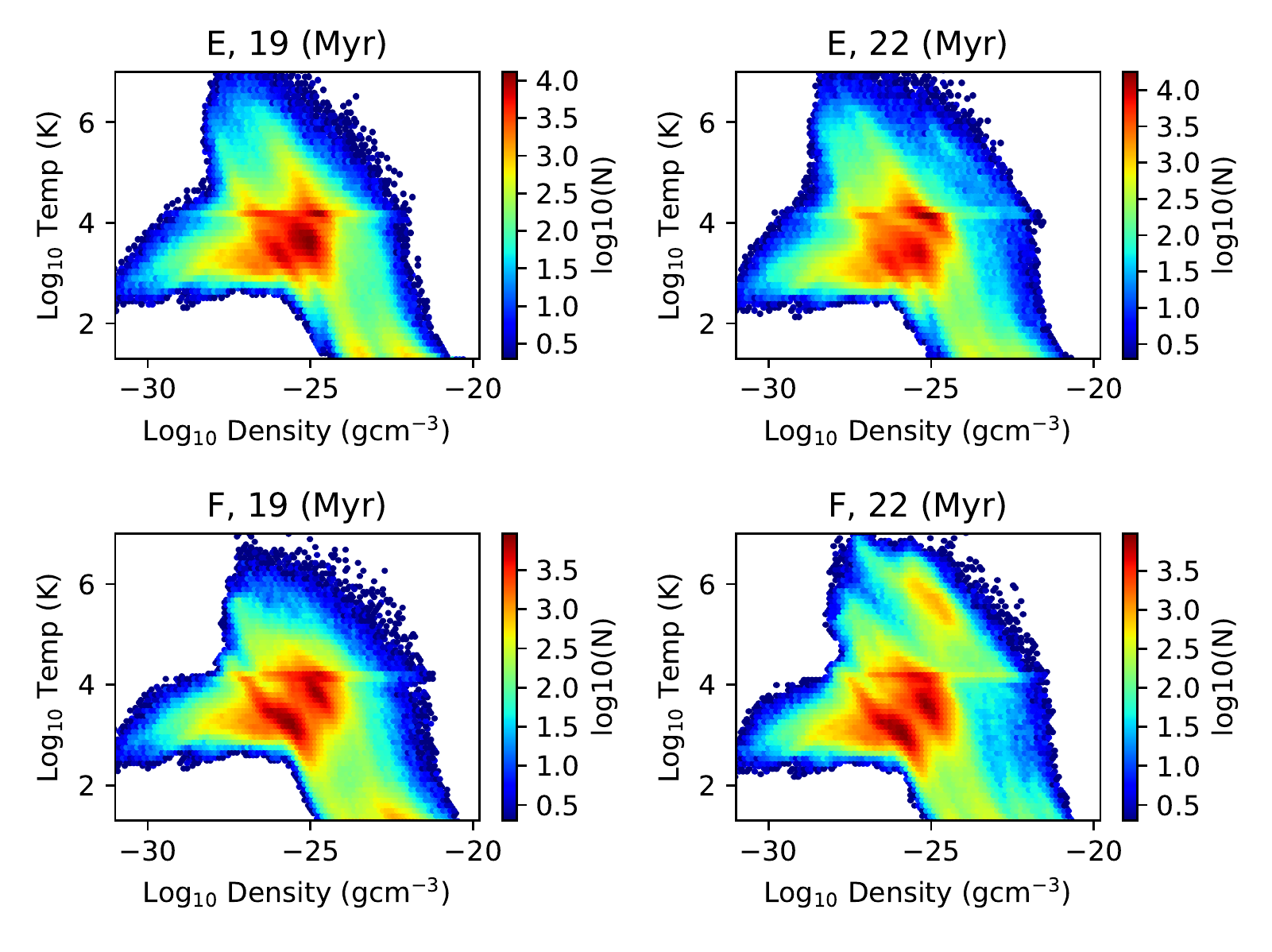,width=0.5\textwidth,angle=0}
  \caption{Temperature versus density plot, rendered according to gas particle number, for Runs E and F, 19 Myr and 22 Myr into each simulation. } 
  \label{fig:Zsol_TvsD_EF} 
\end{figure}
\begin{figure*}
\includegraphics[trim={0 0 9cm 0},clip, width=\textwidth]{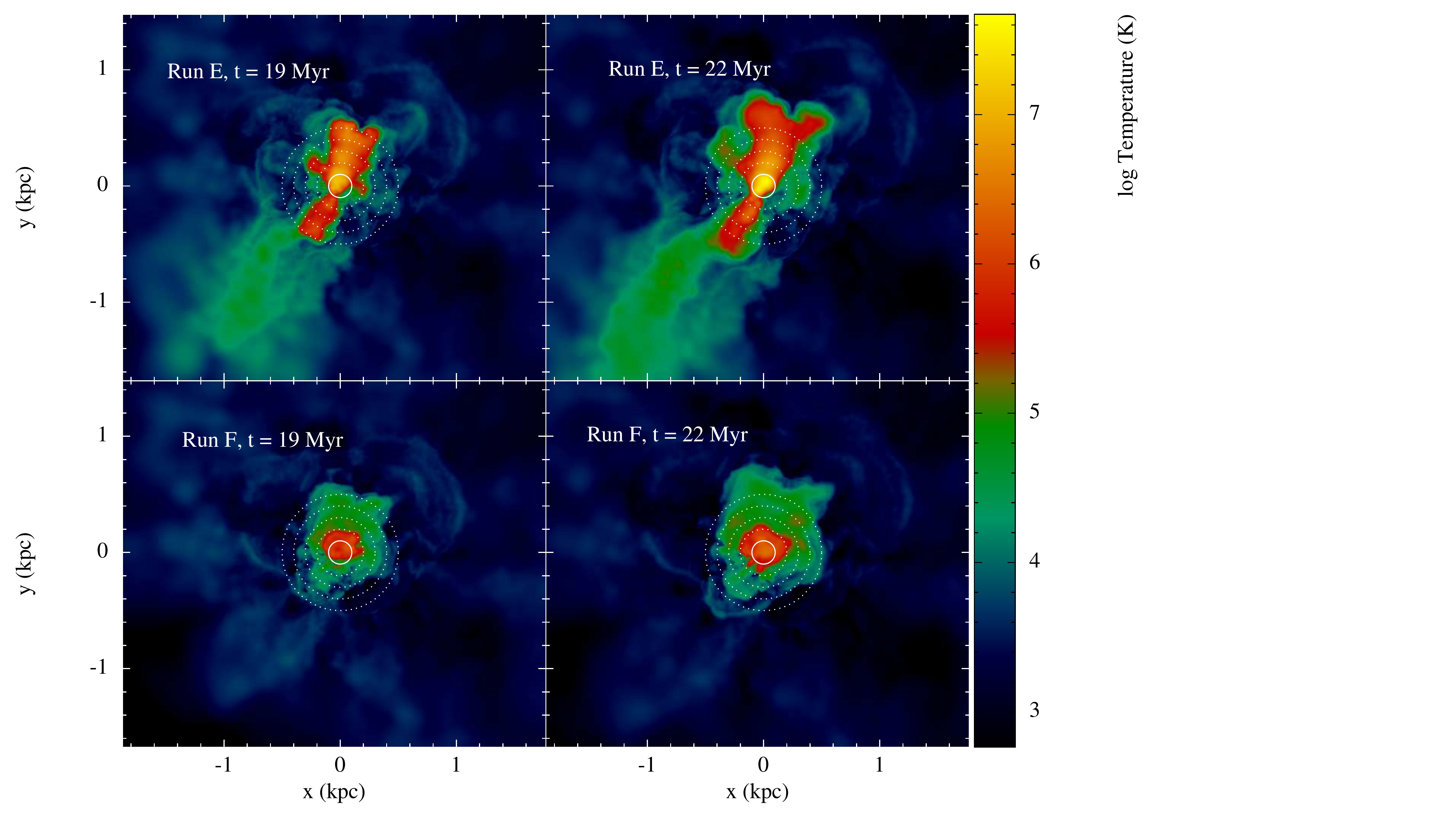}
  \caption{Temperature slices taken at z = 0 in the x-y plane, for Runs E and F, at 19 Myr and 22 Myr. The solid white circle indicates a radius of 100 pc from the origin, while the successive dotted circles beyond this indicate radii at 100 pc intervals (until 500 pc.)} 
  \label{fig:tslice_RunsEF} 
\end{figure*}

However, looking instead at the time evolution of the total energy in both Runs E and F (Fig. \ref{fig:Etot_Zsol}), the picture is more complicated than in Runs A and C. This is due to the fact the total energy of Run E initially exceeds that of the gas in Run F, until 22 Myr, when this situation is reversed. Additionally, Fig. \ref{fig:Zsol_pnet} shows the mean temperature and net radial momentum of the gas in Run F is less than that of Run E prior to 22 Myr and then exceeds it afterwards. At the end of Run E and F, Run F has $\sim$ 25 $\%$ more energy than Run E. In order to investigate this large energy jump in Run F, we plot temperature versus density at 19 Myr and 22 Myr for Runs E and F in Fig. \ref{fig:Zsol_TvsD_EF}. We see the energy jump in Run F is caused by approximately 5000, high density (>10$^{-25}$ gcm$^{-3}$), gas particles, located inside the inner 200 pc of the cloud, being heated to above 10$^6$ K (shown in the top right hand corner of plot). Fig. \ref{fig:Zsol_HMO} shows there are only $\sim$ 5 SNe during this simulation snapshot, however Fig. \ref{fig:Etot_Zsol} shows the rise in total energy is equivalent to $\sim$ 50 SNe. In order to ascertain the underlying mechanism driving this leap in energy, we ran Run F again between 19 Myr to 23 Myr at higher time resolution between snapshots ($\sim$ 0.02 Myr). We then plotted the properties of $\sim$ 4600 of the particles that were heated at $\sim$ 22 Myr (see Appendix \ref{appendix:RunF}), before and after the jump in thermal and kinetic energy. We found the jump in energy is due to the gravitational collapse of dense material in the centre of the cloud, which shock heated the gas within the central 200 pc.

Looking instead at the z = 0 temperature slice in the x-y plane, for Runs E and F at 19 Myr and then at 22 Myr (shown in Fig. \ref{fig:tslice_RunsEF}), we see Run E in the process of funneling large amounts of hot gas away from the central 100 pc of the molecular cloud, inflating the two high temperature lobes between 19 Myr and 22 Myr. However, Run F only shows a small increase in the amount of hot gas outside 200 pc, indicating the gas heated by SNe/gravitational collapse inside the inner region of the cloud has not escaped this central 200 pc. Instead, it has shock heated a large amount of gas in this region and resulted in the high temperature, high density region of gas visible in Fig. \ref{fig:Zsol_TvsD_EF}, along with the total energy jump seen in Fig. \ref{fig:Etot_Zsol}. These results point to the effectiveness of HMXBs at using the SNe-generated low density chimneys to funnel hot gas away from the central $\sim$ 200 pc of the cloud. 

As a result of the energy jump in Run F seen in  Fig. \ref{fig:Etot_Zsol}, the unbound mass fraction for Run F converges on $\sim$ 0.5, indicating the majority of the remaining gas in the cloud (that has not been accreted onto/ become sink particles) has been unbound (Fig. \ref{fig:Zsol_unbound}). Comparing with Runs A and C, it is clear the increase in the initial virial parameter of the cloud has resulted in a greater fraction of the initial gas mass in sink particles, along with a decrease in the mass fraction of gas being unbound. As well as this, the period of sink particle formation has increased in both Runs E and F, when compared with the bound cloud in Fig. \ref{fig:Zsol_SFE}. This appears to be largely due to a lower number of HMXBs and SN events (see Fig. \ref{fig:Zsol_HMO} and Table \ref{tab:Zcomparison}) acting across the simulation. Looking at Table \ref{tab:Zcomparison}, the mean sink particle mass in Runs E and F is lower than Run A, while the fraction of sinks with a mass above 180 M$_\odot$ is also smaller in these runs. However, the mean mass accreted by sinks with masses above 180 M$_{\odot}$ is comparable with Run A.

\begin{figure} 
	\psfig{file=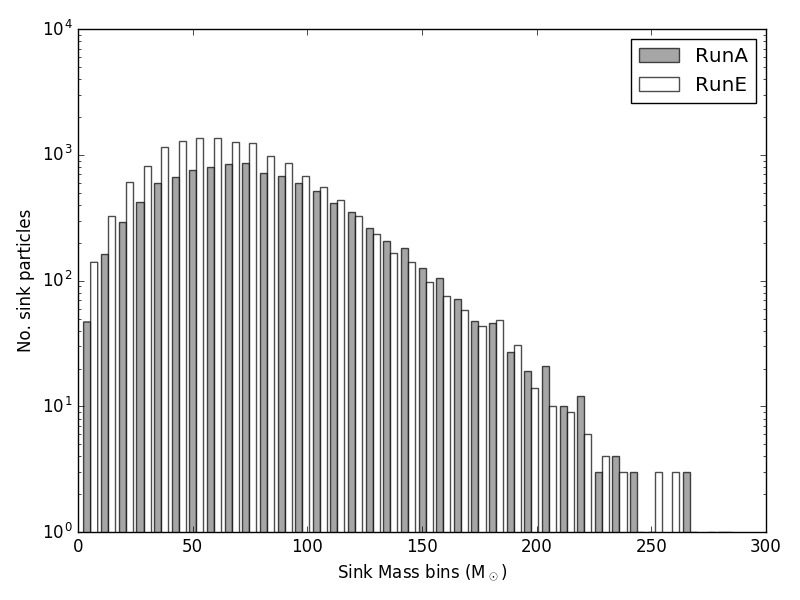,width=0.5\textwidth,angle=0}
  \caption{The number of sink particles occupying different mass bins in Runs A and E.} 
  \label{fig:Sink_Hist_AE} 
\end{figure}

We compare the distribution of sink masses between Run A and E in Fig. \ref{fig:Sink_Hist_AE}. Here we see Run E has a stronger peak at 70 M$_\odot$, along with a sharper negative gradient beyond its mean sink particle mass than Run A. As such, Run E has consistently more sink particles than Run A in mass bins below 125 M$_\odot$, however this situation is more or less reversed beyond this sink mass. Therefore, by increasing the turbulent velocity of the gas initially, the net effect has been to reduce the number of massive stars ($>$ 180 M$_{\odot}$) and increase the number of stars with masses less than 125 M$_\odot$, reducing the mean sink mass in these clouds. Moreover, the higher virial parameter has also resulted in a marginal difference between the run containing HMXB feedback (E) and the run containing just SN feedback (F).

We next investigate the ineffectiveness of the chimneys in Run E at prolonging, as well as increasing, star formation with respect to Run F, as was seen in Run A (and C). To do this we focus on Fig. \ref{fig:theta_bins}; at 12 Myr ($\sim$ the free-fall time of the clouds) we see the density and temperature versus $\theta$ profiles are remarkably similar between Runs E and F. The peaks and troughs in both density and temperature are at the same locations in both clouds. However, the two runs diverge significantly at t = 24 Myr, where a large fraction of the gas has been blown out in Run F, corresponding to the large regions of lower density (which indicate a larger maximal radius). In contrast, two clear chimneys are present in Run E. These are still present at t = 35 Myr, meanwhile, although there two wide-angle chimney-like structures in Run F, from the density scale we can see the density has dropped by at least an order of magnitude when compared with Run E.

These results point to the fact the chimneys, though present and working efficiently beyond 11.7\,Myr in Run E, did not form early enough to have an impact on the star formation and gas expulsion in that crucial free-fall timescale. The most likely cause for this delay is the reduced number of SNe and HMXBs active prior to the free-fall timescale, as seen in Fig. \ref{fig:Zsol_HMO}. Comparing Run E to Run A in the table, we see Run A has approximately 1/3 more SNe, although a similar number of HMXB events. However, in Fig. \ref{fig:Zsol_HMO} we see the HMXB feedback is initiated sooner ($\sim$ 9 Myr) in Run A than Run E ($\sim$ 11 Myr). These results suggest SNe and a higher number of HMXBs are crucial in the process of chimney formation in the cloud. This is likely due to the fact multiple SNe help to carve the chimneys, while HMXBs act to increase the temperature above 10$^7$ K (as in Run A, figure \ref{fig:Temp_Zslices_A}), preventing the collapse of the chimneys via efficient cooling. Furthermore, the gravitational collapse-generated outflow seen in Run F indicates the cooling and subsequent gravitational collapse of material inside the central region of the cloud can be more effective at heating the highest density gas than SNe. It is likely subsequent SNe are then more efficient at maintaining hot gaseous outflows since their environment has a lower density, which decreases the internal energy losses of the gas particles by limiting radiative cooling.

\subsubsection{Changing the size of the molecular cloud: Runs  R, S, V, W}
This sections looks at the results of increasing (Runs V and W) and decreasing (Runs R and S) the size of the molecular cloud in the simulation. Firstly, using the z=0 density slices in Fig. \ref{fig:Dens_Zsol} we can immediately see the gas inside the inner 2 kpc of both cloud R and S has been efficiently blown out. However, the run with HMXB feedback included (R) has more high density gas inside a radius of $\sim$ 2 kpc, compared with Run S (just SN feedback). Similarly to previous results at solar metallicity, this is likely due to multiple low density chimneys, extending from the inner kpc to the outskirts of the cloud. However, these chimneys are only just visible in the temperature slice for Run R (Fig. \ref{fig:Temp_Zsol}) and the temperature of the gas being funneled is between 10$^{6-6.5}$ K, as opposed to the 10$^{7-8}$ K gas seen in earlier runs. Although the gas is hotter in Run S, extending to $\sim$ 10$^7$ K, the temperature distribution is more isotropic and concentrated at the centre of the cloud. This suggests the SNe-heated gas in the inner regions of Run S has been unable to escape to the outskirts of the simulation and has instead collided with, and shock heated, the surrounding high density gas. This has raised the temperature in the central 2 kpc more or less uniformly. 

Instead comparing the density slices of Run V (HMXB and SN feedback) and W (just SN feedback), we can see, contrary to previous results, Run W contains two low density chimneys, while Run V only one. Similarly to Runs E and F in section \ref{sec:EF}, the left-hand chimney of Run W corresponds with the chimney seen in Run V. This also points to the fact that the initial SN feedback events in both clouds V and W (prior to the onset of HMXB feedback) determine the chimney locations.

Focusing instead on the temperature slices of Runs V and W (Fig. \ref{fig:Temp_Zsol}), we see the left hand chimney visible in Fig. \ref{fig:Dens_Zsol} in both clouds V and W, has hotter gas in Run V ($\sim$ 10$^{7}$ K). However, the right-hand chimney, which is only present in Run W, is efficiently funneling hot gas from the inner regions. 

From Fig. \ref{fig:Etot_Zsol}, the total energy of Run R at the end of the simulation is less than Run S. However, for the larger cloud the situation is different; at $\sim$ 20 Myr the total energy of Run V converges with Run W and beyond this, the total energy of Run V is larger than in Run W. Additionally, Fig. \ref{fig:Zsol_pnet} shows the net radial momentum of the gas in Run W is consistently higher than Run V beyond 12 Myr. However, beyond 15 Myr the mean temperature of the gas is lower in Run W than Run V. This helps to explain the lower total energy seen in Fig. \ref{fig:Etot_Zsol} for Run W beyond 20 Myr. Finally, the maximum and mean radius of the gas in Runs V and W follow the same values across each simulation, indicating the bulk of the gas mass is distributed similarly across cloud radius in each run.

It is also interesting to note the number of SN/HMXB feedback events active between snapshots for Runs V and W -- shown in Fig. \ref{fig:Zsol_HMO}. Like in Run E, the HMXB feedback kicks in at a time of 11 Myr, 2 Myr later than in Run A. The effect of the delayed HMXB feedback on the effectiveness of the chimneys is perhaps negated due to the longer free-fall time (13.6 Myr), which is due to a higher initial density. There are also a higher number of SNe and HMXBs active throughout Runs V and W, compared with the other runs at solar metallicity. Looking at Table \ref{tab:Zcomparison}, we can see Run V contains almost twice as many SNe and HMXBs throughout its lifetime as Run A. On the other hand, Run R also shows the delayed HMXB phase and only contains 2 HMXB events, along with 15 SNe across the simulation. 

Despite only few HMXBs acting at 12 Myr, Run V shows a clear chimney towards the edges of the $\theta$ versus temperature/density plot in Fig. \ref{fig:theta_bins}. This was likely carved by the $\sim$ 20 SNe seen in Fig. \ref{fig:Zsol_HMO}. This chimney is also visible in the corresponding plot for Run W. At later times (24 and 35 Myr) a second prominent chimney can also be seen at $\sim$ 0.5 radians in Run W. This chimney can also be seen in Run V in the density profile - there is a corresponding density drop across 3 orders of magnitude. However, there is no clear corresponding peak in temperature, as is seen in cloud W. It is also narrower in $\theta$ than the chimney present in Run W. These results point to the fact HMXBs are not necessary for chimneys to form. They also support the assertion the efficiency of a chimney is determined by the feedback power; HMXBs are not necessary to keep the gas hot inside the chimneys if the power required to do so is supplied by SNe. The gradual heating of HMXBs simply offers a ready power source to keep the gas in the chimneys hot and working efficiently.
\begin{figure} 
 \psfig{file=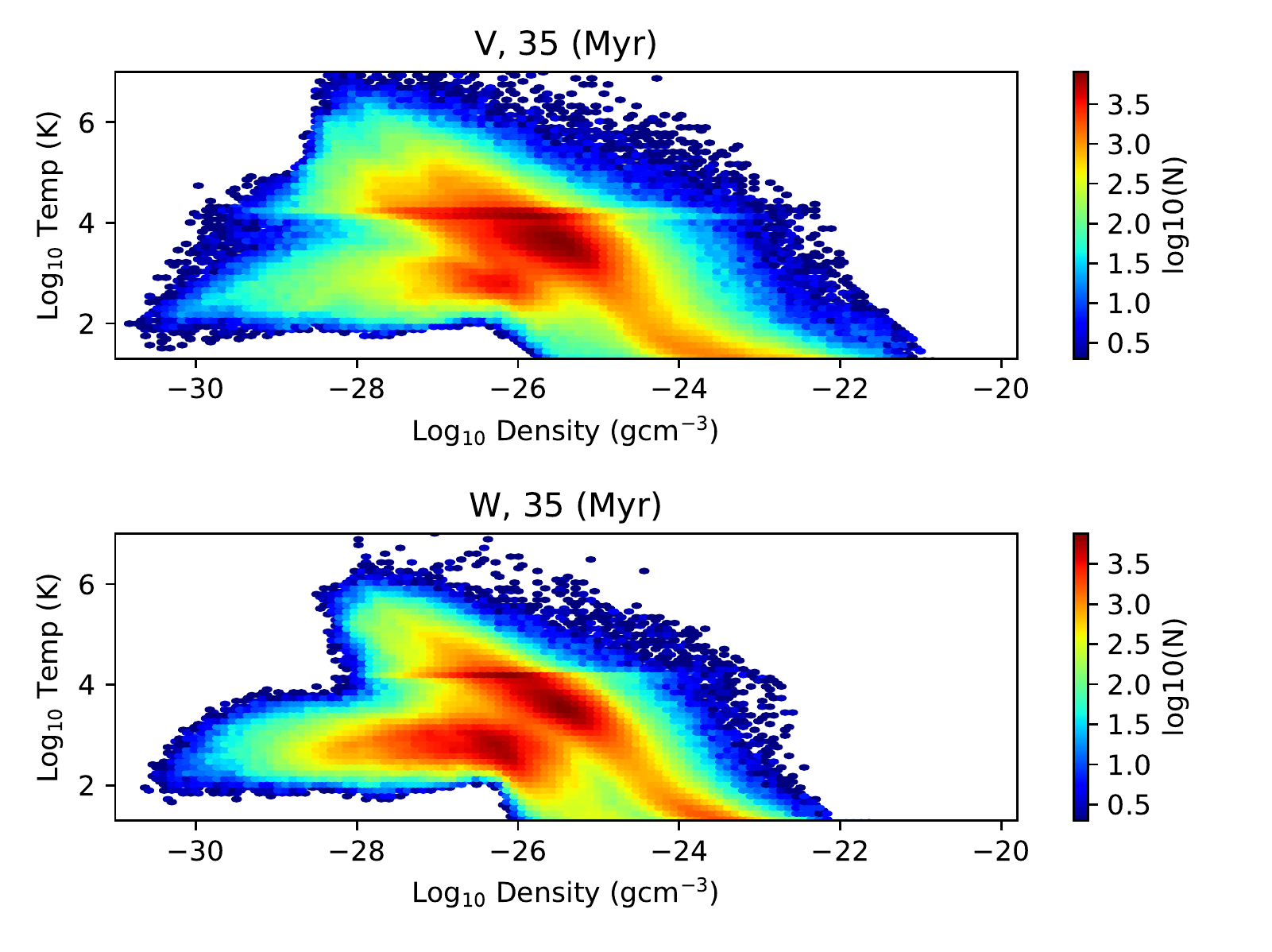,width=0.5\textwidth,angle=0}
  \caption{Temperature versus density plot, rendered according to gas particle number, for Runs V and W, at 35 Myr. } 
  \label{fig:Zsol_TvsD_VW} 
\end{figure}

Focusing on Fig. \ref{fig:Zsol_SFE}, the creation of sink particles in both Runs R and S occurs at around 10 Myr and beyond this there is no further star formation in either simulation. However, Run V continues to produce 10-100 sink particles between snapshots from 15 Myr up until the end of the simulation. In contrast, sink particle formation continues sporadically up until 29 Myr in Run W, with fewer sink particles being produced beyond 13.6 Myr (the free-fall time) than in Run V.

Fig. \ref{fig:Zsol_unbound} looks at the unbound mass fractions for Runs R,S,V and W. A large fraction of the initial gas mass ($>$ 90 $\%$) is unbound in Runs R and S as soon as feedback is turned on at around 10 Myr into each simulation. This is due to the lower binding energy of the gas in clouds R and S and is despite only 15 SNe being present throughout the simulation. As a result, the 2 HMXBs, which were switched on after the majority of the star formation (and unbinding) had occurred, have had a negligible impact on the fraction of gas that is star forming or expelled. Despite the total energy of Run V being greater than Run W beyond 20 Myr, a larger fraction of the gas is unbound in Run W, while a smaller fraction becomes star forming (i.e. goes into sink particles). 

\begin{figure} 
\psfig{file=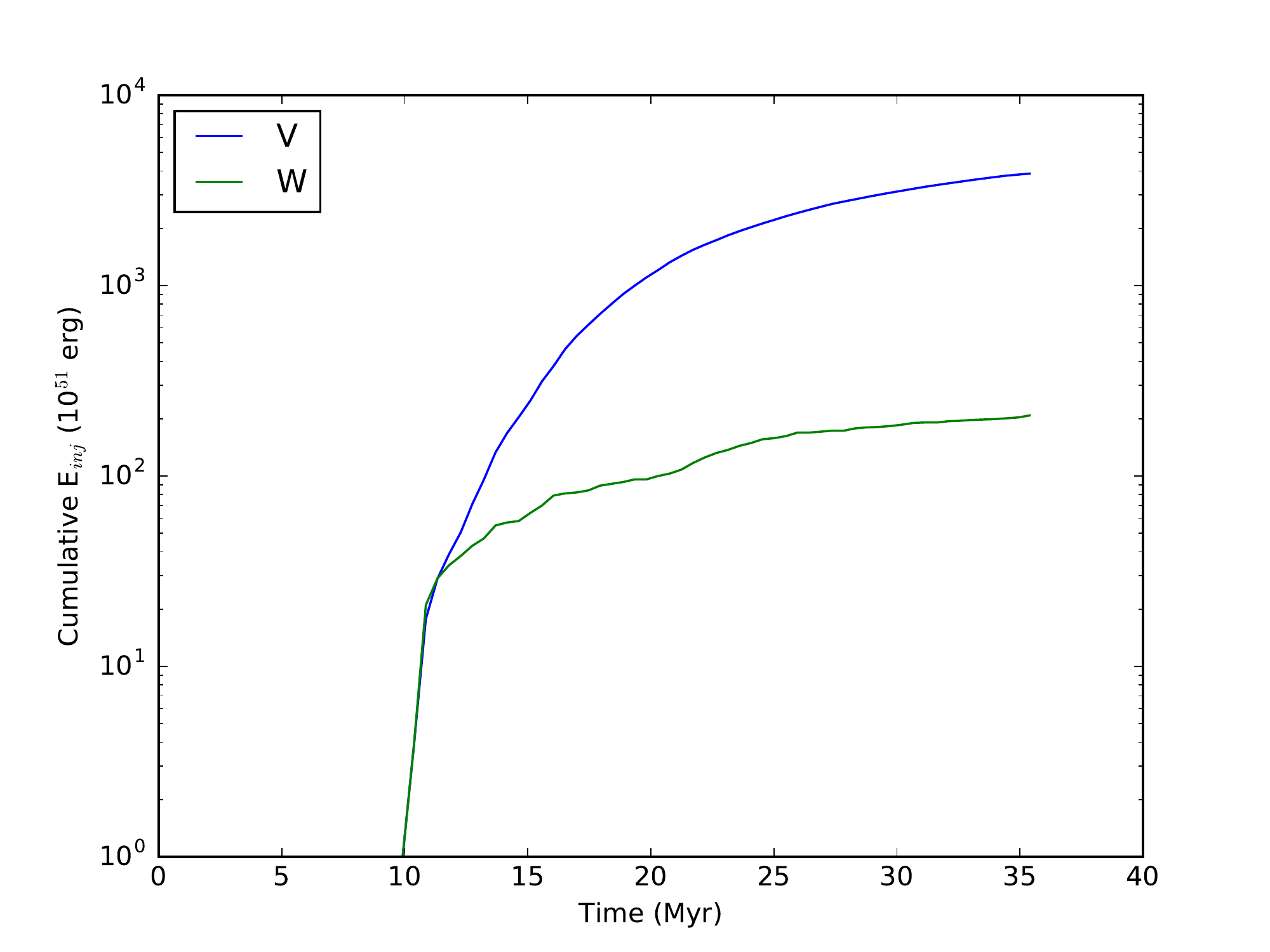,width=0.5\textwidth,angle=0}
  \caption{The cumulative energy injected via SNe and HMXBs in Runs V and W, against time. } 
  \label{fig:VW_Einj} 
\end{figure}

In Fig. \ref{fig:Zsol_TvsD_VW} we plot the temperature of the gas in Runs V and W, versus density at t $=$ 35 Myr. From this plot, we can see although there is a larger amount of high temperature ($>$ 10$^5$ K), low density ($<$ 10$^{-28}$ gcm$^{-3}$) gas in Run V than in Run W, there is also a larger amount of gas with temperatures $<$ 100K and densities $>$ 10$^{-24}$ gcm$^{-3}$. Moreover, in Fig. \ref{fig:VW_Einj} we plot the cumulative feedback injection energy in both runs against time. We see despite the fact less gas is unbound in Run V, the energy injected across the simulation is an order of magnitude higher than Run W through the addition of HMXB feedback. 

\begin{figure} 
\psfig{file=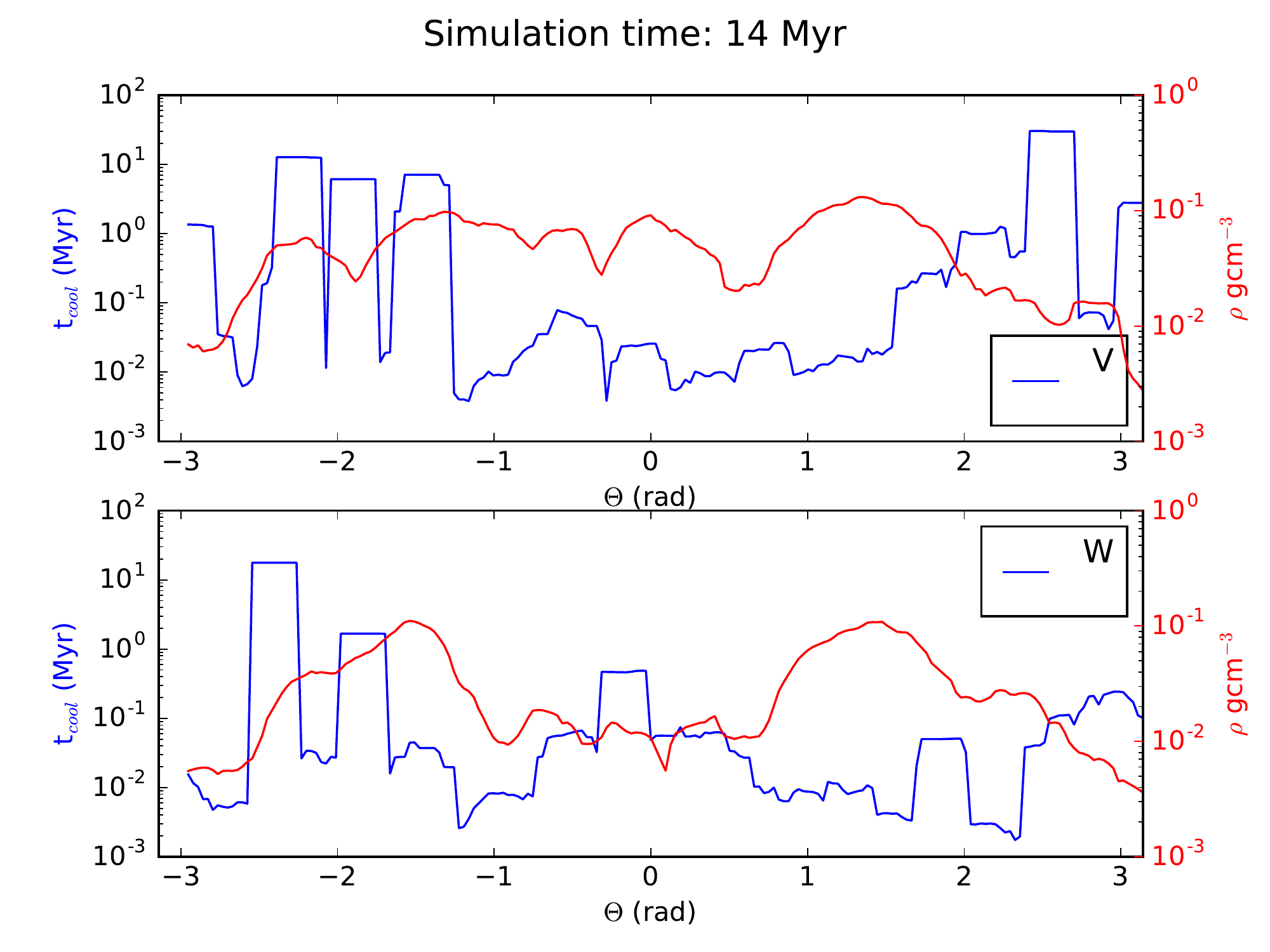,width=0.5\textwidth,angle=0}
  \caption{Blue lines - the mean cooling times in 200 $\theta$ bins at t = 14 Myr, for both Run V (top plot) and W (bottom plot), out to a radius of 200\,pc. Red lines - the mean density in each $\theta$ bin. A chimney corresponds to an order of magnitude drop in density. One can be seen at the far ends of both plots.} 
  \label{fig:VW_tcool} 
\end{figure}

In order to help ascertain the cause of the higher sink mass fraction in Run V compared with Run W, despite an order of magnitude more energy being injected into the simulation, we use Fig. \ref{fig:VW_tcool} to plot the mean cooling time in 200 different $\theta$ bins 14 Myr into Runs V and W. We only use particles within a 200\,pc radius of the centre of the cloud. We chose 14 Myr since this represents the free-fall time of both clouds and where the star formation histories of both clouds diverge; from Fig. \ref{fig:Zsol_SFE} it is at this point Run V continues to form between 10-100 sink particles between snapshots and Run W drops to to between 1-10. Looking at Fig. \ref{fig:VW_tcool} we can see the order of magnitude drop in density towards both -$\pi$ and $\pi$ radians, which corresponds with the low density, hot chimney also seen in Fig. \ref{fig:theta_bins}. Looking at the cooling times, we can see the cooling time is generally higher inside the chimney for Run V: reaching as high as 10 Myr at $\sim$ 2.5 rad, as well as 1 Myr at -$\pi$ rad (compared with $\sim$ 10$^{-2}$ Myr in Run W). This provides evidence for the hypothesis that the HMXB feedback in Run V has acted to increase the cooling times of the gas in the chimneys, keeping it hot and flowing outwards. This has allowed star formation to continue in the inner parts of the cloud. It is, however, worth noting there is a second chimney visible in Run W, that is not present in Run V. This was also visible in figures \ref{fig:Temp_Zsol} and \ref{fig:Dens_Zsol}. The gas in this chimney peaks at a cooling time of around 1 Myr, which indicates SNe are also acting to efficiently keep hot gas flowing outwards from the centre of the cloud. 
The presence of the second chimney in Run W is perhaps a product of a higher number of SNe within 14 Myr -- 171 exploded within cloud W prior to this time, as opposed to 153 in cloud V.  

In summary, the effect of adding a prescription for HMXB feedback on top of SN feedback is largely washed out in smaller molecular clouds, where the number of massive stars is smaller and the binding energy is lower. However, in larger molecular clouds the combination of HMXB and SN feedback can lead to an increase in star formation efficiency and period, compared with the clouds which just include SN feedback. Just as for Runs E and F, the results from the larger molecular cloud also suggest it is the SN feedback that determines the underlying density profile of the cloud, while in order for the chimneys to be effective, they must form within the crucial free-fall timescale of the cloud. Moreover, from Fig. \ref{fig:VW_tcool} there is evidence to suggest HMXBs act to increase the efficiency of chimneys at funneling hot gas away from star-forming material by increasing its cooling time to between 1-10 Myr. As well as this, from the additional chimney present in Run W, there is also evidence to suggest a higher number of SNe leads to a higher number of chimneys.

\subsubsection{Thermal vs Kinetic HMXB feedback: Runs A and B}\label{sec:AB}
\begin{figure} 
\psfig{file=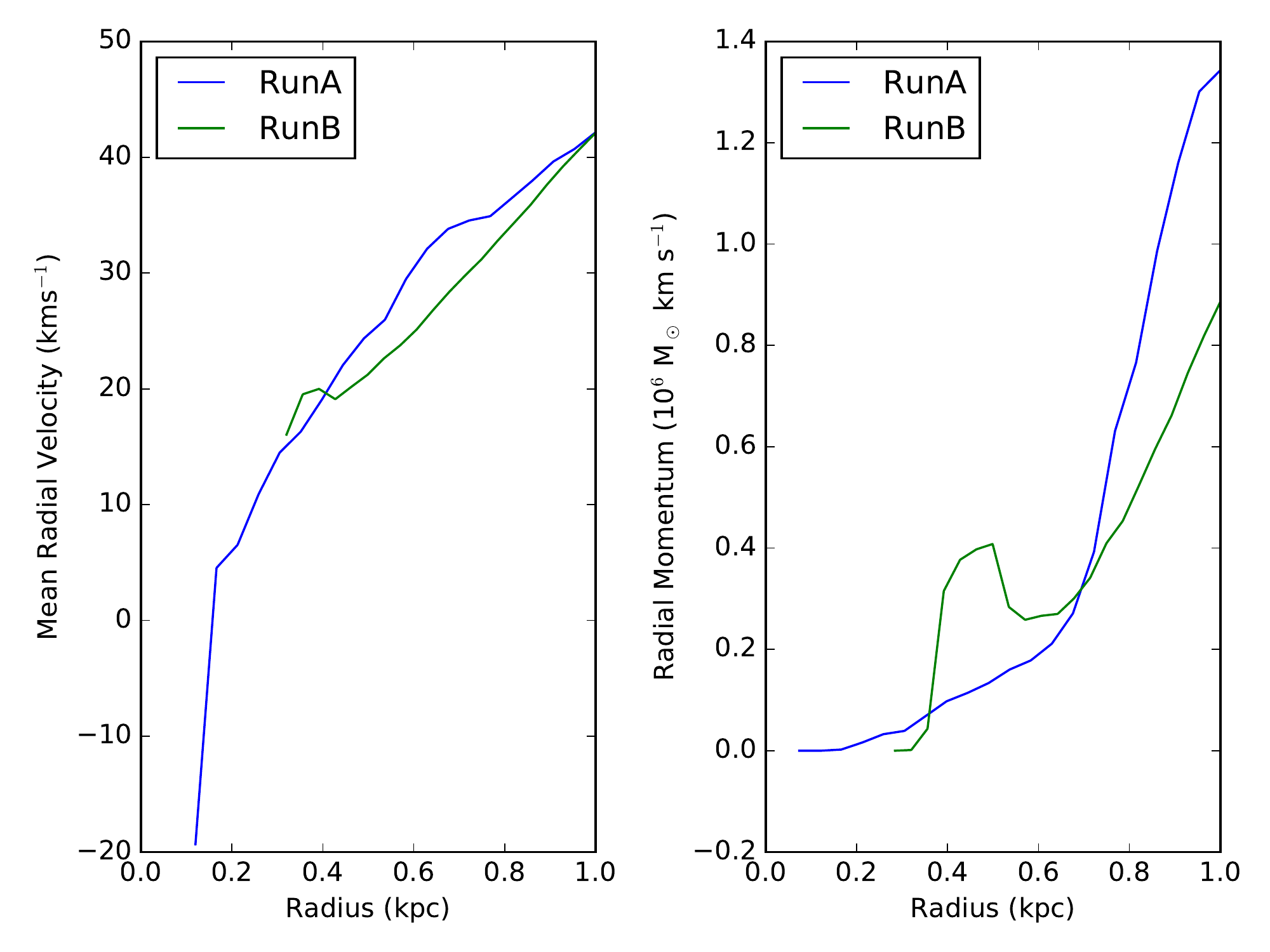,width=0.5\textwidth,angle=0}
  \caption{The mean radial velocity and momentum versus radius taken at t = 35 Myr into Runs A (blue lines) and B (green lines).} 
  \label{fig:Radmom_AB} 
\end{figure}

Firstly, looking at the density slices for both Runs A (thermal HMXB feedback) and B (kinetic HMXB feedback) in Fig. \ref{fig:Dens_Zsol}, Run A appears to have a more anisotropic gas distribution than Run B out to 2\,kpc, with long extended filaments of high density gas and multiple bubble-like cavities filled with lower density gas. However, there is a cavity spanning $\sim$ 200\,pc present in Run B, indicating the feedback has effectively blown out this inner region of gas. The cavity has been maintained by the additional thermal energy injection of SNe and the continuous energy injection from HMXBs (see Fig. \ref{fig:Zsol_Einj}), similar to the bubbles described in \citealt{Weaver1977}. This cavity is also present in the x-z and y-z planes, excluding the possibility the cavity seen in the x-y plane is a chimney viewed outside its principal axis. No gas above 10$^{6}$ K can be seen inside Run B, indicating any thermalisation of the kinetic feedback, along with the internal energy inputted into the ISM via SNe, has been rapidly lost via radiative cooling. Additionally, in Fig. \ref{fig:Radmom_AB} we plot the mean radial velocity and the net radial momentum in radial bins up to 1 kpc for both Runs A and B. Looking at Run B, we see the kinetic feedback has resulted in a momentum-driven outflow, which can be seen at 400 pc. This also corresponds to a small dip in the mean radial velocity, indicating the outflow has been slowed by the swept up mass.

Next, comparing the temperature slices of Runs A and B in Fig. \ref{fig:Temp_Zsol} the temperature of the gas cloud is lower in Run B, while Fig. \ref{fig:Zsol_pnet} shows the mean temperature of Run B is lower than Run A across the the majority of the simulation. This indicates efficient cooling by any shock-heated gas in the central 100 pc, along with possible 'leaking' of high velocity gas through low density cavities such as the one visible in Fig. \ref{fig:Dens_Zsol}. Two such cavities are seen to be working in the temperature slice for Run B. They are chimney-like, however the gas inside these regions is cooler than is seen in the chimneys for the other runs, indicating a momentum-driven flow as opposed to a pressure-driven outflow. Additionally, Fig. \ref{fig:Zsol_pnet} shows the net radial momentum is consistently lower in Run B than Run A, while the mean radius of the gas is smaller across the simulation. This points to the fact the stellar feedback in Run B has been less effective at driving large-scale outflows.

Looking instead at Fig. \ref{fig:Etot_Zsol}, the total energy of Run B is consistently lower than Run A, with two jumps in energy, initially when feedback kicks in at $\sim$ 10 Myr, and again after 20 Myr. The lower total energy of Run B is reflected in Fig. \ref{fig:Zsol_unbound}, where Run A contains a higher unbound mass fraction than Run B, along with a smaller sink particle mass fraction. 

Fig. \ref{fig:Zsol_SFE} shows the number of sink particles formed is not only higher in Run B, but occurs over a single longer period of star formation (between 6 - 23 Myr). This is in contrast to the `bursty' sink particle formation seen in Run A. However, despite 27$\%$ more sinks forming in Run B (see Table \ref{tab:Zcomparison}), the number fraction of these that have masses above 180 M$_\odot$ is half that of Run A, resulting in $\sim$ 45$\%$ fewer SNe and $\sim$ 60$\%$ fewer HMXBs than Run A. Hence, by including momentum-driven HMXB feedback the mean mass of the sink particles has been reduced, resulting in fewer massive stars and hence feedback events. A contributing factor to the lower sink masses in Run B is likely to be the lower mean temperature seen in Fig. \ref{fig:Zsol_pnet}. This will lower the Jeans mass (M$_J$), which scales as T$^{3/2}$, resulting in smaller sink particles masses forming from our Jeans criterion.

Another contributing factor to the lower total energy seen for Run B in Fig. \ref{fig:Etot_Zsol} is the `leaky' nature of the kinetic feedback; high velocity gas particles will escape through the path of least resistance, while others are expected to hit the surrounding high density gas and thermalise, as well as subsequently cool, efficiently. The dissipation of the feedback energy via cooling can be artificially fast due to the high density of the surrounding material. This is a common problem encountered when modelling SN feedback using radial kicks of the surrounding particles. In their 2003 paper (\citealt{Springel2003}), Springel and Hernquist provided a solution to this problem by kinetically decoupling the wind particles until their density has dropped to a threshold value or a set amount of time has elapsed. The cold, low density central region of cloud B (see Fig. \ref{fig:Dens_Zsol}) indicates the rapid dissipation of the energy injected by SNe has manifested inside Run B. This may have artificially limited the effectiveness of the feedback in quenching star formation beyond this radius.

To conclude, implementing HMXB feedback as kinetic energy has resulted in a momentum-driven outflow which has been slowed at a radius of $\sim$ 400 pc due to the large swept up mass. Furthermore, it is likely this 'stalling' has resulted in the thermal SN events occurring in dense environments, which has caused the rapid dissipation of thermal energy via radiative cooling, perhaps artificially limiting its effectiveness at reducing star formation at radii beyond $\sim$ 200 pc. As well as this, the efficient cooling seen in Run B, compared with Run A, has resulted in lower sink particle masses, due to a reduction in M$_J$, which in turn has led to fewer HMXB/SN events (as a consequence of our numerical method).

\subsection{[Fe/H]$=$-1.2 Runs}
Overall, the lower metallicity runs contain more SNe and HMXBs than their solar metallicity counterparts due to a higher average sink particle mass (see Table \ref{tab:Zcomparison}) - resulting from a higher M$_{J}$ (see section \ref{sec:AB}). 
\begin{table*}\label{tab:Zcomparison}
 \caption{A table listing the total sink mass as a fraction of the initial gas mass at the end of each simulation ($mf_{\rm sink}$), the number of sink particles present at the end of each run ($N_{\rm sink}$), the total number of SNe ($N_{\rm SNe}$) and HMXBs ($N_{\rm HMXB}$) across the whole simulation, the mean sink particle mass ($mp_{\rm sink}$), the fraction of the total number of sinks with a mass $>$ 180 M$_\odot$ and finally the mean mass accreted by sinks with masses above 180 M$_\odot$.} 
  \begin{tabular}{|l|l|l|l|l|l|l|l|l|}\label{tab:Zcomparison}
    Run&$mf_{\rm sink}$&Total $N_{\rm sink}$&N$_{\rm SNe}$&$N_{\rm HMXB}$&Mean $mp_{\rm sink}$ (M$_\odot$)&\multicolumn{1}{|p{3cm}|}{Number fraction of sinks $>$ 180 M$_\odot$ (\%)}&\multicolumn{1}{|p{3cm}|}{Mean mass accreted by sinks $>$180 M$_\odot$ (M$_\odot$)}\\
    \hline
    A&0.39&9898&119&17&78.5&1.4&23\\
    \hline
    B&0.46&12615&67&7&72.2&0.72&25\\
    \hline  
    C&0.33&8372&133&21&79.0&1.9&26\\
    \hline
    E&0.49&14376&86&15&68.7&0.9&24\\
    \hline
    F&0.50&14246&92&14&69.9&0.9&24\\
    \hline
    R&0.02&88&15&2&104.4&19.3&61\\
    \hline
    S&0.02&82&15&2&109.4&20.7&62\\
    \hline
    V&0.44&31501&209&32&70.4&0.9&27\\
    \hline
    W&0.42&28652&208&30&72.7&1.0&28\\
    \hline
    H&0.16&2739&334&45&116.3&14.9&36\\
    \hline
    I&0.17&2835&389&74&118.1&15.5&42\\
    \hline
    K&0.11&1913&184&25&113.6&12.6&35\\
    \hline
    L&0.11&1905&190&24&114.3&12.5&38\\
    \hline
    T&0.02&105&16&2&104.2&19.0&112\\
    \hline
    U&0.02&84&30&6&141.8&38.1&65\\
    \hline
    X&0.08&3681&418&65&114.7&11.9&35\\
    \hline
    Y&0.10&4557&452&71&112.0&11.9&33\\
    \hline
  \end{tabular}
\end{table*}

\subsubsection{The injection of HMXB: Runs H, I}\label{sec:Zlow_HMXB}
\begin{figure*} 
 \psfig{file=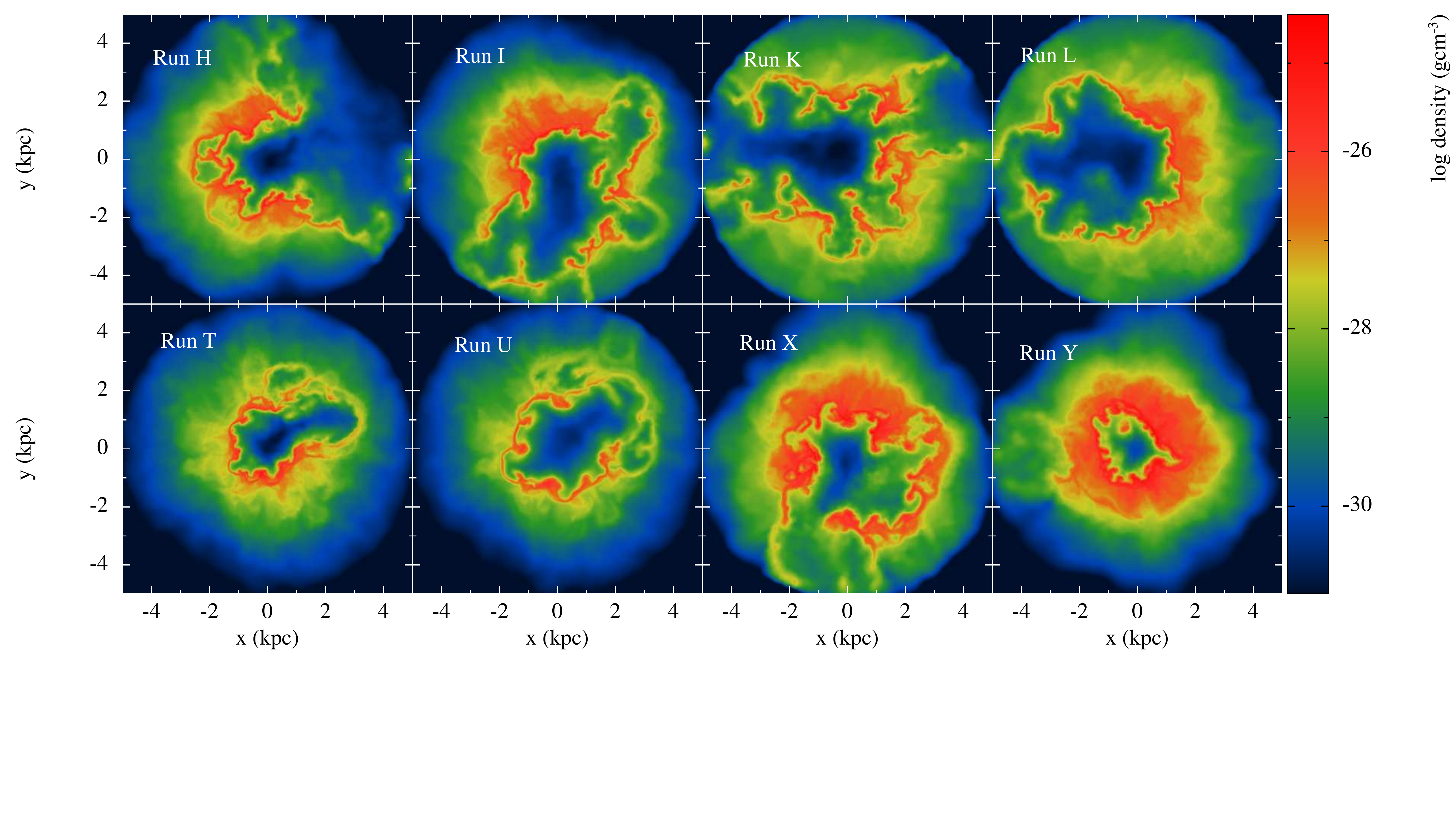,width=\textwidth,angle=0}
  \caption{Density slices taken in the x-y plane at z=0 for runs at a metallicity of [Fe/H] = -1.2, taken 35 Myr into each simulation.} 
  \label{fig:Dens_Zlow} 
\end{figure*}

\begin{figure*} 
 \psfig{file=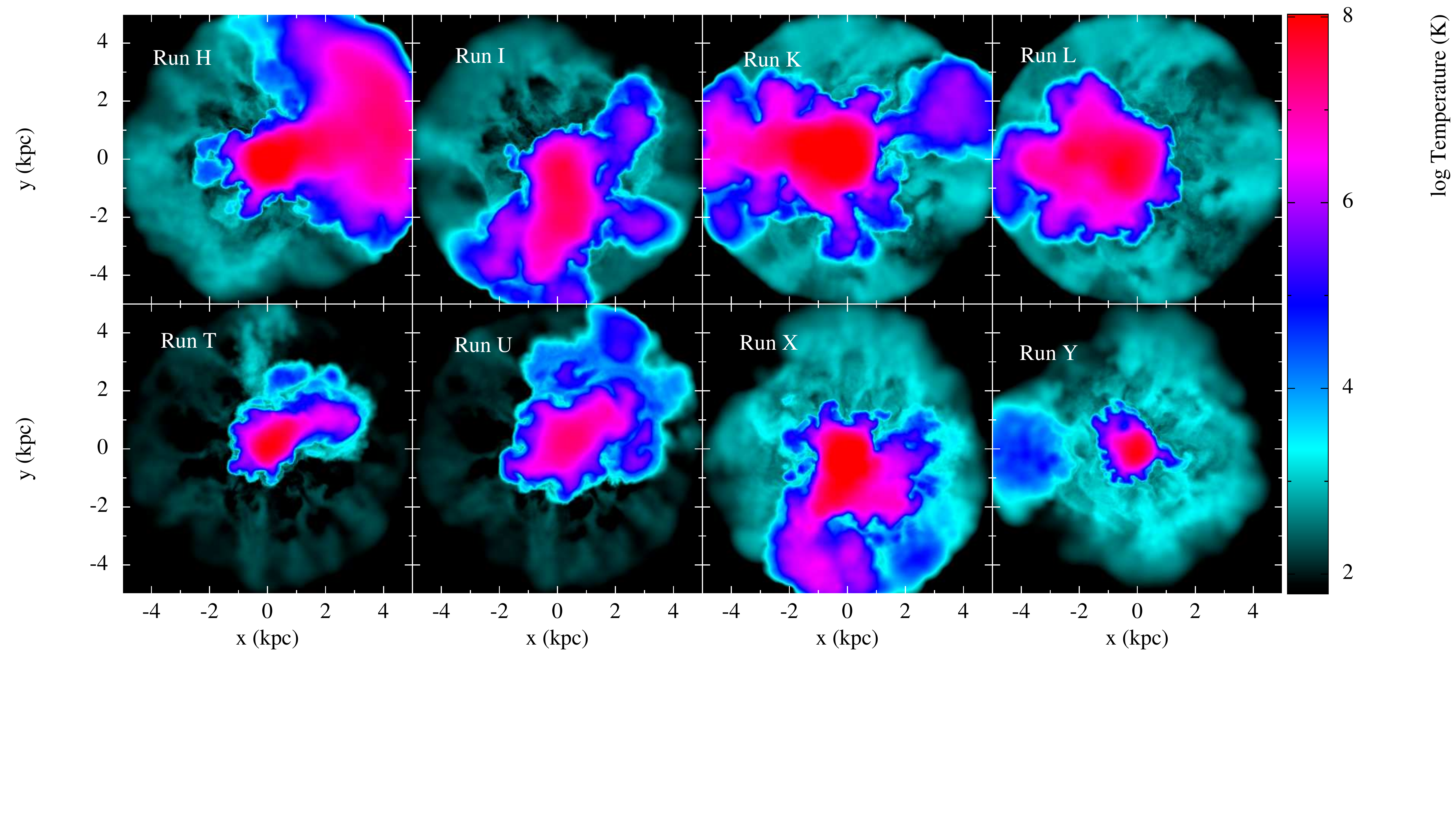,width=\textwidth,angle=0}
  \caption{Temperature slices taken in the x-y plane at z=0 for runs at a metallicity of [Fe/H] = -1.2, taken 35 Myr into each simulation.} 
  \label{fig:Temp_Zlow} 
\end{figure*}

Figures \ref{fig:Dens_Zlow} and \ref{fig:Temp_Zlow} show the density and temperature slices for the low metallicity runs ([Fe/H]$=$-1.2) 35 Myr into each simulation. Looking at the density and temperature slices for Run H, we see the inner region of the cloud has been effectively blown out to a radius of $\sim$ 1.5 kpc, whereupon the remaining hot gas has carved a low density `hole' on the right of the slice. This hole cannot be considered to be a chimney since it does not extend and narrow towards the central kpc of the cloud. The morphology does, however, resemble a champagne flow \citep{Tenorio-Tagle1979} - in this case formed by the hot pressurised gas generated by SNe and HMXBs escaping into the lower density ISM of the cloud's outer regions. As in the case of champagne flows, the anisotropic morphology of this outflow is most likely due to the fact the HMXB and SN events are not spatially isotropic and hence will be closer to one edge of the cloud than another. The temperature of the gas within $\sim$ 1.5 kpc of cloud H is uniformly $10^8$ K (which is radiatively inefficient), indicating there is no cool dense gas remaining in this region which can collapse to form stars within the timescale of the simulation. Similarly in cloud I, the gas has been blown out to 1 kpc and the remainder of the feedback-heated gas has carved a hole towards the bottom of the density slice. Once again the inner 1.5 kpc of cloud I has been uniformly heated, ceasing all star formation in this region. However, the temperature of the hot gas is slightly cooler than in cloud H - between 10$^{7-8}$ K. 
\begin{figure*} 
\psfig{file=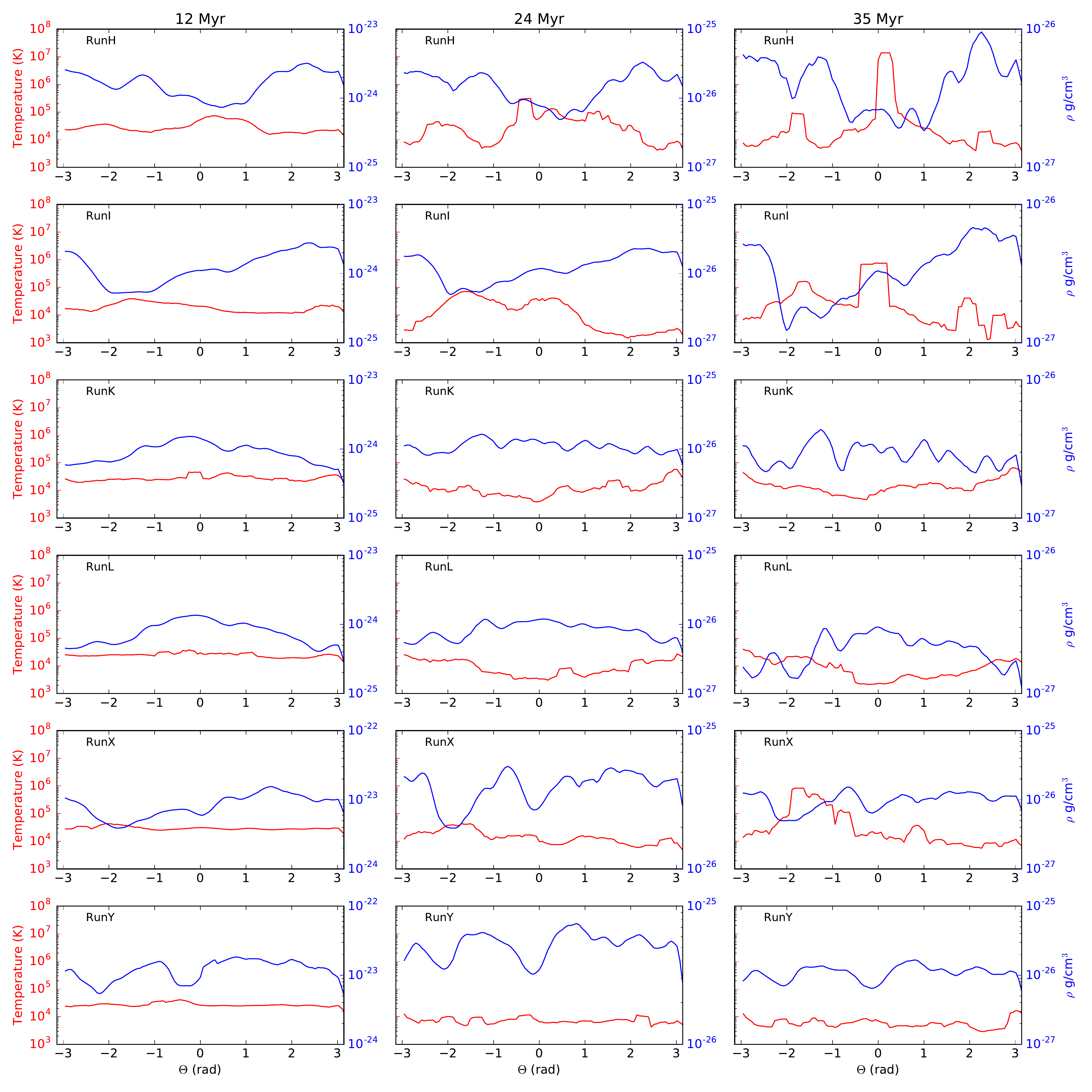,width=\textwidth,angle=0}
  \caption{Plots to show the mean temperature (red) and density (blue) in $\theta$ bins ranging from -$\pi$ to $\pi$ radians for runs at [Fe/H] = -1.2. The maximal radius of each $\theta$ bin is set to the maximum radius of the gas in the individual bin. The left column is for snapshots taken at 12 Myr, the middle column is for 24 Myr and the right column shows snapshots at 35 Myr. The name of the corresponding run is in the upper left hand corner of each plot.} 
  \label{fig:theta_bins_lowZ} 
\end{figure*}

Looking instead at Fig. \ref{fig:theta_bins_lowZ}, which plots the mean density and temperature across various $\theta$ bins for the [Fe/H] $=$ -1.2 runs, there is a chimney-like structure in the $\theta$-density profile of Run H, which develops from 12 Myr to 35 Myr. However, looking at the density scale at 12 Myr, it is clear this trough in the density profile represents less than an order of magnitude drop in density. This is in stark contrast to the two orders of magnitude density drop seen in the corresponding plot for Run A (Fig. \ref{fig:theta_bins}). Even 35 Myr into the simulation, the `chimney' seen in Run H only represents $<$ 1 order of magnitude density drop. The corresponding peak in the mean gas temperature, however, is 3 orders of magnitude larger in Run H (10$^8$ K) than Run A (10$^5$ K). Moreover, the rise in temperature in Run H spans 3-4 orders of magnitude, while the chimney in Run A spans $\sim$ 2. Run I follows a similar trend in density across different $\theta$ bins as Run H; the `chimneys' only span $\sim$ 1 order of magnitude in density. However, the gas in Run I has a lower peak temperature than Run H at 35 Myr.
\begin{figure} 
\psfig{file=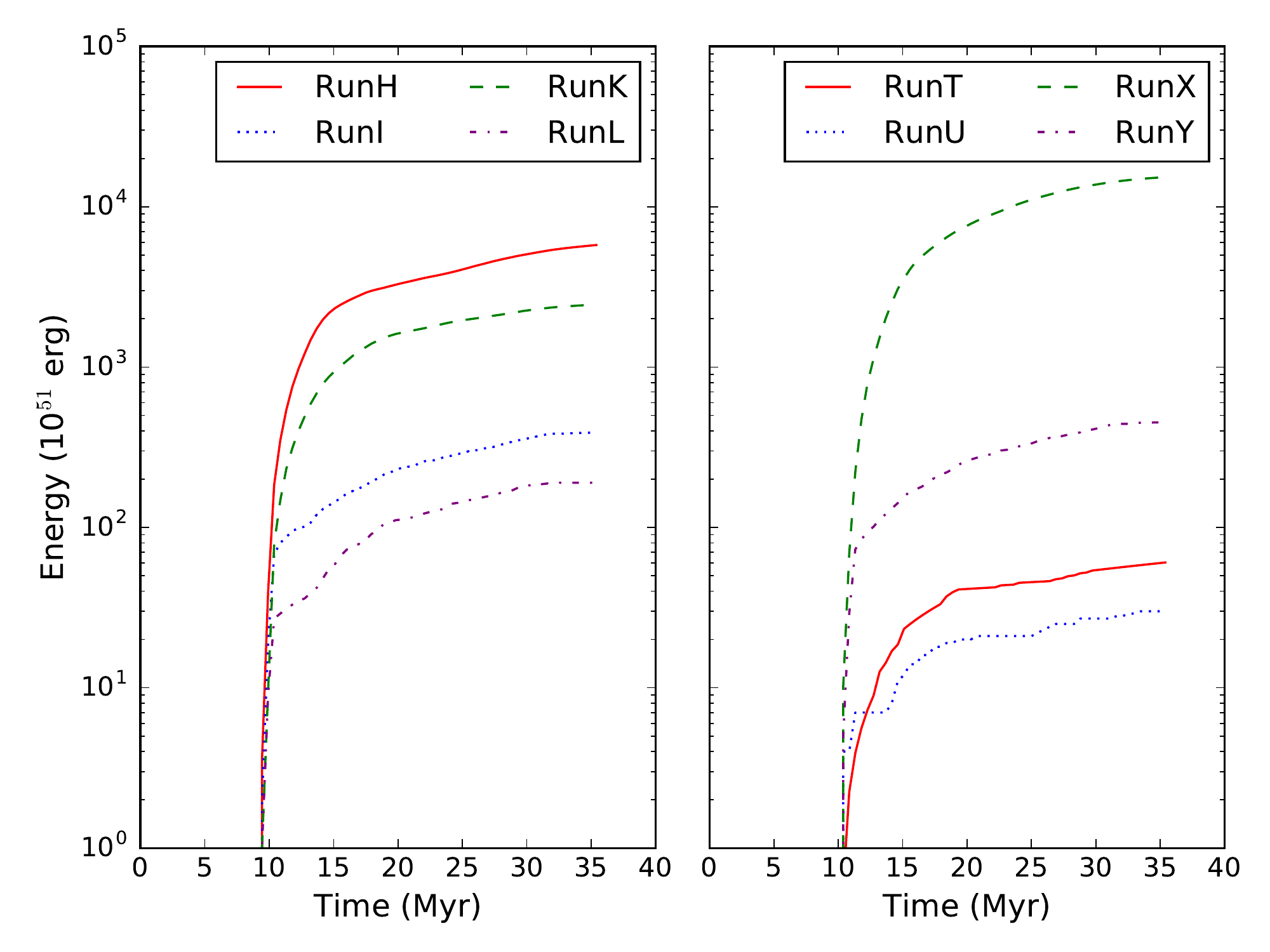,width=0.5\textwidth,angle=0}
  \caption{The cumulative energy injected across the simulation for the [Fe/H] = -1.2 runs which include feedback.}
  \label{fig:Zlow_Einj} 
\end{figure}

\begin{figure} 
\psfig{file=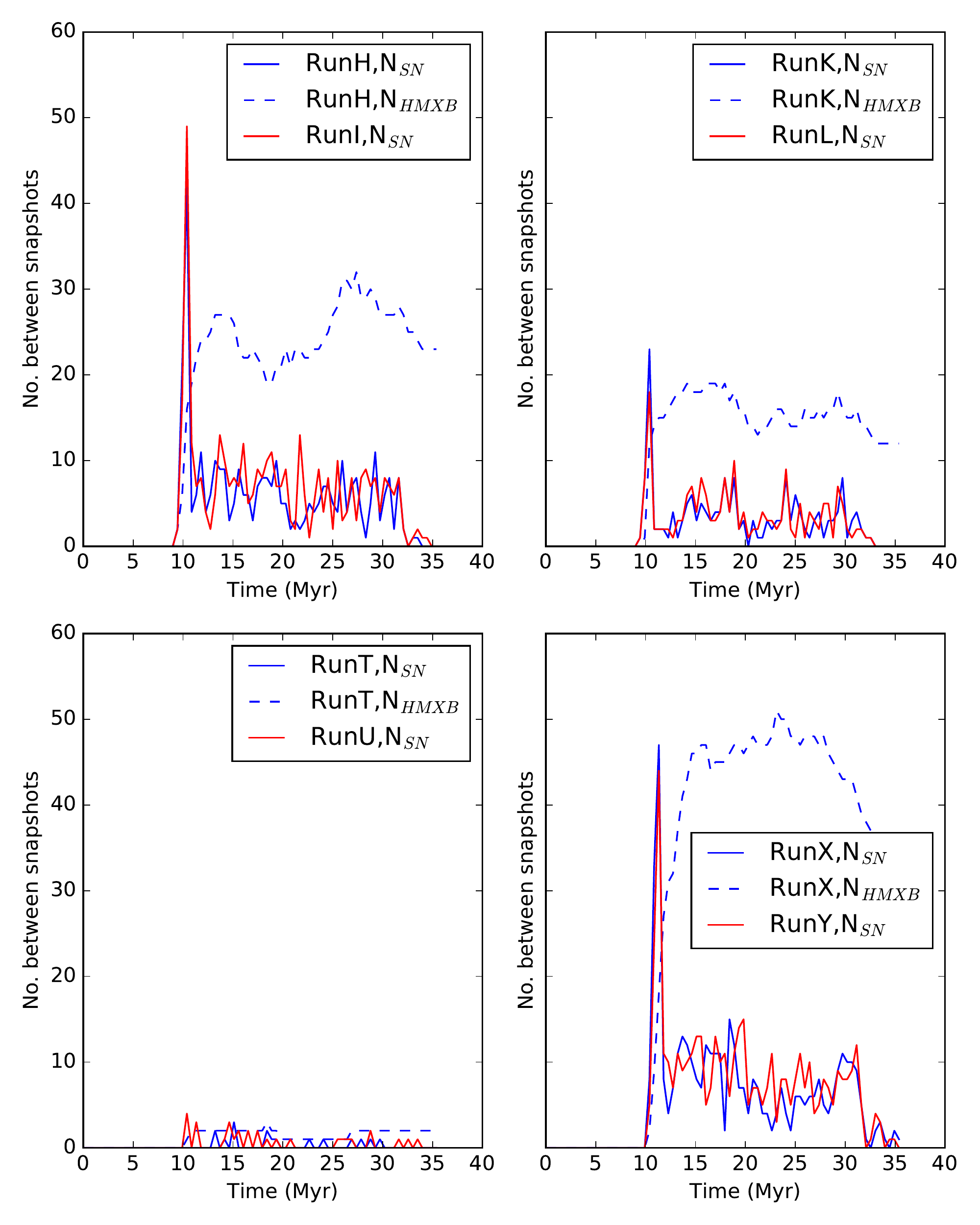,width=0.5\textwidth,angle=0}
  \caption{Plots to show the number of SNe (N$_{SN}$, solid lines) and HMXBs (N$_{HMXB}$, dashed lines) between snapshot times across simulations at [Fe/H] = -1.2.} 
  \label{fig:Zlow_Nos} 
\end{figure}

Fig. \ref{fig:Zlow_Einj} plots the cumulative injected energy for each [Fe/H]$=$-1.2 run, while Fig. \ref{fig:Zlow_Nos} plots the number of SNe and HMXBs between snapshots. The injected energy for Run H is greater than Run I, due to a similar number of SN events in each run and an additional $\sim$ 10-50 HMXBs active between snapshots. 
\begin{figure}
\psfig{file=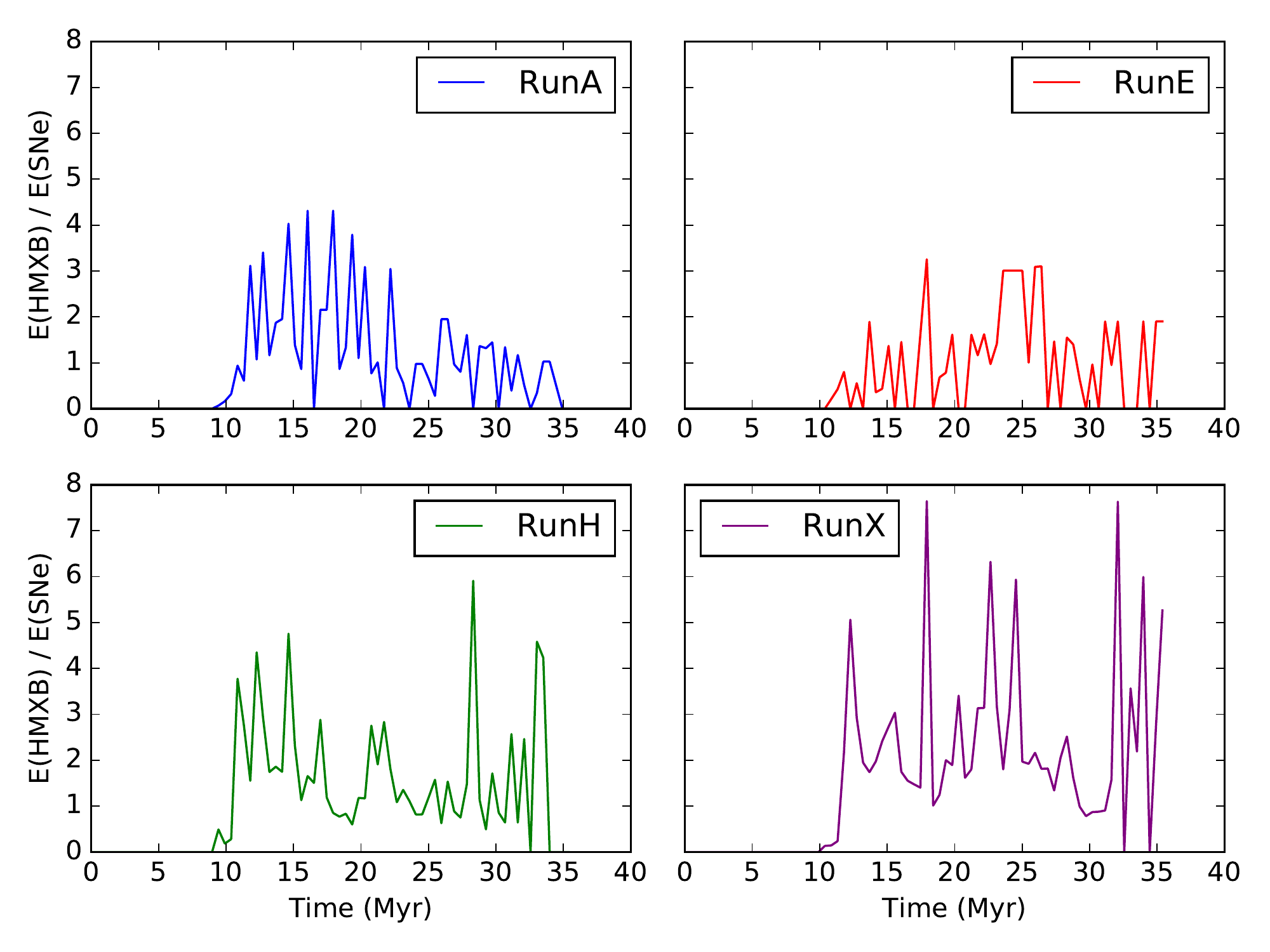,width=0.5\textwidth,angle=0}
  \caption{The time evolution of the ratio between the energy injected by HMXBs - E(HMXBs) - and the energy injected by SNe - E(SNe) - between snapshot times for Runs A (top left), E (top right), H (bottom left) and X (bottom right).}
  \label{fig:Zlow_HMXBvSNe} 
\end{figure}

Moreover, if we compare the number of HMXB events and SN events in Runs A and H (Table \ref{tab:Zcomparison}), we see Run H has nearly 3 times as many SNe and over 2.5 times as many HMXBs as Run A. Also the number of SNe is higher in Run I than Run A. In Fig. \ref{fig:Zlow_HMXBvSNe} we plot the time evolution of the ratio between the energy injected by HMXBs and the energy injected by SN events for a selection of runs. From Fig. \ref{fig:Zlow_HMXBvSNe}, we can see this ratio is typically above 1 in all runs plotted. However, the ratios are particularly high in Run X, while Run H also has peaks above 5, where Runs A and E do not. This helps explain the high peaks in gas temperature seen for Run H in Fig. \ref{fig:theta_bins_lowZ}).  
\begin{figure} 
\psfig{file=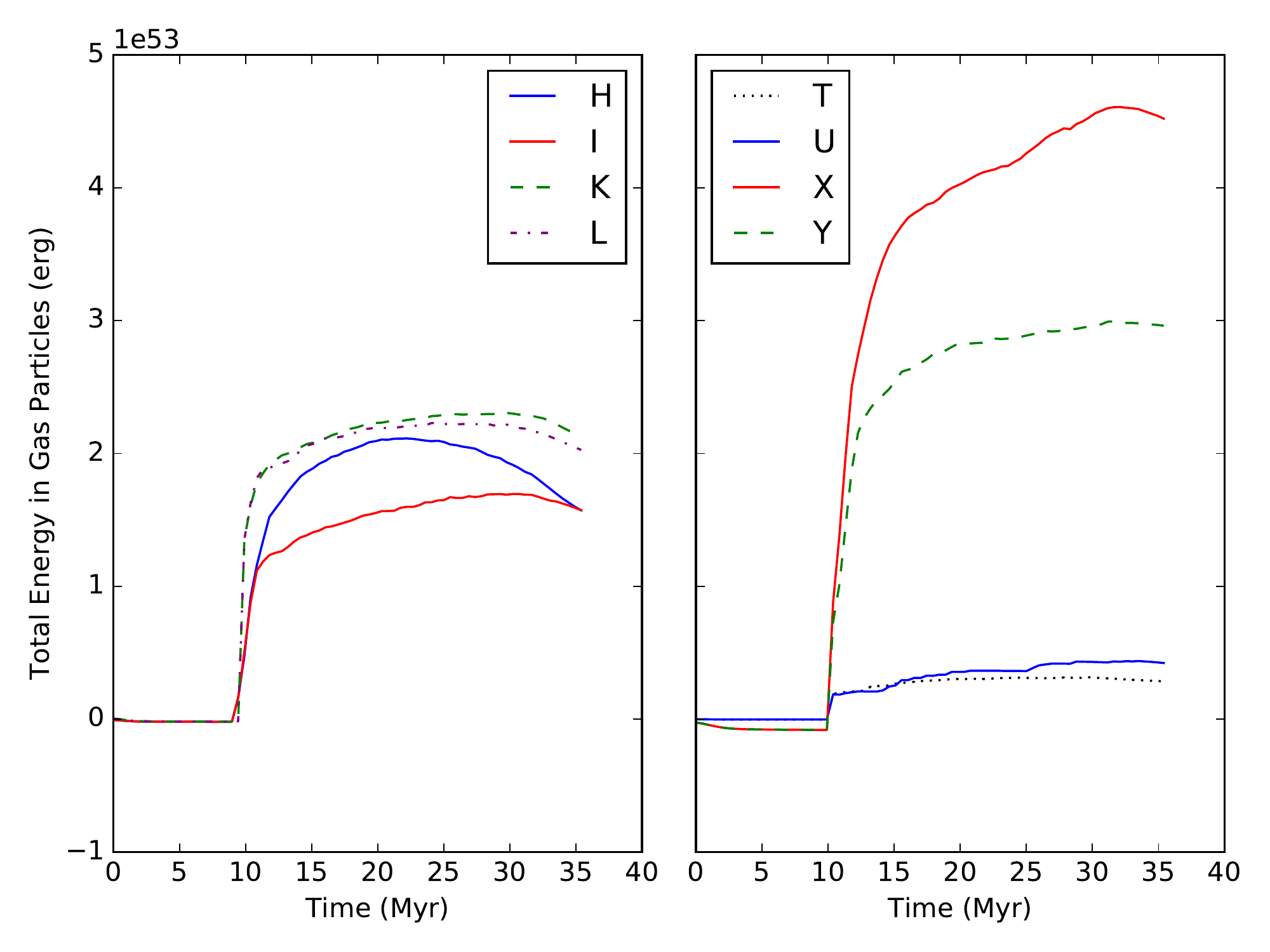,width=0.5\textwidth,angle=0}
  \caption{Plots to show the time evolution of the total energy of each run at a metallicity of [Fe/H] $=$ -1.2} 
  \label{fig:Zlow_Etot} 
\end{figure}

Looking at the total energy of the gas in Runs H and I, plotted in Fig. \ref{fig:Zlow_Etot}, we see between 10-35 Myr the total energy of Run H is greater than Run I, due to the higher amount of energy injected through feedback (see Fig. \ref{fig:Zlow_Einj}). However, at 35 Myr the two total energies of the two simulations converge to the same value. This suggests the high temperature, feedback-heated gas has cooled faster in Run H. 
\begin{figure} 
 \psfig{file=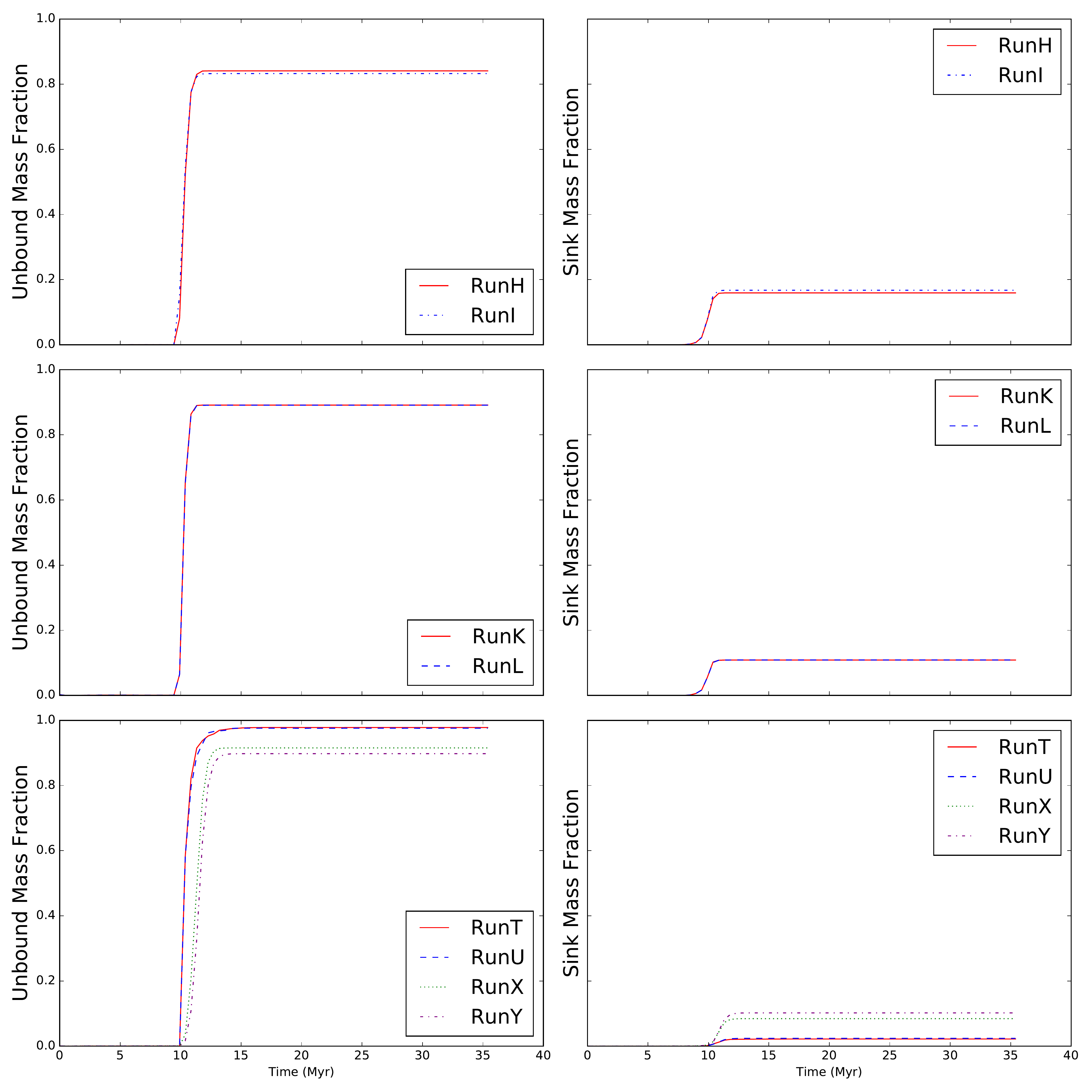,width=0.5\textwidth,angle=0}
  \caption{Plots to show the fraction of the initial gas mass in each [Fe/H] = -1.2 run that has (a) been unbound (left column) and (b) been accreted onto/ become sink particles (right column).}
  \label{fig:Zlow_unbound} 
\end{figure}

\begin{figure} 
 \psfig{file=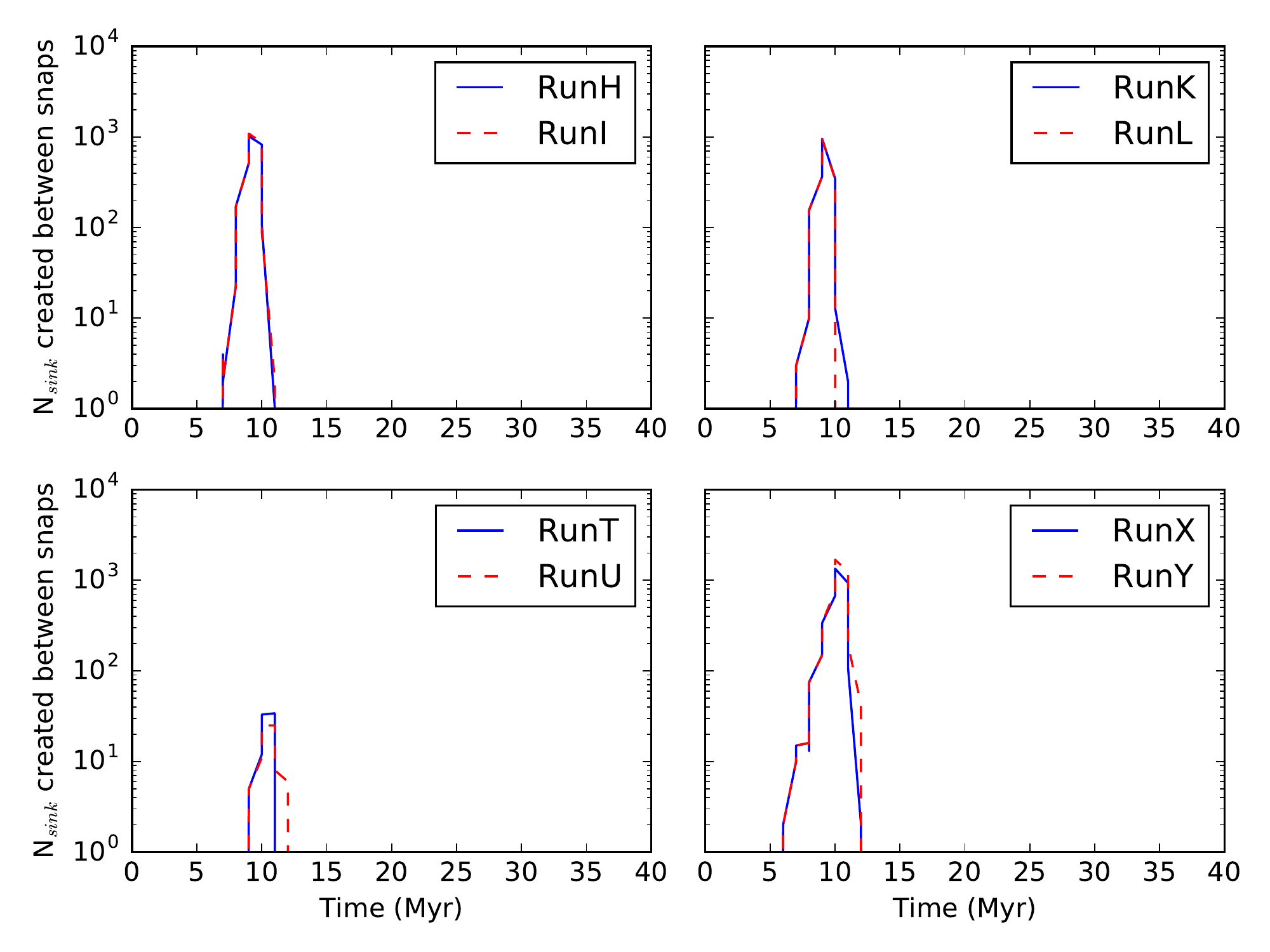,width=0.5\textwidth,angle=0}
  \caption{Plots to show the number of sink particles produced between snapshots for the runs at a metallicity of [Fe/H] $=$ -1.2.} 
  \label{fig:Zlow_SFE} 
\end{figure}

Fig. \ref{fig:Zlow_unbound} plots the fraction of the initial gas mass that has been unbound (left column) or formed sinks (right column). We can see there is marginal difference between these fractions for Runs H and I, with a slight increase in the sink mass fraction in Run I, with a corresponding decrease the mass unbound. Since these fractions are so similar, we can rule out significant mass loss in Run H compared with Run I. Overall, the mass fraction in sink particles (which is also displayed in Table \ref{tab:Zcomparison}) in Run H is much lower than for Run A. This is true of all the low metallicity runs and is down to inefficient cooling due to a relative lack of heavier ions (e.g. OII, OIII).

Fig. \ref{fig:Zlow_SFE} plots the number of sinks produced between snapshots versus time. Similar to the solar metallicity runs, the majority of the sinks are formed around 10 Myr into the simulation (i.e. at the cloud's free-fall time). Comparing figures \ref{fig:Zlow_SFE} and \ref{fig:Nofb_sink}, we see the addition of feedback has not altered the sink particle formation time significantly in either Run H and Run I, compared with Run G (the Z=0.001 run without feedback).

\begin{figure} 
 \psfig{file=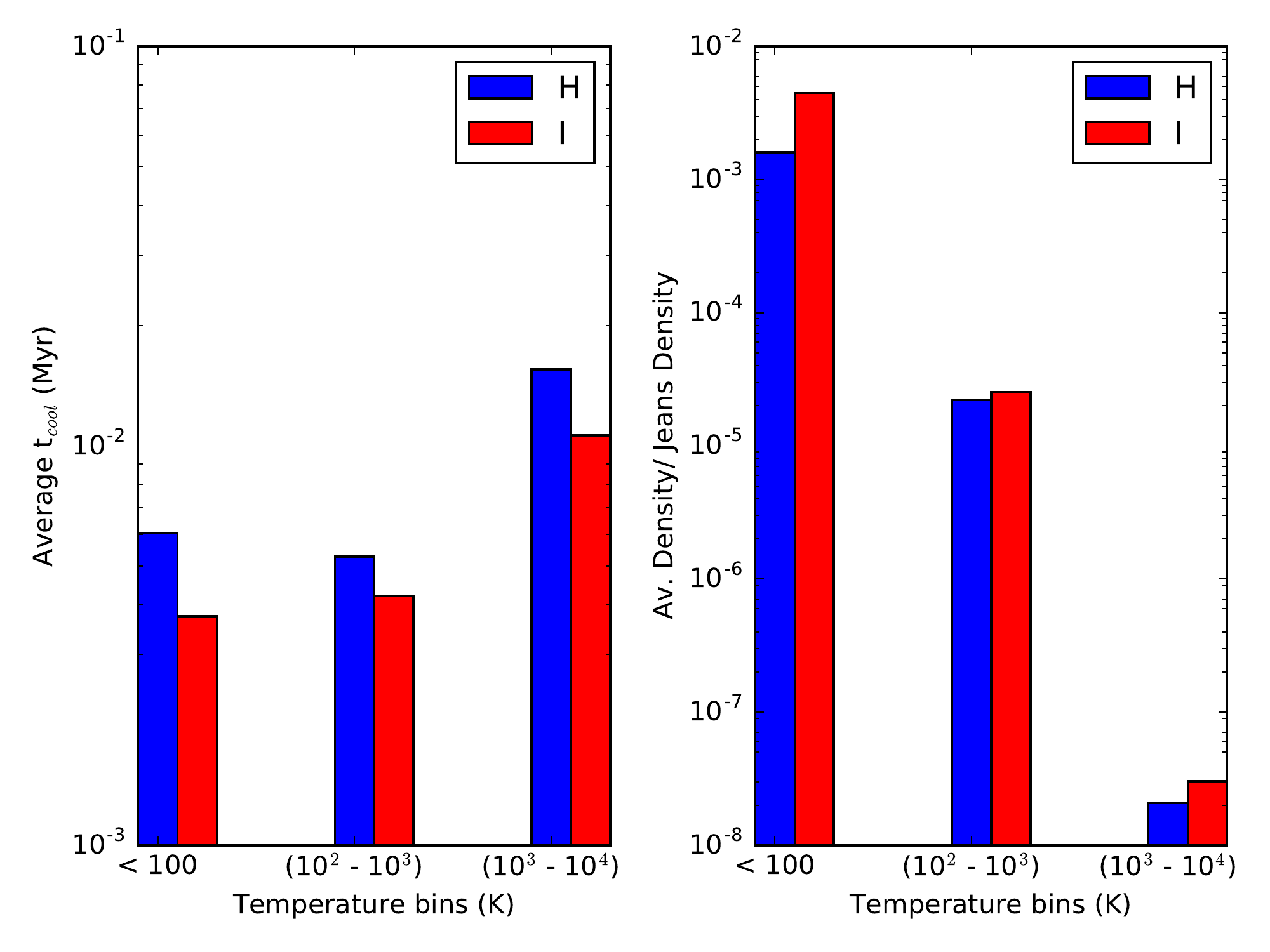,width=0.5\textwidth,angle=0}
  \caption{Left: plot to show the average cooling time in three different temperature bins ($<$ 100 K, 100-10$^3$ K and 10$^3$-10$^4$ K) for runs H and I. Right: plot to show the ratio between the average density in the different temperature bins, divided by the corresponding Jeans density for the average temperature in that bin in runs H and I.} 
  \label{fig:tcool_bins} 
\end{figure}

Since the fate of these gas clouds appears to be decided at the free-fall time of each cloud (11.7 Myr into each simulation), we use Fig. \ref{fig:tcool_bins} to plot the average cooling time and the average value of the ratio $R = \rho/\rho_{\rm Jeans}$ in three different temperature bins, all under 10$^4$ K, at 10 Myr. Only bound gas is included in these plots. From Fig. \ref{fig:tcool_bins}, we can see the average cooling time, predominantly through excitation of the fine structure levels of ions such as OI \citep{Mo2010}, is lower for Run I than Run H in each temperature bin. We can also see from Fig. \ref{fig:tcool_bins}, the ratio $R$ is higher in Run I, indicating the lower temperature gas in Run I is denser than the corresponding gas in Run H, leading to the marginal increase in star formation efficiency in Run I seen in Fig. \ref{fig:Zlow_SFE}. 

Overall, compared with Runs A and C, the overall effect of adding HMXB feedback on top of SN feedback in Run H (compared with Run I) is far less pronounced. This is likely due to the larger numbers of both HMXBs and SNe seen in Table \ref{tab:Zcomparison}, which are in turn a result of the higher Jeans masses present at low metallicity and the resulting top-heavy sink particle mass distribution compared with solar metallicity. These additional HMXB and SN events prevent the gas in the cloud from cooling and forming any more stars.

\subsubsection{$\alpha_{\rm vir}=1.2$ Runs: K (HMXBs and SNe), L (SNe), M (no feedback)}
Runs K ($\alpha_{\rm vir} = 1.2$, with HMXB and SN feedback) and L ($\alpha_{\rm vir}=1.2$, with SN feedback) show very similar density structures in Fig. \ref{fig:Dens_Zlow}, while there are no obvious chimneys present in Fig. \ref{fig:theta_bins_lowZ}. As seen in the corresponding temperature slices in Fig. \ref{fig:Temp_Zlow} and again in Fig. \ref{fig:theta_bins_lowZ}, the temperature of Run K is marginally higher the Run L at three free-fall times, however very similar at 12 Myr and 24 Myr. This similarity in temperature has resulted in a similar total energy in Runs K and L (see Fig. \ref{fig:Zlow_Etot}). This contrasts with the large energy gap seen when $\alpha_{\rm vir} =$ 0.7 (Runs H and I). Looking at Fig. \ref{fig:Zlow_Etot}, we see the total energy of cloud K is consistently higher than cloud L between 10-35 Myr.

Moreover, approximately the same mass of gas has been unbound/formed sinks in Runs K and L - Fig. \ref{fig:Zlow_unbound}. Comparing figures \ref{fig:Nofb_sink} and \ref{fig:Zlow_SFE}, we can see the duration of star formation has been shortened by the addition of feedback in Runs K and L (where Run M had the same initial conditions as K and L, with no feedback included). From Fig. \ref{fig:Zlow_SFE}, sink particle formation ceases at 10 Myr in Run L and 2 Myr later in Run K. Hence the addition of both HMXB and SN feedback in the gas cloud has acted to increase the period of sink particle formation, compared with the purely SN feedback case, similar to previous results (e.g. Runs A and C, along with V and W).

Fig. \ref{fig:Zlow_Einj} shows the injected energy (top plot) of Run K is significantly higher than Run L, due to a similar number of SNe (middle plot) and around 10 HMXBs being active at any point in the simulation beyond 11 Myr (bottom plot). Furthermore, looking at Table \ref{tab:Zcomparison}, the number of HMXBs and SNe in Run K is just over half of the corresponding numbers in Run H. This difference in the number of feedback events is also evident in Fig. \ref{fig:Zlow_Einj}, which shows the cumulative injected energy of Run H is approximately twice that of Run K. However, Fig. \ref{fig:Zlow_Etot} shows the total energy (of the gas) in Run H is lower than Run K. This is in part due to the fact Run H contains 5$\%$ more mass in sink particles, lowering the total energy of the gas in the system. Also, it indicates the gas in Run H is cooling more efficiently than the gas in Run K.

Overall, contrary to the results at solar metallicity, by increasing the virial parameter at low metallicity we have decreased the star formation efficiency of the cloud. This is likely to be due to inefficient cooling, which, given the higher initial energy of the gas particles, leads to less sink particle formation due to fewer gas particles meeting the required Jeans density criterion.  The same result is seen when comparing Runs G and M in Fig. \ref{fig:Nofb_sink}, where Run G(M) has the same initial conditions as Run E(K), only without feedback. In this figure, the mass fraction in sink particles is marginally less for Run M than Run G by the end of the simulation. Another important factor is the number of HMXBs and SNe - Run K contained fewer of each and, as we have seen previously, this can lead to a reduction in effectiveness of the SNe and HMXBs at producing and utilising chimneys in the gas to funnel hot gas outwards, leading to a reduction in star formation.

\subsubsection{Changing the gas cloud size: $5\times 10^5$ M$_\odot$ -- T, U, $5\times 10^6$ M$_\odot$ --  X, Y}\label{sec:Zlow_sizeMC}

Firstly, the density slices (Fig. \ref{fig:Dens_Zlow}) for the smaller molecular cloud (T - with HMXB feedback, U - just SN feedback) follow a similar structure with or without HMXB feedback. Again, this suggests that it is the SN feedback that determines the underlying density structure. The main difference is cloud U is slightly more spatially extended than Run T and contains more 10$^{-24}$ gcm$^{-3}$ density gas. Furthermore, from Fig. \ref{fig:Temp_Zlow}, the cloud containing HMXB feedback along with SN feedback, Run T, has a higher global temperature of around 10$^8$ K within the central kpc, while the same region in Run U has a temperature of $\sim$ 10$^7$ K.

The temperature slices for the larger molecular cloud (X -- with HMXB feedback, Y -- just SN feedback) show a much greater contrast than those for the smaller cloud. Cloud X shows a much larger fraction of hot (10$^8$ K) gas, with a limb extending to the outer edge of the simulation. On the other hand, Run Y contains a single sphere of uniform 10$^8$ K gas extending to approximately 1\,kpc. 

Moreover, focusing on the density slices in Fig. \ref{fig:Dens_Zlow}; Run X is more spatially extended than Run Y and more rarefied in places. Run Y is contains a more-or-less isotropic shell of higher ($\sim$10$^{-24}$ gcm$^{-3}$) density gas, with a region of lower density in the centre. 

Focusing on Fig. \ref{fig:Zlow_Etot}, we see the total energies for the smaller molecular clouds, T and U, are similar, with the cloud without HMXB feedback included finishing the simulation with a higher total energy. On the other hand, the larger molecular clouds, X and Y, show the opposite -- the run which included HMXB feedback (X) has a much higher total energy than the cloud just including SN feedback (Y). 

Fig. \ref{fig:Zlow_Einj} shows that the injected energy for Run T is initially lower at the beginning of feedback ($\sim$ 11 Myr), however the simulation ends a factor 2 higher than Run U. This difference in injected energy can be explained by Fig. \ref{fig:Zlow_Nos}, where 1-2 HMXBs are active at any one time in Run T, beyond 11 Myr. In contrast, Run X contains 30-50 active HMXBs between snapshots. As well as this, the injected energy in Run X is at least two orders of magnitude higher than Run Y, most likely due to the 65 HMXBs present in Run X (see Table \ref{tab:Runs}). 

The sheer amount of excess injected energy in Run X compared with Run Y has led to the increased unbound mass fraction seen in Fig. \ref{fig:Zlow_unbound}, coupled with a decrease in the sink particle mass fraction. On the other hand, both fractions are approximately the same in Runs T and U. As was seen for the smaller molecular cloud at solar metallicity, 90$\%$ of the gas in Runs T and U is unbound within the free-fall time, due to the lower binding energies. As such, the addition of 1-2 extra HMXB events on top of the 16 SNe has made little difference to the star formation efficiency of the cloud. 

Looking at Fig. \ref{fig:theta_bins_lowZ} we can see tentative evidence for a chimney at 24 and 35 Myr in Run X. However, as has been seen previously it only spans less then 1 order of magnitude in density and also appears beyond the free-fall time of the cloud, making it unlikely to influence star formation in the cloud. 

Finally, comparing the numbers of sink particles formed between snapshots as a function of time in Fig. \ref{fig:Zlow_SFE}, we see the duration of star formation is largely unchanged between Runs Y and X, while Run T has a marginally smaller star formation period than Run U. This is interesting, since from Fig. \ref{fig:Zlow_Einj} we can see the injected energy for Run T is smaller than Run U at earlier times, suggesting there is a degree of positive feedback present.

To conclude, while the a larger molecular cloud increases the efficiency of both SN and HMXB feedback by increasing star formation duration and efficiency (compared with the purely SN case) at solar metallicity, this is not the case at lower metallicity. Instead, the larger number of HMXBs present (due to larger sink particle masses) results in an order of magnitude jump in energy between the cloud that includes just SN feedback (Y) and the cloud that includes both SNe and HMXBs (X). Although this number is highly stochastic, it is still likely the larger energy injection of both SNe and HMXBs (due to the higher numbers of massive stars seen in Table \ref{tab:Zcomparison}) would wash out any effects of the combination of the two types of feedback, were the runs to be repeated with a different random seed for the binary population synthesis model. On the other hand, the smaller molecular clouds at low metallicity have the same pitfalls as the smaller ones at solar metallicity; the lower binding energies and smaller number of HMXBs make their effect negligible when compared with molecular clouds just containing SNe.

\section{Discussion}\label{sec:discussion}
In this paper we investigate the effects of including a prescription for HXMB feedback (a gradual heating source) on top of SN feedback in star-forming giant molecular clouds. Our simulations follow the molecular cloud over 3 free-fall times. We ran simulations which varied metallicity, cloud size, and the way the HMXB feedback was implemented.  Below we summarise our main results. 
\begin{itemize}
\item In the clouds studied here, the addition of feedback does not change the fact the majority of star formation occurs within the free-fall time of the clouds.
\item However, the combination of SN and HMXB feedback can lead to efficient use of low density `chimneys' in the gas cloud, funneling hot gas from the central regions of the cloud in order to maintain cold, high density gas to fuel further star formation at later times.
\item At solar metallicity, primarily through the action of chimneys, the combination of HMXBs and SNe can extend the period star formation as well as increase the star-formation efficiency compared with clouds that just include SNe.
\item Due to the similarity between density profiles of runs with just SN feedback and runs with HMXB feedback included on top, it appears that the initial SN events (prior to the onset of HMXB feedback) set the preferential direction of the hot gaseous outflows from the centre of the cloud. In other words, SNe set the locations of the chimneys and the addition of HMXB feedback increases their effectiveness at funneling hot gas outwards. This is despite the addition of HMXBs increasing the energy injected into the cloud.
\item There is evidence to suggest HMXBs increase the efficiency of chimneys at removing energy from the cloud by increasing the temperature of the gas inside to beyond 10$^7$ K, where the cooling is predominantly Bremsstrahlung dominated and relatively inefficient. 
\item The number of HMXBs and SNe in each simulation seems to be the defining factor of the fate of the gas in the cloud. This can vary between clouds due to the inherent stochasticity in HMXB formation. For example, simulations A, C, H and I were re-run with different seeds for the turbulence generator (which introduced stochasticity into the massive star properties that were assigned to sink particles). We found the defining factor in these runs was the inherent stochasticity in the number of massive stars and hence feedback events. We also found the same result when conducting convergence tests (see appendix \ref{appendix:Nyquist}). In this way, clouds with the same initial conditions can have different fates based purely on the random sampling of the underlying IMF.
\item However, the main factor in determining the number of stellar feedback events in these simulations was the average sink particle mass. Inefficient cooling led to higher Jeans masses and hence sink particle masses (this was true for the runs at [Fe/H] = -1.2),  which resulted in a higher number of HMXB and SN events. On the other hand, efficient cooling (as was seen in the kinetic HMXB feedback run; Run B) led to a higher number of sink particles with lower mean masses and hence ultimately fewer HMXBs/ SNe (see Table \ref{tab:Zcomparison}).
\item At solar metallicity, the positive feedback effects of combining SNe and HMXBs are more apparent in larger molecular clouds ($>$10$^6$ M$_\odot$). This is because the gravitational binding energy of the smaller clouds is small enough that $\sim$ 10 SNe can unbind the majority of the gas in the cloud. Also, the rarity of HMXB events means there are typically just $\sim$ 2 HMXBs acting in the lower mass clouds.
\item Overall, the differences between runs that include HMXB feedback and those that just include SN feedback is far less pronounced in runs at low metallicity. This is due to larger numbers of SNe and HMXBs, caused by larger sink particle masses which originate from a higher Jeans mass. Furthermore, the higher Jeans mass is due to relatively inefficient cooling below 10$^4$ K, compared with solar metallicity runs. 
\item Kinetic HMXB feedback resulted in a momentum-driven outflow, which was slowed at $\sim$ 400\,pc due to the large swept up mass. The continuous injection of energy within the central 200\,pc of the cloud resulted in a low density, cold cavity. Furthermore, the effects of feedback beyond $\sim$ 400\,pc were limited by the efficient cooling of injected SN energy and thermalised HMXB kinetic energy input.
\item Our results are comparable to the case when the stellar wind feedback from massive stars is included on top of SN feedback. The key element in chimney efficiency is the presence of a constant heating source of sufficient power. 
\end{itemize}

\subsection{Our set-up in context}
Our initial conditions, while typical of those in other works investigating feedback effects in molecular clouds \citep[e.g.][]{Dale2012,Federrath2014}, represent an idealised set up. In reality, molecular clouds are dynamic objects which are often unbound \citep{Dobbs2011}. Also the galactic environment likely plays an important role in cloud evolution (\citealt{Rey-Raposo2017}). However, to isolate the effects of different processes and parameters a relatively simple set of initial conditions is required. Furthermore, in this paper we employ a basic method to produce a population of massive stars with a set of lifetimes and masses that are physically motivated. A limitation to our method is that we do not know when the first massive star forms within the sink as we do not resolve the formation of individual stars. It may occur near-simultaneous with the sink particle formation, as we have assumed here, or some time later - however this is equivalent to forming the sink slightly late, while the formation time and location of sinks is already highly stochastic (see appendix \ref{appendix:Nyquist}). An alternative method to produce a population of binary stars would be to use the outputs from the existing binary population code BPASS (\citealt{Eldridge2009}, \citealt{Eldridge2017}). This would allow us to follow the properties (e.g. temperature, luminosity) of individual stars, taking into account those in binaries. This would be particularly useful in future work where we want to take into account the radiative luminosity of the stars alongside their mechanical luminosities. 

In our simulations we have followed the evolution of the clouds for 35 Myr. This is longer than the lowest estimates for molecular cloud lifetimes which range from $\sim 1$ Myr to $10^2$ Myr. Thus our simulated cloud lifetimes represent an intermediate value (\citealt{Heyer2015}, \citealt{Cohen1980}). From observations of the solar neighborhood, shorter lifetimes are favoured due to a missing population of older (over 3 Myr) stars inside molecular clouds (e.g. \citealt{Ballesteros-Paredes2006}). This suggests molecular clouds are destroyed before this, whether by photo-ionising radiation, stellar winds, SNe or a combination of these effects (e.g. \citealt{Rahner2017}, \citealt{Skinner2015}, \citealt{Dale2011}, also for a review see \citealt{Krumholz2014}). Furthermore, work by \cite{Elmegreen2000} and \cite{Hartmann2001} suggests molecular clouds should be destroyed within 1 free-fall time. However, the clouds we study here are of masses comparable with GMCs and their free-fall time is $\sim$ 12 Myr, which is comparable to the lifetimes of massive stars. This means they exist towards the upper limit of molecular cloud masses and in a regime where the feedback from massive stars becomes important. Moreover, \cite{Dale2012} found that such clouds are unlikely to be significantly disrupted by photo-ionising radiation within 3 Myr, while \cite{Matzner2002} predicts cloud lifetimes between 10-30 Myr on the grounds this represents the time it would take HII regions generated by the photo-ionising flux, predominantly from massive stars, to evaporate the cloud. Furthermore, work by \cite{Murray2011}, which cross-correlates 32 star-forming complexes identified by WMAP in the Milky Way with a GMC catalogue, found the mean free-fall time of the massive GMCs in the Milky Way to be 27 $\pm$ 12 Myr.

All of the clouds in our simulations exist below the upper mass limit $6\times 10^6$ M$_\odot$, which represents the largest of the molecular clouds in the Milky Way (e.g. \citealt{Williams1997}). As the upper limit of the expected lifetimes is comparable to the length of our simulations, it would be interesting, in future work, to investigate the effect of including photo-ionising radiation from massive stars on top of HMXBs and explore how this influences the formation of chimneys and ultimately the star formation efficiency of the clouds. The impact of prior feedback mechanisms on the way SN interact with their environment has been investigated in a number of works; for example \cite{Walch2015} (photo-ionisation), \cite{Fierlinger2016} (winds) and \cite{Kim2016} (radiation pressure).  

We have described our simulated clouds as either 'bound' or 'unbound' with virial parameters ranging from 0.7 to 1.2. However, this relates to the initial conditions. In reality both clouds are free-falling by the time feedback kicks in (see section \ref{sec:Nofb}). This is due to the fact the initial turbulent velocity field is quickly thermalised and lost due to subsequent cooling, as was also seen in work by \cite{Dale2012}. Therefore, in reality the higher virial parameter in our initial conditions manifested as an alteration to the distribution of sink masses (see Fig. \ref{fig:Sink_Hist_AE}), with a lower mean sink particle mass (see Table \ref{tab:Zcomparison}) and hence fewer HMXB and SN events. However, how HMXB and SN feedback would interplay in an unbound cloud remains an interesting and important question, hence a method of modeling an unbound cloud would be of interest the future.  

Despite focusing on HMXB feedback in this paper, our results are also comparable to including massive stellar winds in the GMC. Table \ref{tab:Zcomparison} shows the number of HMXBs in each simulation is between 10$\%$ and 20$\%$ of the number of massive stars leaving the main sequence. Given the 35\,Myr timescale of the simulations, this means the energy injected via stellar winds would be comparable to the power input from HMXBs. Crucially, it would also mimic the gradual heating from HMXBs and hence would likely lead to the formation of the kpc-scale chimneys (as was seen in previous work by \citealt{Rogers2013}). In future work we will aim to include both stellar winds and HMXBs, as well as scale the energy injected by HMXBs according to the mode of accretion and mass of the companion star. 

Beyond the alternative modes of accretion in HMXBs (as well as the several orders of magnitude increase in luminosity this can produce), another key difference between HXMB feedback and stellar wind feedback is that, on the spatial scale of the jets, the energy input is delivered to the medium an-isotropically in the HMXB case and isotropically by stellar winds. We have shown chimneys can be produced without any directionality to the feedback, however in future work it will be interesting to investigate the interaction of directional HMXB jet feedback with an inhomogeneous ISM, along with how this affects chimney formation.   

\subsection{Our results: chimneys}
\cite{Justham2012} explore the idea that the combination of SN and XRB feedback produces `chimneys' leading to an increase in the star formation efficiency. However, while they suggest XRBs might help create chimneys in the gas, which would help funnel subsequent SNe-heated gas, we find it is the SN feedback that helps evacuate the chimneys and HMXBs instead of increasing their efficiency. This result comes about because HMXB feedback begins after SN feedback. This is set by our feedback scheme, which requires one SN in the binary system before the HMXB feedback can begin. In reality there are a variety of XRB systems ranging from low to high mass and also a variety of lifetimes. Consequently, it is possible XRB feedback may be present before the first SN and may alter the locations of the chimneys. \cite{Justham2012} also discuss this, referencing the population of lower metallicity stellar mass black holes which have progenitor masses of $>$ 40 M$_\odot$ and formed via direct collapse, without forming a SN shock. Further, \cite{Eldridge2004} find at a metallicity of 0.001 (the same as we use in our lower metallicity simulations), the transition between partial and direct collapse of black hole progenitors occurs at a lower value of $\sim 35$ M$_\odot$ (see figure 5 in \citealt{Eldridge2004}).

Therefore it is possible that HMXBs can precede the first supernovae, allowing HMXB feedback to significantly alter the gas cloud before SN feedback kicks in. In order to quantify whether ignoring this population has affected our results, particularly those at low metallicity (where the number of HMXBs was very high) we found the time of the first SN with a progenitor with an initial mass greater than 35 M$_\odot$ contained in an HMXB (hereafter we'll refer to these as SNegt35) in Runs H and X, as well as the times of the first SN events in each simulation, along with the number of SNe prior to the first SNegt35 event. For run H the first SN is at 7.6 Myr and the first SNegt35 is at 9.2 Myr, which occurs after 24 further SNe. As such, during Run H, the SNegt35 events would be unable to affect the ISM prior to SNe. However, 25$\%$ of the HMXBs in Run H contain a primary with an initial mass greater than 35\,Myr. Therefore, these systems represent a significant fraction of the HMXBs operating in Run H and in future work this could be factored in when considering the number of SNe. Moreover, 20$\%$ of the HMXBs in Run X contain a $>$ 35M$_\odot$ primary star, however 22 SNe occur before the first SNegt35 event, mirroring the results of Run H. 

The idea of SN feedback carving chimneys (or channels) in the surrounding gas has been studied in the literature over the past twenty years \citep[e.g.][]{DeYoung1990,MacLow1999}. In particular, recent work by \cite{Iffrig2015} looked at the effect of SNe in turbulent molecular clouds, finding the hot SNe-heated gas was able to escape through low density channels and subsequently form super-bubbles. Moreover, work by \cite{Martizzi2015} and \cite{Kim2015} also look at the interaction of a single SN in an inhomogeneous medium, focussing on the use of low density channels at smaller scales. Furthermore, \cite{Kimm2015} hypothesise the use of low density channels by SNe-heated gas may be crucial in modeling star formation in galaxies. On the other hand, other papers focus on the interaction of stellar winds and SN feedback and the subsequent formation of chimneys in molecular clouds; \cite{Rogers2013} find stellar winds help to remove gas from low density channels, through which hot gas can escape and potentially globally affect the host galaxy. Other works which include chimneys are; \cite{Rosen2014} \cite{Fierlinger2016}, \cite{Ibanez2016}.

Additionally, the fact that the mean sink masses in our simulations are higher at low metallicity (see Table \ref{tab:Zcomparison}) is consistent with other investigations in the literature. For example, \cite{Jappsen2005} and \cite{Bonnell2006} both reference the drop in Jeans mass (and therefore greater fragmentation) seen with lowering the temperature of the gas through additional cooling. Since we do not resolve star formation, we cannot say the top-heavy fragmentation we see in the lower metallicity runs occurs down to small scales, however numerous papers have found massive stars tend to form in isolation at lower metallicity (or higher redshift equivalently); for example  \cite{Abel2002}, \cite{Bromm2002}, \cite{Oshea2007} and \cite{Yoshida2007}. 

Finally, the hot gaseous chimneys seen in our results could have implications during the Epoch of Reionization; by increasing the UV photon escape fraction from regions of star formation. At z$>$6, the contribution of ionizing photons from stars is thought to outweigh that from quasars (e.g. \citealt{Madau1999}, \citealt{Fan2002}, \citealt{Srbinovsky2007}). There is a possibility the kpc-scale chimneys seen in this paper could act to enhance the UV photon escape fraction by providing low density 'holes' in the ISM. This would lower the star formation efficiency required to reionize the intergalactic medium at high redshift. The enhancement in escape fraction could be investigated further by including the ionizing photons produced by the stellar population and post-processing the simulations using a radiative transfer model, however this is beyond the scope of this paper. 

\section{Conclusions}\label{sec:conclusions}
We have explored the effect of gradual heating feedback, as well as SN feedback, in giant molecular clouds of varying metallicity, size and virial parameter. This gradual heating can arise from different mechanisms and in this work we focussed on feedback from HMXBs. Our primary result was that the two types of feedback combine to produce kpc scale chimneys of low density, hot gas. These chimneys help funnel energy away from the inner regions of the cloud, allowing star formation to continue there. The chimneys are present in runs that just include SN feedback, however the addition of the gradual power input from HMXBs can help to prevent the gas inside the chimneys from cooling by increasing its temperature to beyond $10^7$ K, where the cooling is Bremsstrahlung dominated and relatively inefficient. The chimneys are more prevalent in larger clouds at higher metallicity. We found the combined effects of SNe and HMXBs are largely washed out in smaller clouds and runs at lower metallicity due to the lower binding energy of the gas and larger population of massive stars respectively. 

In future work we will investigate the interplay of SN and HMXB feedback in dwarf galaxies at high redshift. Furthermore, we will investigate the inclusion of photo-ionising radiation and winds from the massive OB type stars and their subsequent effects on chimney formation and ultimately the star formation history of the cloud.

\section*{Acknowledgements}
The authors would like to thank Martin Bourne and Jim Pringle for useful and interesting discussions, along with the anonymous referee for their insights and suggestions. LGS is supported by a Science and Technology facilities council (STFC) PhD studentship.  CJN is supported by the Science and Technology Facilities Council (STFC) (grant number ST/M005917/1). CP  is  supported  by  Australia Research Council (ARC) Future  Fellowship FT130100041. This work used the DiRAC Complexity system, operated by the University of Leicester IT Services, which forms part of the STFC DiRAC HPC Facility. This equipment is funded by BIS National E-Infrastructure capital grant ST/K000373/1 and STFC DiRAC Operations grant ST/K0003259/1. DiRAC is part of the National E-Infrastructure. Figures \ref{fig:Dens_Zsol}, \ref{fig:Temp_Zsol}, \ref{fig:RunA_xz}, \ref{fig:Dens_Zslices_A}, \ref{fig:Dens_Zslices_C}, \ref{fig:Temp_Zslices_A}, \ref{fig:Temp_Zslices_C},  \ref{fig:tslice_RunsEF}, \ref{fig:Dens_Zlow}, \ref{fig:Temp_Zlow}, \ref{fig:Grav_Collapse},\ref{fig:Dens_Nqy}, \ref{fig:Temp_Nqy}, \ref{fig:Dens_conv} and \ref{fig:Temp_conv} were produced using SPLASH (\citealt{Price2007}). 

\bibliographystyle{mnras}
\bibliography{Slow_and_Steady}

\appendix

\section{Sedov-Taylor Comparison}\label{appendix:Sedov}
In order to ascertain whether our SN implementation in GADGET-3 can capture the Sedov-Taylor phase of the shock expansion, we performed a simulation of a single supernova explosion in a homogeneous gaseous sphere and compared our results with the Sedov-Taylor solution. The particle resolution was set to 2 $\times$ 10$^6$ (lower than our result runs), while the SN injected 2 $\times$ 10$^{51}$ erg of thermal energy into the surrounding 100 SPH neighbours according to the kernel weighted scheme we descried in section \ref{sec:nummodel}. The radius of the cloud was set to 60\,pc and the initial temperature of the gas was set to the virial temperature, which was 534 K. We are unable to resolve the initial free expansion phase of the shock, since this breaks down when the swept-up mass equals the mass ejected during the supernova explosion; a condition which is instantly met when we inject the SN energy into the surrounding 100 SPH particle neighbours.

\begin{figure}
 \psfig{file=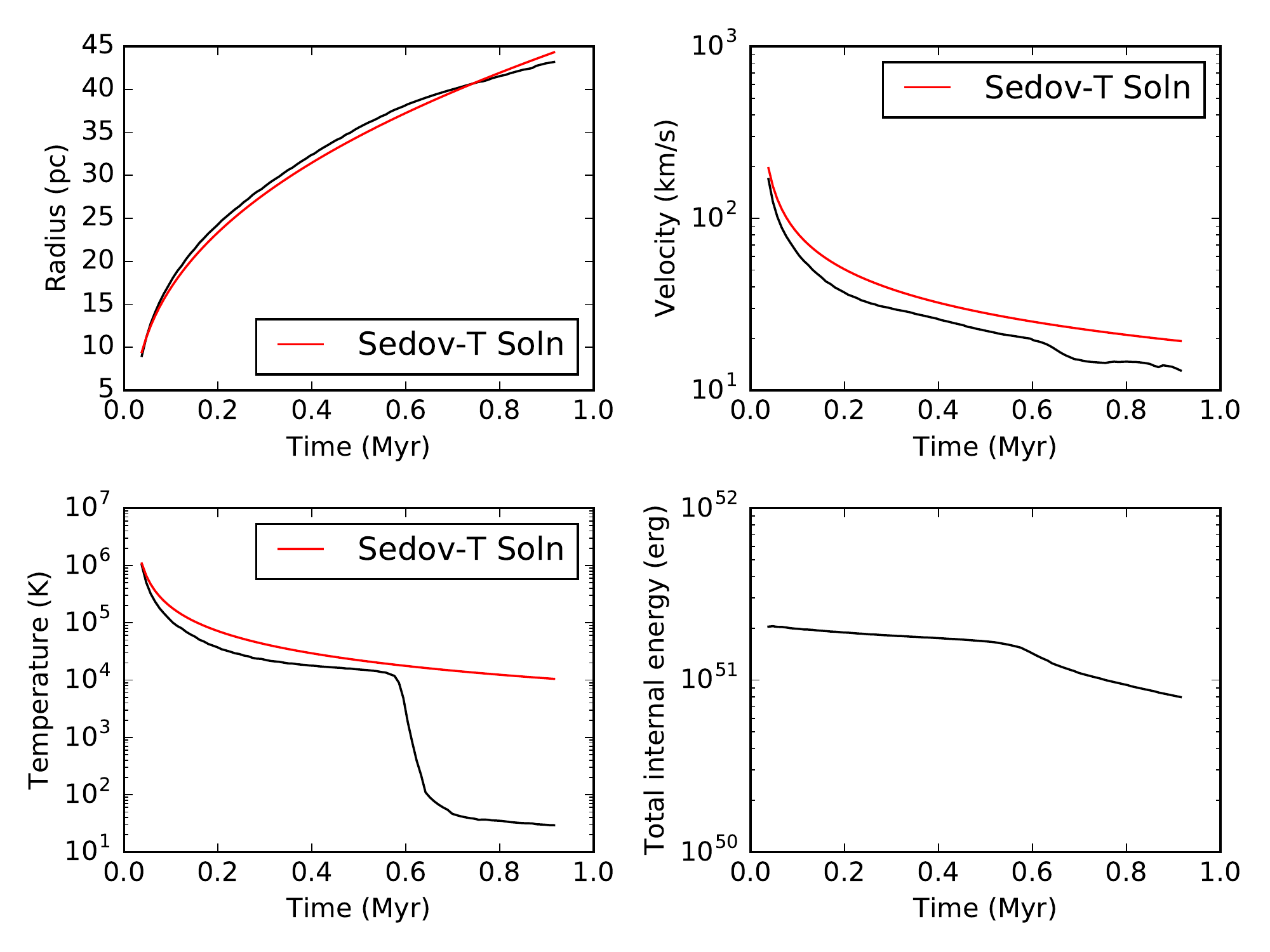,width=0.5\textwidth,angle=0}
  \caption{Plots to show the time evolution of the shock radius (in pc, upper left), velocity (in km/s, upper right), the post-shock temperature (in K, lower left) and the total internal energy of the gas inside the shock front (in erg, lower right). The Sedov-Taylor analytic solution for the radius, velocity and temperature of the shock is plotted in red. At $t\approx 0.6$\,Myr the shell is able to cool, meaning the expansion is no longer adiabatic and the solution deviates from the Sedov-Taylor result.} 
  \label{fig:Sedov} 
\end{figure}

Figure \ref{fig:Sedov} plots the shock radius, velocity, post-shock temperature and total internal energy inside the supernova remnant (SNR), against the analytic Sedov-Taylor solution. The shock radius, velocity and temperature were defined as the mean values of particles in a 2 pc radial bin, centered on the position of maximum density. The total internal energy is just the sum of the internal energies of the gas particles located inside the spherical shock front.  The SN energy was injected at $\sim$ 0.01 Myr. From \ref{fig:Sedov}, it can be seen the shock radius, velocity and temperature align well with the analytic solution until $\sim$ 0.6 Myr, when the temperature of the particles in the 2 pc radial bin drops steeply. There is also a corresponding drop in shock velocity and total internal energy inside the remnant. The drop in temperature indicates the post-shock medium immediately behind the shock-front has been able to cool efficiently, forming a cool shell. This shell is being pushed forwards by the hot, lower density gas in the centre of the remnant and indicates the SN remnant (SNR) has left the Sedov-Taylor phase. As a result, the expansion of the SNR is no longer adiabatic, which can be seen in the total internal energy plot in Fig. \ref{fig:Sedov}; at $\sim$ 0.6 Myr, the total internal energy of the gas inside the shock front begins to drop. Once the gas inside the centre of the remnant has cooled, it will have entered the `Snow plow', or momentum-conserving phase. 

The roughly constant radial power-law is likely due to the fact the simulation volume is finite and the shock has reached the outer-parts of the gaseous sphere. The lack of gas outside the shock causes a break-down in the analytic progression of the shock, however it would be expected once the SNR enters the Snow-plow phase, the radius would increase as t$^{1/4}$ instead of the t$^{2/5}$ power law seen in Fig. \ref{fig:Sedov}. 

The fact we can resolve the Sedov-Taylor phase of the SNR expansion indicates our resolution and SN feedback implementation are both sufficient to capture the effect of SN explosions on the ISM down to tens of parsec scale. 

\section{The energy jump seen in Run F at 22 Myr}\label{appendix:RunF}
\begin{figure*}
\includegraphics[trim={0 0 0 0.12cm},clip, width=\textwidth]{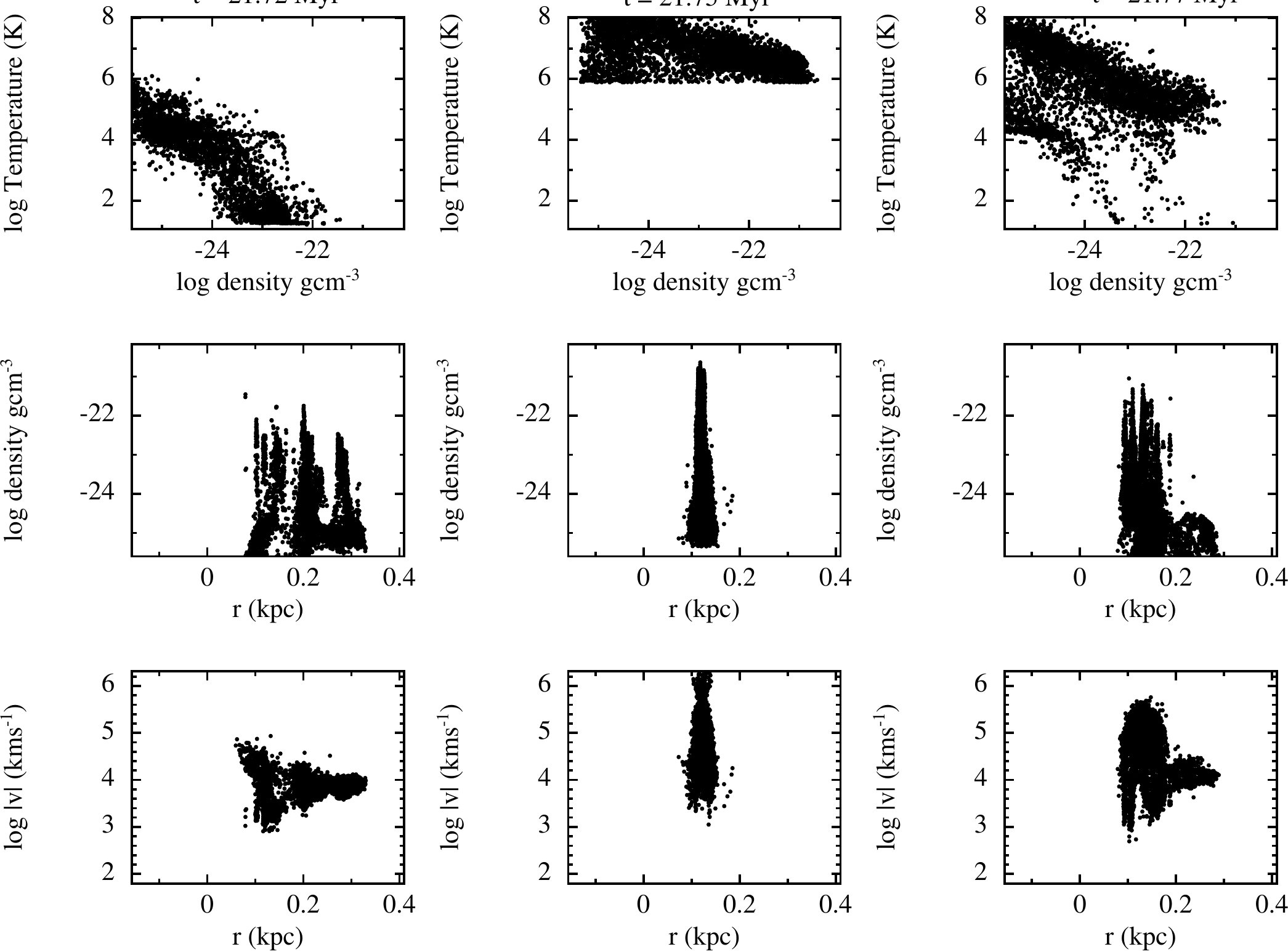}
  \caption{Plots to show the time evolution of 4604 high density ($>$ 10$^{-25}$ gcm$^{-3}$) particles that were heated to above 10$^6$ K 21.75 Myr into Run F. The first column shows the properties of the gas particles at 21.72 Myr, the second column shows the same at 21.75 Myr and the third column corresponds to t = 21.77 Myr. Top row - plot to show the temperature and density of the particles at 3 different times. Middle row - plots to show the densities and radii of the particles at the 3 different times. Bottom row - plots showing the magnitude of the particle velocities and how this varies with radius.} 
  \label{fig:Grav_Collapse} 
\end{figure*}

In Fig. \ref{fig:Grav_Collapse} we plot the evolution of the temperature, density, radius and velocity of the high density particles that were heated to 10$^{6}$ K, 22 Myr into Run F (see Fig. \ref{fig:Zsol_TvsD_EF}). We can see the particles were heated to above 10$^6$ K at 21.75 Myr and prior to this they occupied radii between $\sim$ 100 -- 350 pc and had correspondingly high densities of between 10$^{-25}$ gcm$^{-3}$ to 10$^{-22.5}$ gcm$^{-3}$. However, at 21.75 Myr we can see the peak density of the gas particles has increased and the radii have converged to a value of $\sim$ 120 pc. Furthermore, the absolute velocities of the particles have increased by an order of magnitude. The increase in density of the particles, coupled with the velocity increase, indicates this region has undergone gravitational collapse, which has resulted in the shock-heating of the gas particles. Furthermore, there were no SN events between 21.72 Myr and 21.75 Myr, indicating this heating was not due to direct heating via SN feedback. Beyond 21.75 Myr the gas particles have begun to cool, however the mean radius of the particles has increased. This, coupled with the fact the velocities are still larger than they were prior to the gravitational collapse, indicates the gravitational shock-heating has generated an outflow.

\section{Resolution Tests}\label{appendix:Res}
A number of resolution tests were run in order to test for convergence in our results. Initially, we varied between 3-12$\times$10$^{6}$ particles whilst keeping the total gas mass 2$\times$10$^6$ M$_\odot$ and found significant differences in the star formation efficiency and unbound gas fractions between resolutions. In order to ascertain the source of this disparity, we first explored whether or not the initial random turbulent velocity field was having a significant impact on the cloud's fate (see section \ref{appendix:Nyquist}).  

\subsection{Nyquist Frequency}\label{appendix:Nyquist}
When we set up the initial turbulent velocity field, the minimum length scale corresponds to the maximum $k$ value ($k_{\rm max}$), which is the Nyquist frequency and is set by the particle resolution of the simulation. In order to investigate whether it is the choice of $k_{\rm max}$ that is determining the results of each simulation, we ran 5 simulations, with 3,5,6,7 and 10 million particles respectively, each with a $k_{\rm max}$ value corresponding to the lowest resolution run ($3\times 10^6$ particles). The total cloud mass in each case was $2\times 10^6$ M$_\odot$, while the radius was 100pc. 

\begin{figure}
  \psfig{file=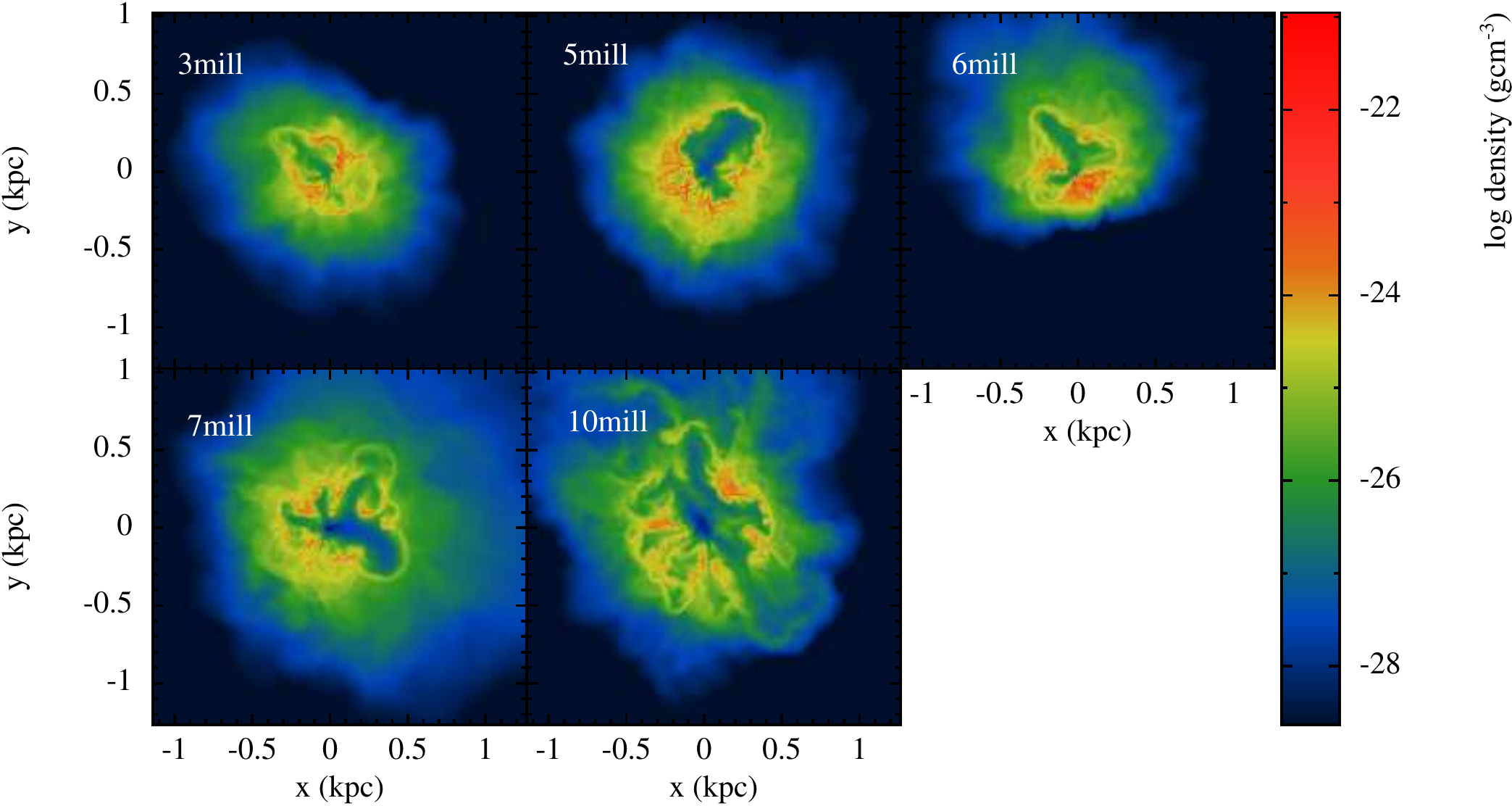,width=0.5\textwidth,angle=0}
  \caption{Plots to show z=0 slices in density, taken at the free-fall time (11.7 Myr) of a cloud with varying resolution (stated on each plot). The Nyquist frequency in each case is set by the value at the lowest resolution (3 $\times$ 10$^{6}$ particles).} 
  \label{fig:Dens_Nqy} 
\end{figure}

\begin{figure}
  \psfig{file=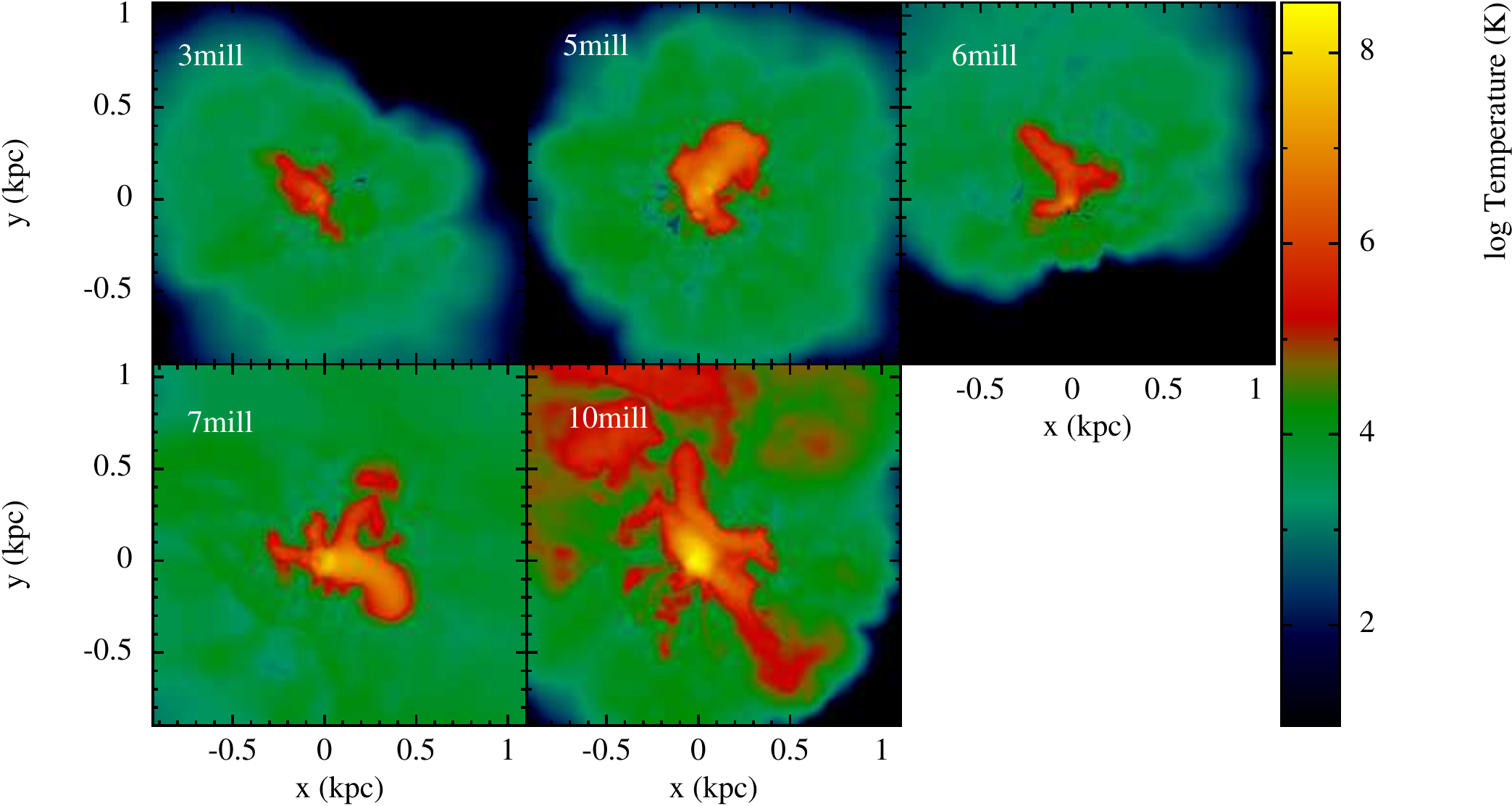,width=0.5\textwidth,angle=0}
  \caption{Plots to show z=0 slices in temperature, taken at the free-fall time (11.7 Myr) of a cloud with varying resolution (stated on each plot). The Nyquist frequency in each case is set by the value at the lowest resolution (3 $\times$ 10$^{6}$ particles).} 
  \label{fig:Temp_Nqy} 
\end{figure}

Figure \ref{fig:Dens_Nqy} shows the density slice in the x-y plane, at z=0 and taken at the free-fall time of the cloud; 11.7 Myr, for each resolution. Low density chimneys can be seen in all runs, however, their spatial extent, location and number do not appear to significantly correlate between resolutions. This points to another factor determining the fate of each cloud. This can also be seen in the corresponding temperature slice (Fig. \ref{fig:Temp_Nqy}). However, the run containing 10$^7$ particles does show a significant increase in the amount of hot gas in the cloud, as well as an increase in the spatial extent of the low density, hot gaseous chimneys/bubbles present; in Fig. \ref{fig:Temp_Nqy} the lower right chimney can be seen to extend to 1 kpc. 

\begin{figure}
  \psfig{file=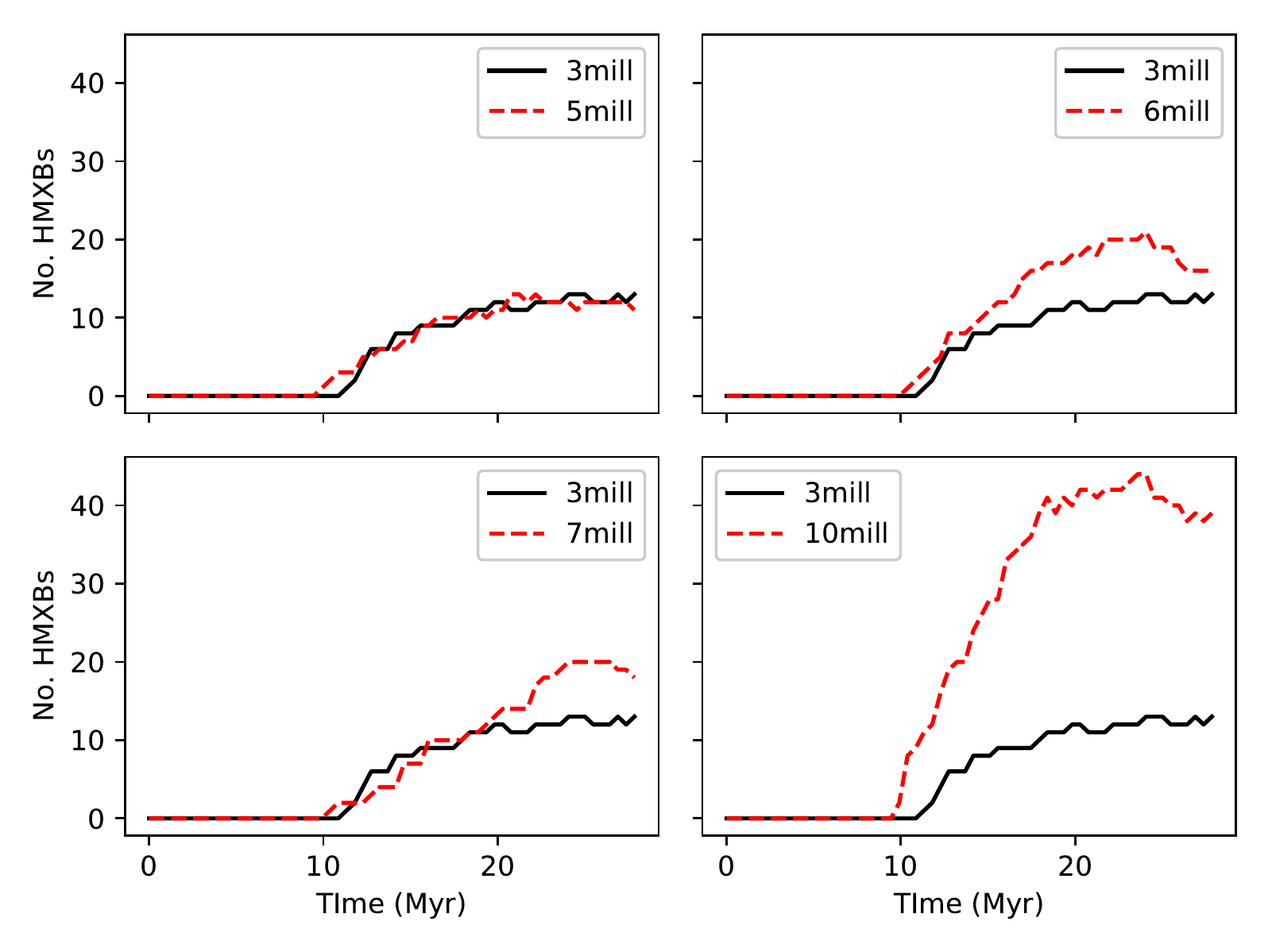,width=0.5\textwidth,angle=0}
  \caption{Plots to show the number of HMXBs active between snapshot times, for particle resolutions varying between 3-10 $\times$ 10$^6$. The lowest resolution run, 3 $\times$ 10$^6$ (labeled 3mill), is plotted in the black solid line on each plot for reference.} 
  \label{fig:NHMXBs_Nyq} 
\end{figure}

\begin{figure}
  \psfig{file=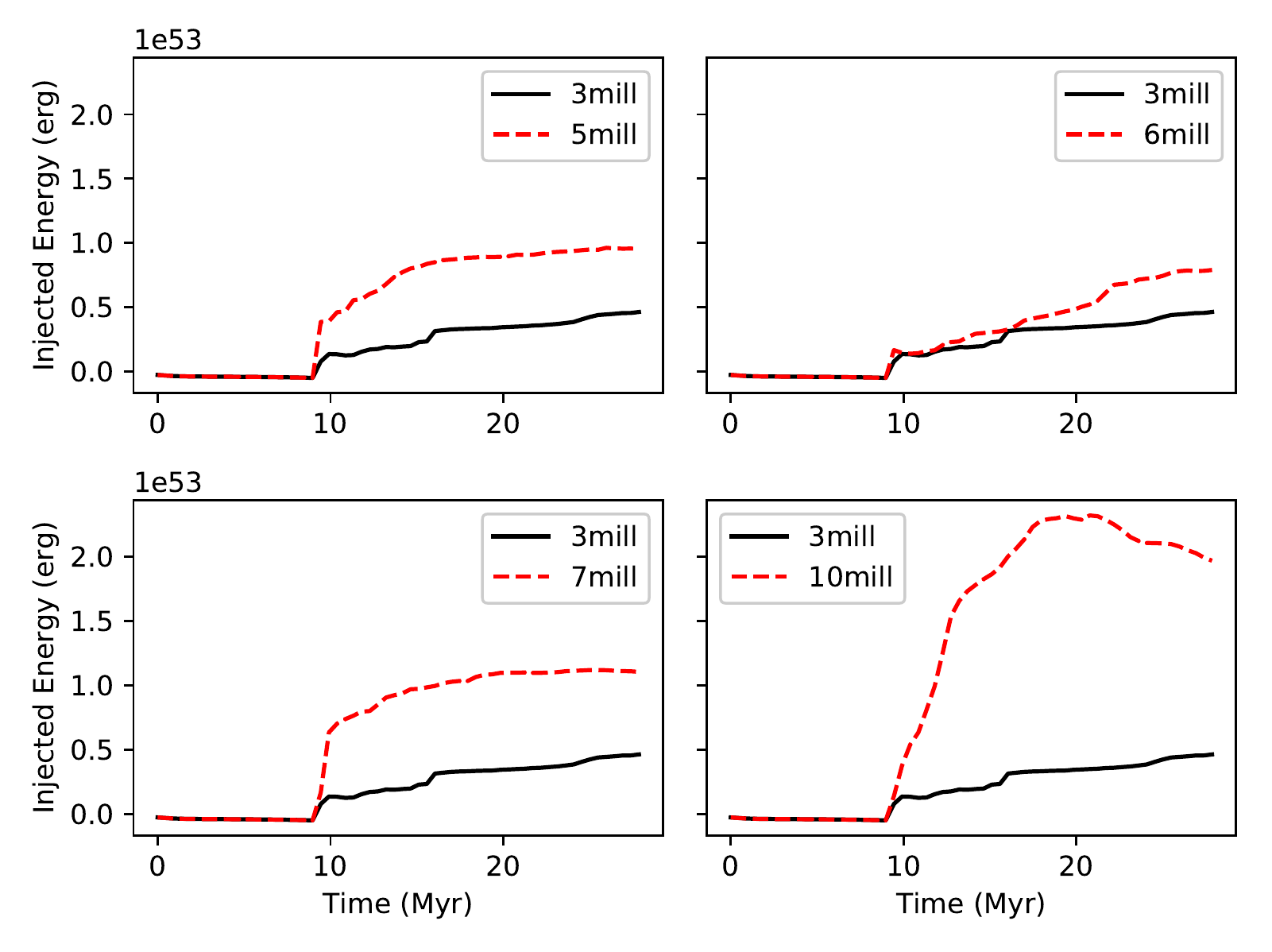,width=0.5\textwidth,angle=0}
  \caption{Plots to show the amount of feedback energy (SNe + HMXBs) injected between snapshot times, for particle resolutions varying between 3 - 10 $\times$ 10$^6$. The lowest resolution run, 3 $\times$ 10$^6$ (labeled 3mill), is plotted in the black solid line on each plot for reference.} 
  \label{fig:Einj_Nyq} 
\end{figure}

In order to find the cause of the differences between each of these simulations, (despite a constant Nyquist frequency), we instead plotted the number of HMXBs active between snapshots in each simulation (Fig. \ref{fig:NHMXBs_Nyq}), along with the amount of energy injected in each snapshot (Fig. \ref{fig:Einj_Nyq}). Fig. \ref{fig:NHMXBs_Nyq} shows the number of active HMXBs varies between resolutions, however there is no clear trend with increasing resolution. The same is true for the overall injected energy in Fig. \ref{fig:Einj_Nyq}. When the two plots are compared, it is apparent there are discrepancies between the number of HMXBs active and the amount of energy injected between snapshots. This is due to varying numbers of SNe, as well as different HMXB lifetimes (and therefore rates of energy injection).  Moreover, the origin of the large amount of hot gas seen in Fig. \ref{fig:Temp_Nqy} can be seen in Fig. \ref{fig:Einj_Nyq}; the energy injected in this run is a factor of $\sim$ 2 greater than the other runs at around the free-fall time, due to a large number of HMXBs being active across the simulation (see Fig. \ref{fig:NHMXBs_Nyq}). The results of these simulations indicate the inherent stochasticity in HMXB feedback is the most significant factor in determining the clouds fate and also an obstacle when attempting to perform convergence tests. This stochasticity was introduced in these simulations, since by altering the realization of the turbulent velocity spectrum and hence the random velocities assigned to the gas particles in the initial conditions, the identities and locations of the gas particles that formed sink particles were altered, along with their formation time. This in turn changed the random seeds utilised when assigning massive star properties (see section \ref{sec:ICs}) between simulations.
\newline \indent In the next section (section \ref{appendix:conv}) we avoid the stochasticity effects associated with Monte Carlo-type HMXB population synthesis method by inserting a pre-determined population of binaries, with set lifetimes, energy injection rates and locations, into each simulation.

\subsection{Convergence Testing}\label{appendix:conv}
For simplicity, in order to ascertain numerical convergence, we set up a population of 20 sink particles at set locations inside each cloud (of total gas mass $2\times 10^6$ M$_\odot$). The location of each sink particle was set throughout the simulation, along with the lifetime of both the primary and secondary. Each sink particle therefore underwent a SN feedback event and a HMXB feedback phase, along with a second SN event. This time we varied the resolution from $3-12 \times 10^6$ particles. 

\begin{figure}
  \psfig{file=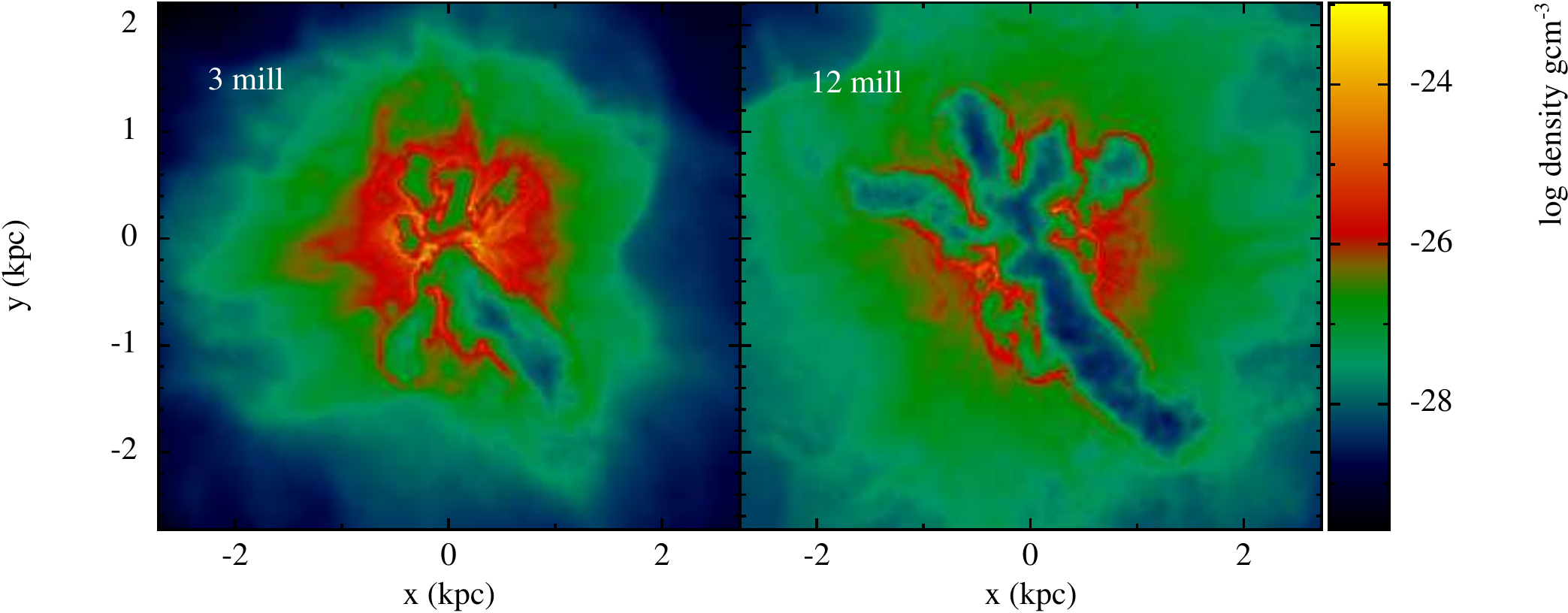,width=0.5\textwidth,angle=0}
  \caption{Density slices taken in the $x$-$y$ plane at $z=0$, showing the convergence tests of varying particle resolution (where the total cloud mass is kept the same and 3mill is $3\times 10^6$ particles) at the free-fall time ($11.7$\,Myr) of the cloud.}
  \label{fig:Dens_conv} 
\end{figure}

\begin{figure}
  \psfig{file=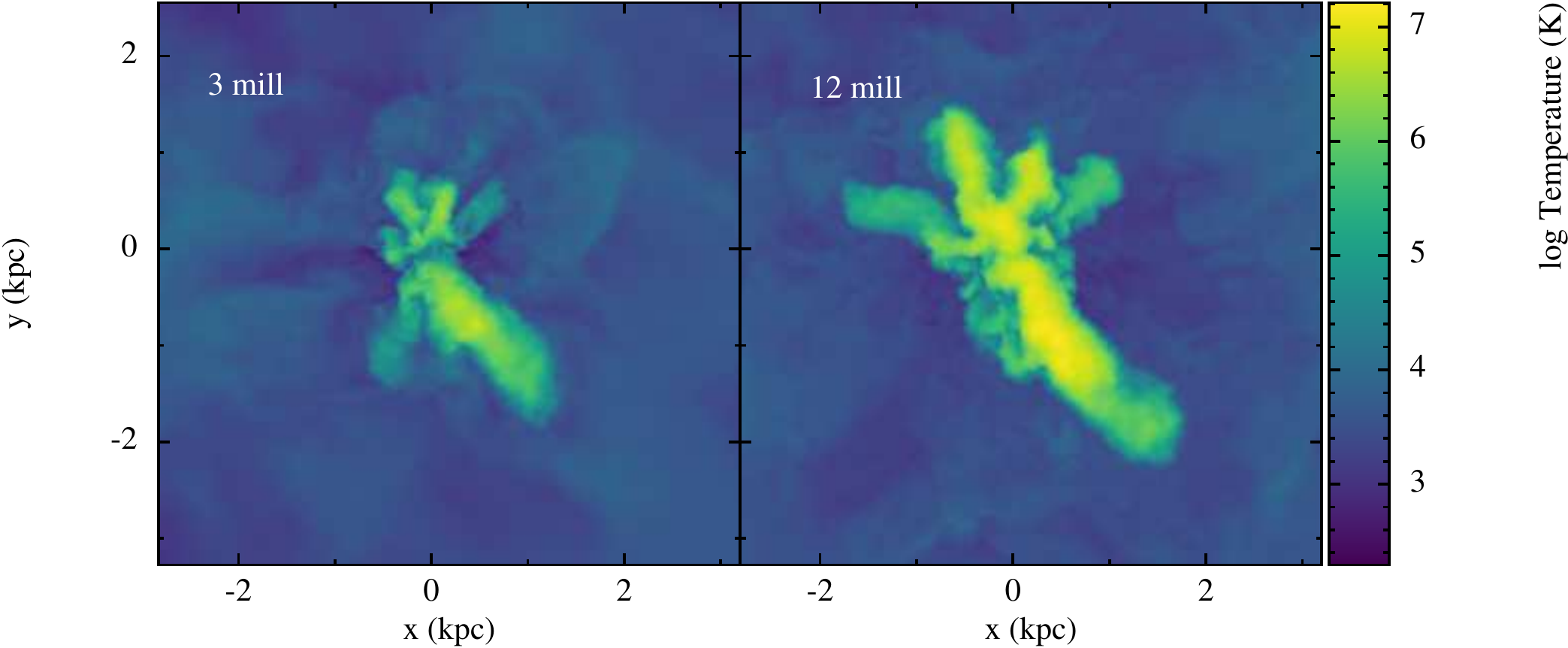,width=0.5\textwidth,angle=0}
  \caption{Temperature slices taken in the $x$-$y$ plane at $z=0$, showing the convergence tests of varying particle resolution (where the total cloud mass is kept the same and 3mill is $3\times 10^6$ particles) at the free-fall time ($11.7$\,Myr) of the cloud.}
  \label{fig:Temp_conv} 
\end{figure}

Figures \ref{fig:Dens_conv} and \ref{fig:Temp_conv} plot the density slices and temperature slices (taken in the $z=0$ plane) of the lowest and highest resolution runs respectively. Comparing between $3\times 10^6$ particles and $12\times 10^6$ particles, both the temperature and density slices show a high level of agreement, particularly in the location and of the hot, low density chimneys. This agreement is in contrast to figures \ref{fig:Dens_Nqy} and \ref{fig:Temp_Nqy}. Given the Nyquist frequency was altered according to the particle resolution in each case, this indicates the initial turbulent velocity field is playing a lesser role in the generation of chimneys and the ultimate fate of the gas in the cloud. The spatial extent of the chimneys appears to increase with resolution. This is explored in figures \ref{fig:unbound} and \ref{fig:Conv_TempVdens} below.

The top plots of Fig. \ref{fig:unbound} show the time evolution of the fraction of the initial gas mass that has been unbound in each simulation, minus the corresponding unbound mass fraction of the lowest resolution run (3 million particles). Moreover the bottom plots of Fig. \ref{fig:unbound} show the fraction of the gas that has been unbound and subsequently expelled from each simulation domain. It is clear the main difference between the lowest resolution run and the others is the time delay in the majority of the gas being unbound once feedback kicks in at 7.5\,Myr. This is indicated by the peak in the top row plots of Fig. \ref{fig:unbound} at this time. It also appears the unbinding occurs faster in runs with higher resolution. However, beyond this time all runs show good agreement. Furthermore, the amount of gas that has been both unbound and ejected from the simulation (by reaching a radius of $5$\,kpc) increases with resolution. This is expected since the higher mass particles of the lower resolution runs will have lower velocities, arising from the fact the kinetic energy is not varying while the gas mass is. Overall, these plots indicate the agreement between the energetics of the gas (thermal, kinetic, potential) is in good agreement across varying resolution.

We also plot the temperature of the gas versus the density and render this according to particle number in Fig. \ref{fig:Conv_TempVdens}. We see the bulk of the gas mass at each resolution is at $\sim 10^4$\,K, ranging between $10^{-30}-10^{-24}$\,gcm$^{-3}$. However, there exists a low temperature ($<100$\,K), high density ($10^{-22}$\,gcm$^{-3}$) tail in the lowest resolution run ($3\times 10^6$ particles), that does not exist in the run with the highest particle resolution (12$\times 10^6$ particles). This tail represents $>1$\% of the gas mass and indicates the higher resolution runs are marginally more effective at quenching star formation. Physically, the difference in the amount of high density, cool gas with resolution can be explained by the higher particle masses in the lower resolution run - which can form higher mass cold clumps that can effectively cool and hence are more resistant to heating than the lower mass clumps in the higher resolution runs.
\begin{figure}
  \psfig{file=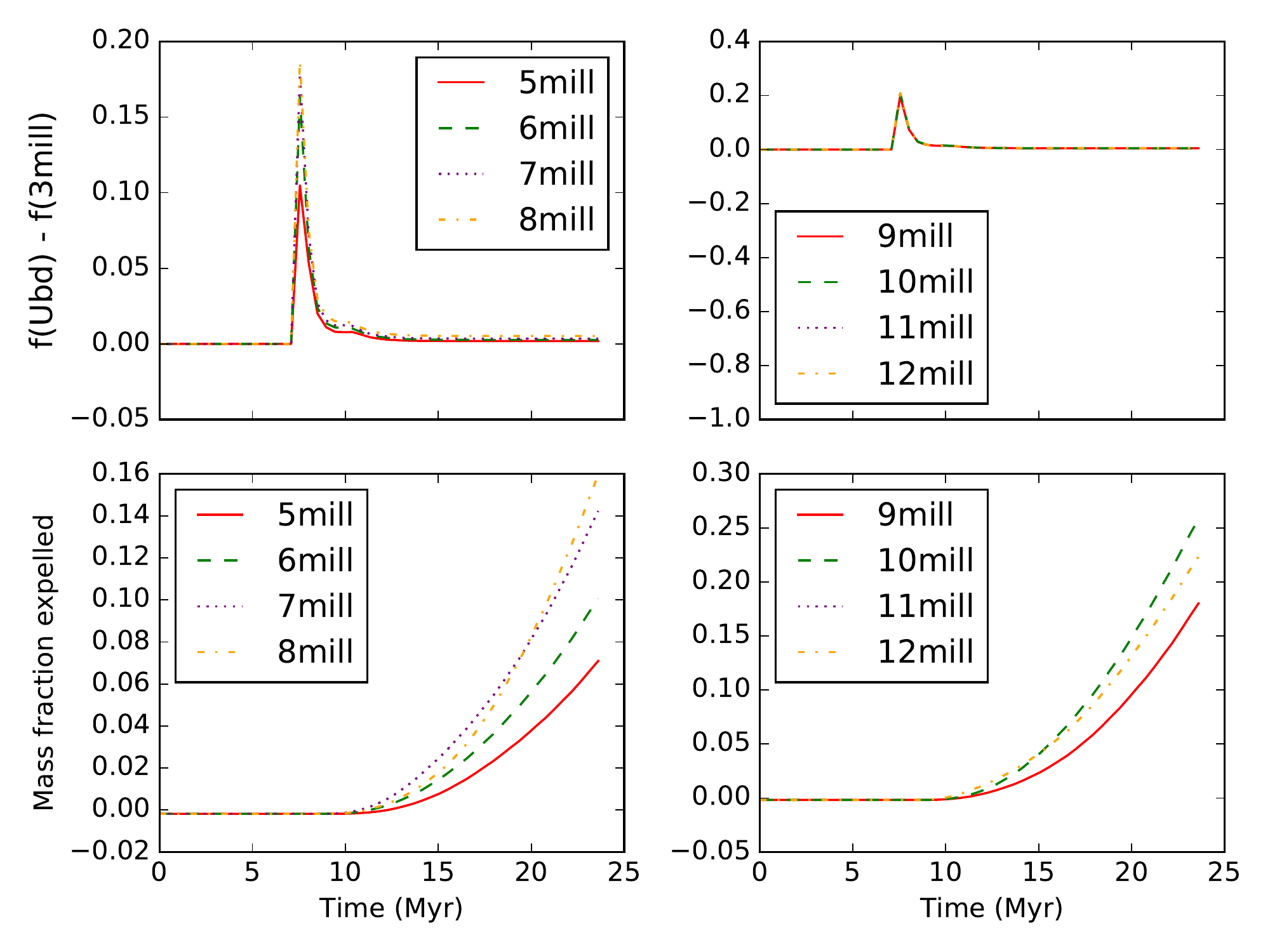,width=0.5\textwidth,angle=0}
  \caption{Top row: the time evolution of the fraction of the initial gas mass that is unbound - f(Ubd) - in the convergence tests of varying resolution minus the unbound mass fraction of the lowest resolution run - f(3mill), 3 million particles. Bottom row: when gas particles are both unbound and reach a radius of $5$\,kpc, they are expelled from the simulation. This plot shows the fraction of the initial gas mass that has been expelled from each simulation. Here 3mill denotes $3\times 10^6$ particles.}
  \label{fig:unbound} 
\end{figure}
\begin{figure}
  \psfig{file=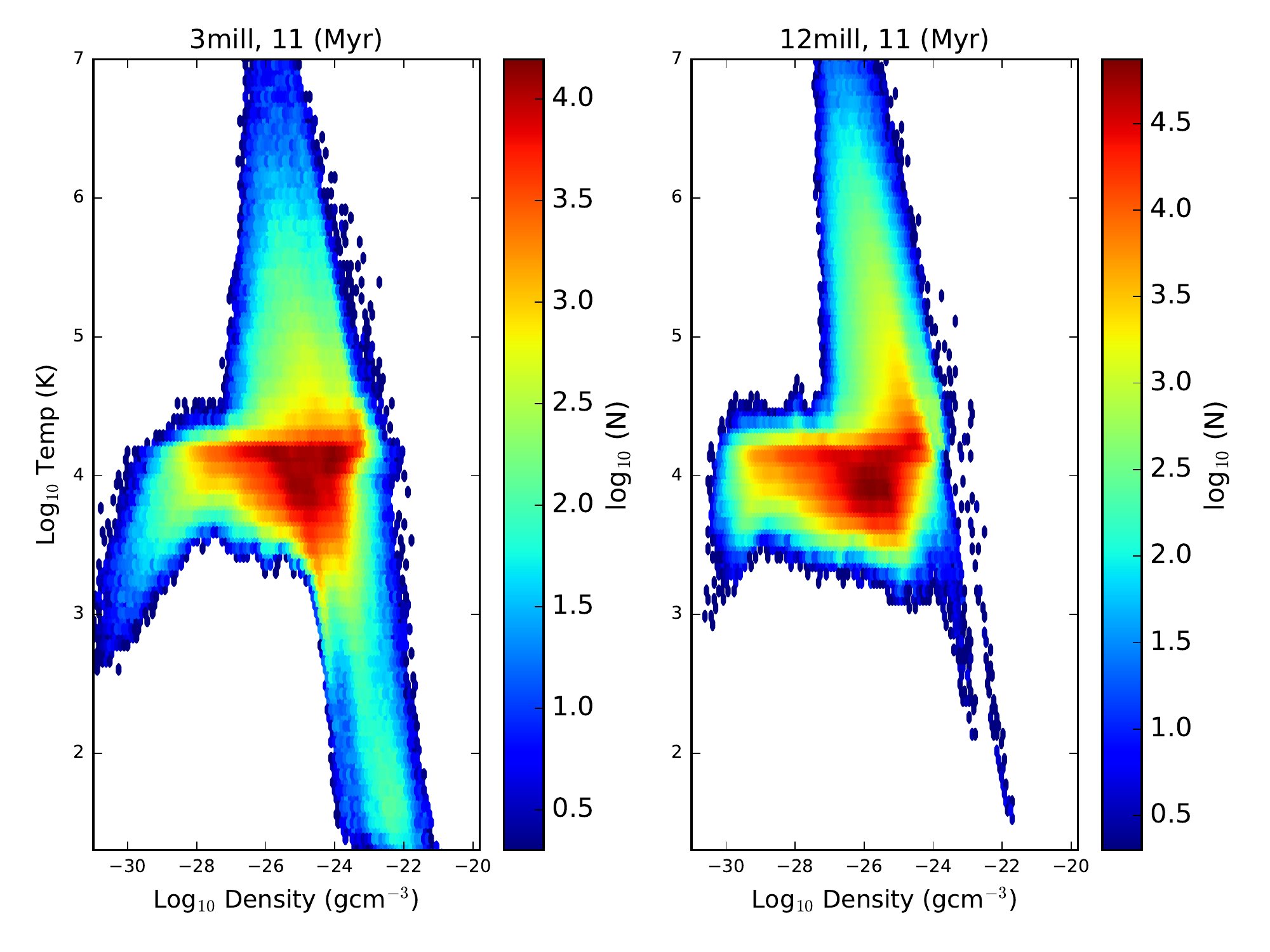,width=0.5\textwidth,angle=0}
  \caption{Plots to show temperature versus density, rendered according to particle number, at the free-fall time ($\sim 11$\,Myr) of clouds of varying resolution. Here 3mill denotes $3\times 10^6$ particles.}
  \label{fig:Conv_TempVdens} 
\end{figure}

Overall, our results suggest good agreement between different resolutions, down to the $\sim 1$\% level (grouping all unbound gas together). We can therefore conclude our results runs are at a resolution that shows a good degree of numerical convergence with runs of higher (and lower) resolution, particularly in terms of the multiphase ISM (as is seen in Fig. \ref{fig:Conv_TempVdens}). 

\bsp
\label{lastpage}
\end{document}